\documentclass[pdftex]{ws-rv9x6}
\usepackage[pdftex,colorlinks=false,bookmarks=true, pdfauthor=Misguich_and_Lhuillier]{hyperref}
\usepackage{rotating_rv}

\newcommand{\La}{\line (1,0  ){12}}
\newcommand{\Lb}{\line (3,5 ){6}}
\newcommand{\Lc}{\line (-3,5 ){6}}
\newcommand{\Ld}{\line (-1,0){12}}
\newcommand{\Le}{\line (-3,-5){6}}
\newcommand{\Lf} {\line(3,-5){6}}
\newcommand{\C} {\circle*{4}}

\newcommand{\pA}{\put(-6,-10)}
\newcommand{\pB}{\put(6,-10)}
\newcommand{\pC}{\put(12,0)}
\newcommand{\pD}{\put(6,10)}
\newcommand{\pE}{\put(-6,10)}
\newcommand{\pF}{\put(-12,0)}
\newcommand{\pZ}{\put(0,0)}

\newcommand{\Hex}{\pA{\C}\pB{\C}\pC{\C}\pD{\C}\pE{\C}\pF{\C}}

\newcommand{\pG}{\put( 18,-10)}
\newcommand{\pH}{\put( 18,10)}
\newcommand{\pI}{\put(0,20)}
\newcommand{\pJ}{\put(-18,10)}
\newcommand{\pK}{\put(-18,-10)}
\newcommand{\pL}{\put(0,-20)}

\newcommand{\KagHex}{\pA{\C}\pB{\C}\pC{\C}\pD{\C}\pE{\C}\pF{\C}}
\newcommand{\KagStar}{\KagHex\pG{\C}\pH{\C}\pI{\C}\pJ{\C}\pK{\C}\pL{\C}}

\newcommand{\Z}{\mathbb{Z}_2}

\usepackage{color}
\usepackage[utf8]{inputenc}

\begin{document}

\title{Two-dimensional quantum antiferromagnets}

\chapter*{TWO-DIMENSIONAL QUANTUM ANTIFERROMAGNETS}
\markboth{G. Misguich and C. Lhuillier}{Two-dimensional
quantum antiferromagnets}

\author{Gr\'egoire Misguich}
\address{
Institut de Physique Théorique,\\
CEA, IPhT, F-91191 Gif-sur-Yvette, France\\
CNRS, URA 2306, F-91191 Gif-sur-Yvette, France\\
E-mail: gregoire.misguich@cea.fr}

\author{Claire Lhuillier}
\address{Laboratoire de Physique Th\'eorique de la matière condensée\\
Universit\'e P. et M. Curie and UMR 7600 of CNRS\\
Case 121, 4 Place Jussieu, 75252 Paris Cedex, France\\
E-mail: claire.lhuillier@upmc.fr}


\setcounter{footnote}{0}

\setcounter{figure}{0}

\setcounter{equation}{0}

\setcounter{section}{0}

\begin{abstract}
This review presents some theoretical advances in the field of quantum magnetism in two-dimensional systems, and quantum spin liquids in particular.
It is to  be published as a chapter in the second edition of the book ``Frustrated spin systems'', edited by H. T. Diep (World-Scientific).
The section (Sec. 7) devoted to the kagome antiferromagnet has been completely rewritten/updated, as well as the concluding section (Sec. 8).
The other sections (Secs. 1-6) are  unchanged from the first edition of the book (published in 2005). 
\end{abstract}

\tableofcontents


\section{Introduction}

In this review we  present some theoretical  advances in the
field of quantum    magnetism in      two-dimensional   (2D)
systems.      The spin-$\frac{1}{2}$ nearest-neighbor
2D Heisenberg models on Bravais lattices
(square\cite{manousakis91}, triangular\cite{bllp94}) are  N\'eel
ordered\footnote{This  generic  kind   of order,  with  a
macroscopic sublattice magnetization  is called in the following
{\it magnetic  }  long-range  order  (LRO),  in contrast to
other ordered phases where the long-range ordered correlations
concern $S=0$ scalar observables (on dimers, quadrumers...)}  at
$T=0$.  Frustration, small coordination  number, competition
between   interactions can  lead  to specific quantum phases
without  magnetic long-range order.   Since a decade this subject
is a highly debated issue in  the field of quantum magnetism.
It   was revived  by   the    discovery  of  high-$T_c$
superconductivity  in  the  doped   cuprates  and  fuelled by
numerous experimental studies of 2D antiferromagnetic
insulators.\cite{HFM2000}

Section~\ref{sec:j1j2} is devoted to the spin-$\frac{1}{2}$ Heisenberg
model     on the  square   lattice   with  first- and  second-neighbor
interactions  ($J_1$--$J_2$ model).  This  model  is one  of the  most
studied  in  the field of  and this  section is a   short guide to the
literature, with a  special emphasis on  the various methods used  for
this problem.

Section~\ref{sec:VBC} deals  with general properties  of  valence-bond
crystals (VBC) and   related  states, the  simplest  phase which  is
commonly realized in frustrated spin systems without magnetic LRO.

In  section~\ref{sec:largeN}  we present  large-$N$
generalizations of the Heisenberg model.  This approach was
extensively developed by Read and Sachdev from the   early 90's
and has  been  the first to give  an insight into  the alternative
between  VBC  and related phases, which have long-range  order in
local  singlet patterns (whence the name of crystals), and
resonating valence-bond  (RVB)\index{resonating valence-bond, RVB}
spin-liquids (SL) which are translationally invariant phases where
the  quantum coherence is a central issue.

Section~\ref{sec:QDM} presents  some  results of quantum  dimer
models\index{quantum dimer models, QDM} (QDM).  These models are
effective approaches to the quantum phases of antiferromagnets
which are dominated by short-range valence-bonds (or singlets).
They have received recently some special  attention  and provide
useful insights onto the phenomenology of VBC and RVB SL.

In Section~\ref{sec:MSE} we review some results concerning models
with multiple-spin  exchange\index{multiple-spin exchange, MSE}
(MSE) (also called ring exchange) interactions.  These
interactions are now recognized to be present in several physical
systems  and appear to  play an important role in the
stabilization of RVB liquid ground states.

The  last section  is devoted to  the Heisenberg  model  on the kagome
lattice  (and  related models).  Despite of  an  important activity on
this subject,  the   understanding of the  low-energy   physics of the
spin-$\frac{1}{2}$  kagome   antiferromagnet remains   a   challenging
problem and we discuss some of the important results and questions.

We should warn the readers that this review is quite ``inhomogeneous''
and            cannot,          of       course,               replace
textbooks.\cite{fradkinbook,auerbachbook,tsvelikbook,sachdevbook}
While some parts deal  with some rather recent  works (QDM or  MSE for
instance), some others are devoted to older results which we think are
still of importance   for current research  (beginning  of the section
$J_1$--$J_2$, large-$N$).  The  final part devoted  to kagome reflects
our   own views  and some  unpublished    material on still  unsettled
issues. Some parts  are intended to  be more pedagogical and  concrete
(QDM and beginning   of large-$N$ section)  while  some others contain
more  qualitative  discussions  of the  physical  issues  (end of  the
section $J_1$--$J_2$, VBC, kagome).

\section{$J_1$--$J_2$ model on the square lattice}\index{J1-J2 model}
\setcounter{footnote}{0}
\label{sec:j1j2}

We consider the following Heisenberg model on the square lattice:
\begin{equation}
    {\cal H} =  2 J_1 \sum_{\langle ij\rangle}  {\vec S}_i\cdot{\vec S}_j
    +  2 J_2 \sum_{\langle\langle ij \rangle\rangle}  {\vec S}_i\cdot{\vec S}_j
\label{J1J2}
\end{equation}
where $\langle ij\rangle$ and $\langle\langle ij \rangle\rangle$
denote pairs of nearest and next-nearest-neighbors respectively.
Although  quite simple in appearance,  this spin model realizes
several interesting   phenomena which are    relevant   to a large
class   of 2D  frustrated    quantum magnets: classical
degeneracy, order by disorder,\index{order by disorder}
destruction of some long-range order by quantum fluctuations,
break down of the spin-wave expansion, opening of a spin gap and
(possibly ?) spontaneous translation symmetry breaking, etc. For
this reason we start with a general overview of  some important
results concerning this system.  We  will   focus on  the
properties related  to {\em frustration}.  A  review  on the
non-frustrated model ($J_2=0$) can be found in
Ref.~\cite{manousakis91}

\subsection{Classical ground state and spin-wave analysis}
It  is  easy  to find    {\em    some} classical ground state  of    a
translation   invariant  Heisenberg  model  on   a Bravais lattice
because the energy can always be minimized by a planar helix
\begin{equation}
    \vec{S}_{\bf    r}=
        \vec{e}_1 \cos ({\bf q} \cdot  {\bf r})
    +   \vec{e}_2 \sin ({\bf q}\cdot {\bf r})
\end{equation}
provided  that the pitch  ${\bf q}$  minimizes  the Fourier  transform
$J({\bf q})$ of the coupling.\cite{lt47}  In the case of the $J_1$--$J_2$
model one has
\begin{equation}
    J({\bf q})=2J_1\left(\cos(q_x)+\cos(q_y)\right)
    +2J_2\left(\cos(q_x+q_y)+\cos(q_x-q_y)\right)
\end{equation}

\begin{itemize}

\item $J_2<0.5J_1$:
$J({\bf q})$ has a single minimum at $(\pi,\pi)$. It corresponds to the
``usual'' N\'eel state.

\item $J_2>0.5J_1$: $J({\bf q})$ has two isolated minima at
$(0,\pi)$ and $(\pi,0)$.  They correspond to ferromagnetic   lines
(resp.  columns)  arranged  in an antiferromagnetic   way.  These
states are    sometimes  called {\em collinear}  (in {\em real}
space).  From these planar helix states one can build many other
ground states  by rotating globally all the spins of one
sublattice with  respect to the other.   Although this costs no
energy  for classical  spins at  zero  temperature, it is known
(order by disorder, see  below)\index{order by disorder} that
configurations where both sublattices have  their staggered
magnetization {\em collinear  in spin space} are selected by
thermal or quantum fluctuations.

\item $J_2=0.5J_1$:
$J({\bf q})$  has  lines of minima  around  the edges of the  Brillouin
zone. At this point the classical ground state  is highly degenerate :
We can write $\mathcal{H}={\rm cst}+J_2 \sum
\left(S_1+S_2+S_3+S_4\right)^2$ where the   sum runs over   all square
plaquettes and any  state where each plaquette  has a vanishing  total
spin minimizes the classical energy.
\end{itemize}

Even   at the  lowest    order   in $1/S$, zero-temperature    quantum
corrections to the  sublattice magnetization (order parameter) diverge
around $J_2=0.5J_1$  (Chandra    and  Dou{\c  c}ot\cite{cd88}).   Such
large-$S$ approximation usually tends to overestimate the stability of
magnetic phases, therefore this breakdown around $J_2\sim 0.5J_1$ is a
strong  evidence for the existence  of quantum  disordered phase(s) in
this region of parameter space.

\subsection{Order by disorder ($J_2>J_1/2$)}\index{order by disorder}

The concept of ``order by disorder'' was introduced in 1980 by Villain
and  co-workers\cite{vbcc80} in the study  of a frustrated Ising model
on the square lattice. In this model the next neighbor couplings along
the rows are ferromagnetic as well as those on  the odd columns (named
$A$ in the following).  The couplings  on the even columns (named $B$)
are  antiferromagnetic.  At  $T=0$   the ground state has   no average
magnetization   and  is   disordered.    This  changes when    thermal
fluctuations are introduced:    a $B$-chain sandwiched  between  two A
chains  with parallel  spins  has {\em lower  excitations}  than a $B$
chain between two $A$-chains with  anti-parallel spins.  This gives  a
larger Boltzmann weight  to ferrimagnetically ordered states.  Villain
{\it  et al.}    have  exactly  shown    that  the system is    indeed
ferrimagnetic at  low temperature.  They were  also able  to show that
site dilution (non-magnetic  sites) selects the  same ordered pattern,
whence the name  of ``order by disorder''.

A  somewhat less drastic  phenomenon  has  been  observed in   quantum
systems.  It is   the  selection  of particular   long-range  ordered
quantum states  among a larger family of   ordered solutions which are
classically  degenerate  at $T=0$.\footnote{In   Villain's  model  the
system   is truly  disordered  at $T=0$   and  an  ordered solution is
entropically  selected at     finite  temperature.   In   the  quantum
$J_1$--$J_2$ model  above, the classical   solutions can adopt various
ordered patterns: quantum fluctuations select among these patterns the
most ordered one, that is the situation  with the highest symmetry and
the  smallest degeneracy.    The   ultimate effect  of  these  quantum
fluctuations can be the destruction of the N\'eel  order in favor of a
fully   quantum  ground state  with  $\mathcal{O}(1)$     degeneracy.}
Consider a spin system in  which the  molecular  field created by  the
spins of one  sublattice on the other  cancels, which is the case when
$J_2>0.5J_1$. Shender\cite{shender82}  showed that if fluctuations are
included,    the system will  select states   in   which all spins are
collinear to each  other.  This follows  from the fact that (moderate)
fluctuations at   one site are  orthogonal  to the mean  value  of the
magnetization at  that site  and  the  system can  gain  some magnetic
exchange energy  by  making such fluctuations coplanar  on neighboring
sites, that  is to making  the spins collinear.   Such a  selection of
order by   quantum   fluctuations  (and dilution)   was   discussed by
Henley\cite{h89}  and appears also   quite straightly  in a  spin-wave
expansion.\cite{mdjr90}

This  selection of the   $(\pi, 0)$ or  $(0,\pi)$ order spontaneously
breaks a four-fold lattice symmetry.  An Ising order parameter is thus
generated.  It  takes two values  depending whether  the ferromagnetic
correlations are locally arranged  horizontally or vertically. Chandra
and co-workers\cite{ccl90a} have  studied this mechanism and predicted
the  existence of   a   finite  temperature   Ising  phase  transition
independent  of   the    subsequent  development  of     a  sublattice
magnetization.  This result has been questioned recently\cite{szosh03}
and  the  transition    has    not been observed   so      far in  the
spin-$\frac{1}{2}$ model.\cite{mbp03,szosh03}  It    has however  been
confirmed by some   recent    Monte Carlo simulations of     the  {\em
classical}   Heisenberg model.\cite{wm03}  Very  similar phenomena are
present in the   $J_1$--$J_2$  quantum Heisenberg  model  on  the {\em
triangular} lattice.\cite{jdgb90,cj92,k93,lblp95}

Melzi   {\it   el   a.}\cite{melzi00,melzi01}  have  studied a   quasi
2D spin-$\frac{1}{2}$ system  which  is believed to be  a
$J_1$--$J_2$ square  lattice Heisenberg antiferromagnet.  They found some
evidence (splitting   of NMR  lines)  for a   collinear ($(\pi,0)$  or
$(0,\pi)$)              magnetic        ordering.              Several
estimates\cite{melzi01,rosner02,rosner03,mbp03} indeed  point to $J_2>J_1$    in
this compound.

\subsection{Non-magnetic region ($J_2\simeq J_1/2$)}

Consider the two classical ``Ising  states'' corresponding to the wave
vectors $(\pi,\pi)$ and $(\pi,0)$.    They  can be taken  as   (crude)
variational states for the Hamiltonian Eq.~\ref{J1J2}.  Their energies
(per  site) are  $E_{\pi,\pi}=-J_1+J_2$   and  $E_{\pi,0}=-J_2$.    As
discussed above, these states cross at $J_2=\frac{1}{2}J_1$.  However,
one  can also consider  any  first-neighbor singlet (or  valence-bond)
covering of the lattice  as   another variational  state.  In such   a
completely dimerized state the  expectation  value of the  energy  per
site is $E_{dimer}=-\frac {3}{4}J_1$,   which is below the  two  Ising
states around $J_2\simeq J_1/2$.  Of course this very simple argument
does not  prove anything  since  ``dressing'' these classical  states
with quantum fluctuations (spin   flips in the N\'eel-like  states  or
valence-bond motions in the  dimerized wave functions) will  lower the
energies of all these trial  states and it  is absolutely not clear
which  one may eventually win.   Nevertheless,  this shows in a simple
way why non-magnetic states  ({\it i.e} rotationally invariant or spin
singlet) such as dimerized states can be a route to minimize the energy in
a   frustrated   magnet.\footnote{Klein\cite{klein82}  introduced   a
general procedure to generate local and $SU(2)$ symmetric Hamiltonians
for which any first-neighbor dimerized state is an exact ground state.
These Hamiltonians   are simply defined  as  sums of  projectors which
annihilate all dimer  coverings.   The Majumdar-Gosh\cite{mg69} chain
is the simplest example of a ``Klein model''.}

\subsubsection{Series expansions}

High-order  series  expansions\index{series  expansions}  can   be
a powerful technique   to investigate frustrated quantum magnets.
The general method to generate zero-temperature perturbation
expansions  in quantum many-body systems was described  by Singh
{\it et al.}\cite{sgh88} and Gelfand {\it et al.}\cite{gsh90} For
instance,  one   can  consider  the following anisotropic model:
\begin{eqnarray}
    {\cal H}(\lambda) =  &&2 J_1 \sum_{<ij>}
    \left[
    S^z_i S^z_j
    + \lambda \left(S^x_i S^x_j + S^y_i S^y_j \right)
    \right] \nonumber \\
    && +  2 J_2 \sum_{<<ij>>}
    \left[
    S^z_i S^z_j
    + \lambda \left(S^x_i S^x_j + S^y_i S^y_j \right)
    \right]
\end{eqnarray}
${\cal H}(\lambda=0)$ is a classical Ising model which ground state is
known. The series expansion about  the Ising limit amounts to  compute
expectation values in the ground state $\left|\lambda\right>$ of ${\cal
H}(\lambda)$ in powers of $\lambda$:
\begin{equation}
    \frac{
    \left<\lambda\right| {\hat O} \left|\lambda\right>
    }{
    \left<\lambda | \lambda\right>}=\sum_n a_n \lambda^n,
\end{equation}
(energy  gaps, dispersion  relations and susceptibilities  can also be
computed in the same approach).  The calculation of $a_n$ requires the
enumeration  and evaluation  of  the {\em connected  clusters} of size
$\sim n$, whose number grows exponentially with $n$.  Depending on the
quantity ${\hat  O}$  and  on the model,    orders from  7 to   20 can
typically be  obtained   on present computers.   The   series  is then
extrapolated to  $\lambda=1$  by  standard  Pad\'e, Dlog   Pad\'e   or
integrated differential approximations.  Such a series expansion about
the  Ising limit was done  by Weihong~{\it et al.}\cite{woh91} for the
first  neighbor   square-lattice   antiferromagnet.     Oitmaa     and
Weihong\cite{ow96}  extended  the  series  to the $J_1$--$J_2$  model,
where  each $a_n$ is   now  a polynomial   in  $J_1$ and  $J_2$.   The
disappearance of N\'eel order in the Heisenberg model manifests itself
by   a  vanishing sublattice magnetization as    well as some singular
behavior of  the series for  $\lambda_c<1$.   The results indicate the
absence of N\'eel long-range  order in the strongly frustrated region
$0.4\le  J_2/J_1\le  0.6$.   Such  an   expansion can  locate  with  a
reasonable accuracy  the phase boundary  of the N\'eel ordered regions
but provides no direct information  on the nature of the  non-magnetic
phase.

To study the model around $J_2\simeq  J_1/2$, several other expansions
have been carried  out.  A dimer  expansion about an exactly dimerized
model was       done       by  Gelfand~{\it    et    al.},\cite{gsh89}
Gelfand,\cite{gelfand90} Singh {\it   et   al.}\cite{swho99} and Kotov
{\it et al.}.\cite{kosw00}  In this approach $J_1$  and $J_2$ are set
to zero everywhere except on isolated bonds arranged in a columnar way
and all  the other  couplings are  treated perturbatively.   At zeroth
order  the  ground state is simply  a  product of singlets.   In these
calculations the dimerized   phase remains stable  in the intermediate
region.  Singh  {\it et al.}\cite{swho99}  also performed a different
kind of zero-temperature series expansion.  They  started from a model
of isolated  4-spin plaquettes in order to  check a prediction made by
Zhitomirsky  and    Ueda\cite{zu96}   that   such   plaquettes  could
spontaneously  form around $J_2\simeq  J_1/2$ to produce a state which
is   invariant    under $\pi/4$  lattice   rotations.     Although the
ground state energy they  obtained is very  close to the  one obtained
from   the  dimerized limit  (within   error bars of the extrapolation
procedure) they observed an instability in the plaquette scenario (the
triplet gap vanishes   before  reaching the isotropic   square-lattice
model)   which suggests  that plaquette  order  is not  the issue (the
analysis of  the exact numerical  spectra for 36  sites confirmed this
result\cite{cbps01}).

Sushkov {\it et al.}\cite{sow02}  (improved numerical results compared
to Ref.\cite{sow01})  computed the susceptibility  $\chi_D$ associated
with the appearance of columnar dimer  order in the $(\pi,\pi)$ N\'eel
phase  by   a series   expansion   about  the   Ising limit.   Such  a
susceptibility  seems    to       diverge  at    $J_2/J_1=g_{c1}\simeq
0.405\pm0.04$.   On the other hand  the  disappearance of the magnetic
LRO is   observed (through the N\'eel  order  parameter or through the
anisotropy   in   spin space    of   the  spin-spin  correlations)  at
$J_2/J_1=g_{c2}\simeq 0.39\pm0.02$.  This point could a priori be {\em
different} from $g_{c1}$.  In such a case the system would first break
the $\pi/4$ lattice rotation  symmetry at $g_{c1}$, while magnetic LRO
remains (gapless  spin waves).   Only  at $g_{c2}>g_{c1}$  the $SU(2)$
rotation  symmetry is restored and the  magnetic excitations acquire a
gap.  From  field  theoretical  arguments based  on  effective actions
valid close to  the critical points,  Sushkov {\it et al.}\cite{sow02}
argue   that  the proximity  (or  possible  equality)  of $g_{c1}$ and
$g_{c2}$ is a general  feature  in frustrated magnets which  originate
from the coupling of triplet and singlet excitations.

Sushkov  {\it   et al.}\cite{sow01}  computed  susceptibility $\chi_P$
associated  to plaquette  order by an   expansion around the dimerized
limit, assuming that  the system  has  columnar dimer  LRO. The result
shows a  divergence of  $\chi_P$ when  $J_2/J_1\to g_{c3}=0.5\pm0.02$.
From these results Sushkov {\it et al.} suggested that the translation
symmetry along the columns is broken  down at $g_{c3}$ (giving rise to
an   eight-fold  degenerate ground state in  the  thermodynamic limit)
before  the $(\pi,0)-(0,\pi)$   magnetically ordered phase  appears at
$g_{c4}\simeq0.6$. This picture  is qualitatively consistent  with the
spin-spin    correlations computed in  a  $10\times10$   system with a
density matrix renormalization group (DMRG) algorithm.\cite{cls00}

Due to the relatively short series (typically of order 7) involved and
the uncertainties in the extrapolation procedures, such results should
be confirmed  by  other methods  but this succession  of quantum phase
transitions  represents a very interesting scenario.   We note that if
the model has a fully symmetric  liquid ground state in some parameter
range, it should be difficult to  capture from series expansions about
limits where some lattice symmetries are explicitly broken.

\subsubsection{Exact diagonalizations}
Exact diagonalizations \index{exact diagonalizations} have a
priori no bias, and were used very early in      this
field.\cite{dm89,fkksrr90,pgbd91} Large-size  computations and
sophisticated finite size scaling analysis are nevertheless needed
to extract   significant information.        Schulz {\it     et
al.}\cite{schulz} performed extensive exact diagonalizations of
the $J_1$--$J_2$ model  for system sizes up to  $36$ sites.  They
analyzed the behavior of several quantities such as structure
factors (N\'eel order parameter), ground state energy, spin-wave
velocities (obtained from the   finite  size corrections   to the
ground state energy), spin  stiffness and uniform susceptibility.
Their analysis, including  quantitative comparisons   with
non-linear sigma   model predictions,\cite{nlsm}  concluded to the
absence of N\'eel long-range order in  the strongly frustrated
region $0.4\le J_2/J_1\le0.6$.   There, they   show  enhanced
columnar dimer-dimer correlations  as  well as  chiral ones but
the size effects were not clear  enough to discriminate between
short  or long-range order for these order parameters.

\subsubsection{Quantum Monte Carlo}

Quantum Monte Carlo (QMC) \index{quantum Monte Carlo} methods have
been extensively applied to the $J_1$--$J_2$  model  in the low
frustration regime giving an highly accurate description of the
N\'eel phase (Sandvik\cite{s97} and Refs. therein).  In the
non-magnetic  and highly frustrated regime a  simple QMC approach
is ineffective due to the  so-called sign problem. The fixed node
approach is the first answer  to this  problem: the exact
imaginary time propagator   $e^{-\tau\mathcal{H}}$ used to  filter
out the ground state from a variational guess $|\psi_g\rangle$ is
replaced by  an    approximate  propagator,  which   has  the same
nodes as $|\psi_g\rangle$.  The quality of the result depends on
the quality of the nodal regions of $|\psi_g\rangle$. Various
schemes have been used to    try to  go  beyond  this limitation:
stochastic reconfiguration (Sorella\cite{s98}), eventually
associated  to  a   few    Lanczos iterations.\cite{s01,cbps01} An
alternative method has been devised by du Croo de Jongh {\it et
al.},\cite{cls00} where the guiding function is replaced  by the
result of  a  DMRG calculation.\cite{w92,w93} Both methods have
their  own bias. Using  the first of them, Capriotti and
Sorella\cite{cbps01} concluded  that  for   $J_2/J_1\sim  0.45$ a
Gutzwiller-projected   BCS  wave function $|p\;BCS\rangle$  was an
excellent guiding wave function:
\begin{eqnarray}
    |p\;BCS\rangle&=&\hat\Pi\left|BCS\right>  \\
    |BCS\rangle&=&\exp{\left(
        \sum_{i,j}f_{i,j}
        c^\dag_{i\uparrow}c\dag_{j\downarrow}
    \right)}|0\rangle
\end{eqnarray}
where      $\left|0\right>$    is     the        fermion        vacuum,
$c^\dag_{i\uparrow}c\dag_{j\downarrow}$ creates  a     valence-bond on
sites  ($i,j$)  and   $\hat\Pi$  projects   out  states   with  double
occupancy. The pairing  amplitude $f_{i,j}$ (often called gap function
$\Delta_k$)\footnote{After the  Gutzwiller projection $\Delta_k$ is no
longer the observable gap.\cite{gl91}} is optimized with a Monte Carlo
algorithm in order to minimize the energy.  Capriotti and Sorella gave
convincing indications that their wave function is quite accurate. The
best variational energies are obtained in the frustrated region with a
pairing amplitude    which     mixes   $d_{x^2-y^2}$ and      $d_{xy}$
symmetries. In particular it reproduces the correct nodal structure of
the ground state in the frustrated region at least for moderate system
sizes where the  variational result can be  checked against the  exact
result.   This is a   subtle and non-trivial  information for  systems
which do not  obey the Marshall's sign rule  as this frustrated model.
They  concluded from these results   that  the system probably had  no
long-range order neither in dimer-dimer correlations nor in four-spin
plaquette correlations.  On  the other hand,  du Croo de Jongh {\it et
al.} argued in favor of columnar dimerized phase  which also break the
translation  symmetry along  the columns (plaquette-like  correlations
similar to those found by series expansions\cite{sow01}).

The comparison of the results of these different approaches shows that
this problem  remains   a very challenging  one.   The  model   in the
frustrated regime  is probably never very  far from a quantum critical
point and in these conditions none of the available methods seems able
to discriminate between a VBC with  tiny gaps both  in the singlet and
triplet sectors, a critical  phase with a quasi   order in dimers  and
gapless singlet excitations, or a  true SL with  gaps in any sector of
spin but no long-range order in any observable. As we will explain in
the following sections some other frustrated models are happily deeper
in the non-magnetic phases and exhibit quantum phases which are easier
to characterize.


\section{Valence-bond crystals}\label{sec:VBC}\index{valence-bond crystals, VBC}
\setcounter{footnote}{0}
\subsection{Definitions}
Among the different quantum  solutions to overcome frustration the VBC
is  the simplest scenario.   In this phase,  neighboring spins arrange
themselves in  a regular pattern  of singlets: dimers,\footnote{whence
the  name   Spin Peierls  phase sometimes  given   to the  VBC phase.}
quadrumers or 2$n$-mers  $S=0$ plaquettes.  The  stability of this phase
comes from the extreme stability of  small $S=0$ clusters (recall that
the  energy  of a  singlet  of two spins  $\frac{1}{2}$  is -3/4 to be
compared to the energy of two classical (or Ising) spins which is only
-1/4), and eventually from the  fact that frustrated bonds between two
different singlets do not contribute to the total energy.

In a  VBC phase there is no  $SU(2)$ symmetry breaking, no long-range
order in spin-spin correlations,  but long-range order in dimer-dimer
or  larger singlet units.  Except  at  a quantum  critical point,  all
excitations  of a VBC are gapped.   Depending on the lattice geometry,
such a wave  function can  spontaneously  break some  lattice symmetry
({\em  spontaneous VBC}) or can  remain fully symmetric ({\em explicit
VBC}).   In a  strict sense,  the name  VBC  should be  reserved  for
systems with a spontaneous lattice  symmetry breaking. However,  since
these two kinds  of systems share   many similarities we  will discuss
both in this section.

When the Hamiltonian has some  inequivalent bonds and an integer  spin
in the unit cell (even number  of spin-$\frac{1}{2}$ for instance) the
system can take  full advantage of the  strong bonds and  minimize the
effects of the frustrating ones.  In that case  the {\em explicit} VBC
is  the ``natural'' strong coupling solution.   One can build a simple
Hamiltonian in which the bonds  which are not  occupied by the singlet
objects  are   turned off.  The   resulting  model is a   set of small
decoupled clusters (dimers  or larger plaquettes) and the ground state
is a trivial product  of singlets.  Importantly, this strong  coupling
limit has  the same lattice symmetry  as the original one.  Going back
to   the  original Hamiltonian {\em  no   quantum  phase transition is
encountered when  going from  the trivial  singlet  product up to real
interacting ground state}. Models with an half-odd-integer spin in the
unit cell cannot realize a VBC unless they {\em spontaneously} enlarge
their unit  cell.   In these  situations  there is no   unique elected
position for the 2$n$-mers and a symmetry  breaking must take place in
order to form a VBC.  Examples of these two kinds of VBC will be given
below.


\subsection{
One-dimensional and quasi one-dimensional examples (spin-$\frac{1}{2}$
systems)}

One of the  simplest example of  (spontaneous) VBC is observed in  the
$J_1$--$J_2$ model  on  the chain for   $J_2/J_1 >  (J_2/J_1  )_c \sim
0.24$.\cite{mg69,ss81,h82,a89,ys97}  For $J_2/J_1=0.5$   the    doubly
degenerate   ground states        are  exact    products            of
dimers:\cite{mg69,auerbachbook}
\begin{equation}
|MG_{\pm}> =\prod_{n=1}^{N/2} |(2n, 2n \pm 1)\rangle.
\label{mg}
\end{equation}
Here   and in the  following we  call ``dimer''  a  pair of spins in a
singlet state, and note it:
\begin{equation}
|(i,j)> =\frac{1}{\sqrt2} \left[ |i, +\rangle|j, -\rangle  - |i, -\rangle |j, +\rangle\right].
\label{dimere}
\end{equation}

For  all $J_2/J_1 >  (J_2/J_1 )_c $,
the ground states are products
of dimers, dressed by fluctuations of valence bonds, dimer long-range
order persists in all the range of  parameters.  This model has gapful
excitations  which  can be   described     as pairs   of    scattering
spin-$\frac{1}{2}$       solitons      separating    the     two exact
ground states\cite{ss81} (these fractionalized      excitations    are
specific of the 1D chain).

The  Heisenberg chain  with alternating  strong and  weak  bonds (Spin
Peierls  instability),  has indeed  a  unique  ground state where  the
dimers  are mainly  located on  the strong  bonds. In  that  case, the
spin-$\frac{1}{2}$   excitations  are   confined  by   the  underlying
potential    and   the   true    excitations   are    gapful   integer
magnons~(\cite{nt97a,asrp99} and refs. therein). It is an explicit VBC.

A    two-leg ladder with   AF rung   exchange has   also a  unique VBC
ground state  and  gapped magnons as  excitations.\cite{dr96}  On  the
other hand   Nersesyan  and Tsvelik\cite{nt97,km98} have   proposed an
example  of frustrated  ladder,  with  a   spontaneously dimerized
ground state, and gapful  excitations.  Excitations of this last model
are identified as pairs of singlet and triplet domain walls connecting
the two ground states, they form a continuum.

As  can be seen  from this  rapid and  non exhaustive enumeration, VBC
ground states are relatively  frequent in  frustrated  one-dimensional
spin-$\frac{1}{2}$  models.  All   these systems are   gapful  but the
excitations could be of different nature emerging as modes (associated
to integer spin excitations) or continuum of pairs of excitations that
could be  fractionalized   (it  is then specific   of  one-dimensional
systems) or not.

\subsection{Valence Bond Solids}\index{valence bond solids, VBS}
\label{ssec:VBS}
The  VBS  wave function was introduced  by
Affleck, Kennedy, Lieb and Tasaki (AKLT).\cite{aklt87,aklt88} It
can be constructed whenever the spin $S$ on a site  is a multiple
of one half the coordination number $z$: $2S=0$ mod $z$.   Let us
consider the  simplest case $2S=z$.   In that case the local spin
$S$ can  be seen as the symmetric combination of $2S$ (fictitious)
spin-$\frac{1}{2}$.    Now on  each bond  of  the lattice    one
can    make    a  singlet    between   two   fictitious
spins-$\frac{1}{2}$.  Such a  product  of singlets does not belong
to the  physical Hilbert space  of the  original spin-$S$ model
but to a much larger space.  The VBS wave function is defined as
the projection of the singlet-product state onto the physical
space.  This projection amounts  to symmetrize (for all lattice
sites) the wave function with respect to the fictitious spins to
force them into a physical spin-$S$ state. A VBS can be viewed as
an {\em explicit  VBC of the fictitious spins}.  Simple
Hamiltonians with  short-range and $SU(2)$-symmetric interactions
for   which  the VBS is  an  exact   ground state can  be
constructed (sum of projectors\cite{aklt87,aklt88}).  By
construction the VBS wave function  is a  spin  singlet and {\em
breaks no lattice symmetry}.  By extension  we may say that a
system is in  a VBS {\em phase} if  its ground state can be
adiabatically transformed into the VBS wave function without
crossing a phase  transition.  As the  VBC, models in  the VBS
phase have a  gap  to all excitations\footnote{This may however
not always  be  true when  the coordination number  of the lattice
is large.\cite{aklt87} In such cases  the VBS wave function is
still a spin  singlet but has  long-range spin-spin correlations.
We do  not consider such  cases  here.}   but  their
wave functions   are slightly more   complex and their order
parameter   is non-local.  The order of VBS is associated to
long-range singlet-singlet correlations in the {\em fictitious
spins}.  Expressing such observable in terms of the physical spins
leads to a  non-local quantity  called {\em string order
parameter}.\cite{nr89,kt92} Contrary to  explicit VBC,  VBS have
fractionalized degrees of freedom at  the edges of the system with
open boundary conditions.  These are simply associated to the
unpaired fictitious  spins.  To our  knowledge  these properties
have not been explored in  quantum 2D systems.

The spin-1 Heisenberg chain is the prototype of VBS in
1D.\footnote{In 1D, some authors call ``Haldane
systems''\index{Haldane gap} all the spin-gapped phases, whatever
their true nature: VBC  or VBS.} Such a  state has a unique
ground state, a  gap in the excitations and exponentially
decreasing spin-spin and dimer-dimer\cite{j02} correlations. See
the chapter by P.~Lecheminant in this volume for more details
about the VBS phase of the spin-1 chain.

A spin-$\frac{3}{2}$ specific $SU(2)$-invariant model on the
honeycomb lattice\cite{aklt87,aklt88} is another  example of 2D
VBS.  The spin-1 Heisenberg  model on  the kagome
lattice\index{Kagom\'e lattice} was proposed  to realize a
VBS-like ground state\cite{h00}  in which singlets  form   on
every hexagon without any spontaneous symmetry breaking (hexagonal
singlet solid).\footnote{Each kagome site belongs to  two
hexagons. Each physical spin-1 can be split into   two
spin-$\frac{1}{2}$, each of them being involved in the formation
of a singlet on one neighboring hexagon.}  A similar approach  was
carried out for the spin-1 pyrochlore Heisenberg model.\cite{yk00}
In that case a lattice distortion was invoked to lift the
degeneracy between the two singlet states of each tetrahedron.


\subsection{Two-dimensional examples of VBC}
\subsubsection{Without spontaneous lattice symmetry breaking}

\begin{figure}
\begin{center}
\includegraphics[height=4cm]{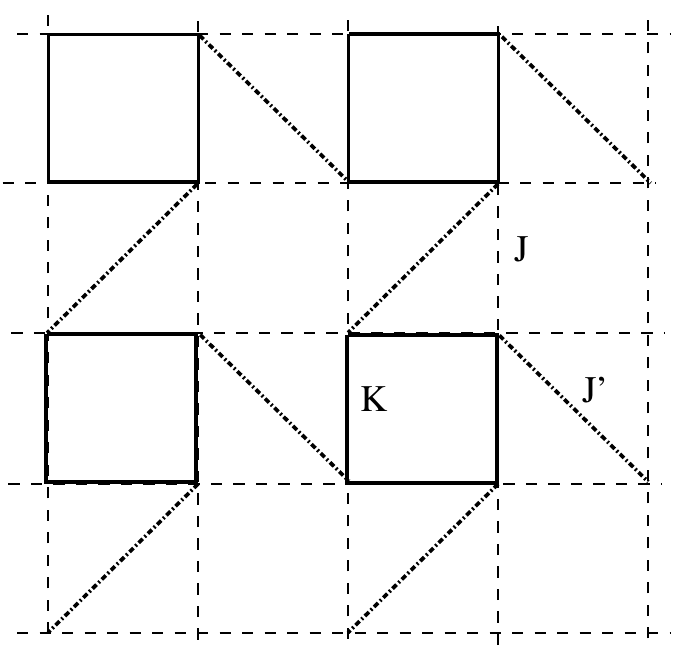}
\end{center}
\caption[99]{
The $\frac{1}{5}$-depleted lattice and the
Shastry  Sutherland  lattice.  The
strong  bonds  of the  Shastry Sutherland  model  are the   bonds $J'$
(dotted dashed lines): they can  accommodate orthogonal dimers  ($J=K$
can be considered as  a ``perturbation''  in  the  real SrCu$_2$(BO$_3$)$_2$).    The
lattice formed by the  strongest exchanges in CaV$_4$O$_9$ is obtained
with $J=0$.    The phase diagram  of   this model\cite{lws02} contains
(at least) collinear N\'eel phases, dimer and 4-spin plaquette VBC.}
\label{qss}
\end{figure}

Two spin-$\frac{1}{2}$ experimental examples of 2D (explicit) VBC have
recently                     attracted                      attention:
CaV$_4$O$_9$\cite{CaV4O9,uksl96,am96b,tku96,sr96,szksu96,wgsoh97}  and
SrCu$_2$(BO$_3$)$_2$.\cite{k99,nkoum99,mu99,k00,kk00,msku00,tmu01,lws02,kthb02,mu03}
In CaV$_4$O$_9$ the V$^{4+}$ ions are on a 1/5 depleted square lattice
and in SrCu$_2$(BO$_3$)$_2$  the exchange couplings  between Cn$^{2+}$
ions realize the    Shastry Sutherland  model.\cite{ss81a} A   lattice
embedding the main couplings  of these two  physical problems is drawn
in    Fig.~\ref{qss}.     Interactions   are    of   the    Heisenberg
type.\footnote{Small  Dzyaloshinsky-Moriya   interactions    have been
identified in SrCu$_2$(BO$_3$)$_2$.\cite{cz02}}

VBC are obvious  ground states in the strong  coupling limits of each
problem.

The  exact ground state of the Shastry  Sutherland model is built from
singlets   on   the  $J'$  bonds.\cite{ss81a,mu99}   For  $J'/J\gtrsim
1.55\pm0.05 $ the  model has a gap in  the spectrum of excitations and
is in a dimerized VBC phase.
\footnote{
Consider a decomposition  of the  Shastry-Sutherland lattice as  edge-
and  corner-sharing  triangles.    For  $J'=2J$  the Hamiltonian    is
accordingly written  as a  sum of $J(\vec{S}_1+\vec{S}_2+\vec{S}_3)^2$
terms for each  triangle ($J'$-bonds are  shared by two triangles) and
each such  term is   minimized by the   dimerized state.   This  shows
rigorously that  the dimerized  state is  the ground state as  soon as
$J'/J\gtrsim 2$.}  For $J'/J\lesssim 1.15\pm0.05$ the system is in the
$(\pi,\pi)$    N\'eel  state  of   the   square   lattice~(results  of
zero-temperature series expansion\cite{zoh02}).  The possibility of an
intermediate phase,  possibly with helical  short-range correlations,
has  been actively discussed in the literature.\cite{am96a,kk00,cms01}
S.~Miyahara and  K.~Ueda have recently written  a review of the theory
of    the    orthogonal     dimer    Heisenberg   spin  model      for
SrCu$_2$(BO$_3$)$_2$.\cite{mu03}

The $\frac{1}{5}$-depleted Heisenberg square lattice model ($J=0$) has
been  studied  as  a  function  of the   ratio  of the  two  different
couplings: bonds   within  a plaquette ($K$)  and   dimer bonds ($J'$)
between  plaquettes.     At isotropic   coupling  ($J'=K$)   collinear
long-range N\'eel order survives the  depletion (the order  parameter
is about $35\%$\cite{tku96} of the  maximum classical value).  A small
unbalance in couplings drives the system either  in a 4-spin plaquette
VBC ($K>J'$)  or in a dimer VBC  ($K<J'$). Both (explicit)  VBC phases
have a spin gap.  A recent generalization of these models by L\"auchli
{\em et al}  encompasses  both the  $\frac{1}{5}$-depleted  Heisenberg
square lattice model and the Shastry Sutherland model\cite{lws02} (see
Fig.~\ref{qss}).  Its phase  diagram exhibits collinear  N\'eel phases
($(\pi, \pi)$ or $(0,  \pi)$) separated from the  VBC phases by second
order phase transitions.  Transition between  the two VBC phases which
have  different  symmetries    occurs   via  a  first     order  phase
transition.\footnote{A recent $Sp(N)$ study  of the Shastry Sutherland
model\cite{cms01}  suggests that a  spin  liquid phase with deconfined
spinons might appear in such a model. No evidence of  such a phase has
emerged from the $SU(2)$ studies so far.}

Excitations in these models come from the  promotion of local singlets
to triplet excitations.  In 2D the ordered dimer background provides a
confining  force  for  the  spin-$\frac{1}{2}$  excitations.   Indeed,
separating two  unpaired  spins (that is   two  {\em spinons}) creates
disruption in the  ordered pattern all the way  from the  first to the
second.  The energy  cost is thus proportional   to the length  of the
string of defaults and  both spin-$\frac{1}{2}$ excitations remains in
fact confined.   Only integer spin excitations   are expected.  On the
other hand  in  these strongly coupled models  single-triplet hoppings
can be difficult and correlated motions might be important, leading to
a large zoology of excited modes  (see Ref.\cite{tmu01} and references
therein).  This potential frustration of the triplet motion favors the
appearance       of      magnetization    plateaus       in        VBC
models.\cite{fo99,k99,mt00,mt00b,mu00,mjg01} This aspect  was  briefly
discussed in the lecture notes published by the authors.\cite{lm02}


\subsubsection{With spontaneous lattice symmetry breaking}

In  the previous models  the (explicit)  VBC \index{VBC} phases do
not break any lattice  symmetry.  They can be directly  related to
the geometry and relative strength of the couplings.  In more
symmetric situations with frustration, spontaneously  symmetry
breaking VBC  can appear as a way to overcome  this frustration by
taking full advantage of the quantum fluctuations.  This is
probably the  case in the $J_1$--$J_3$ model on the square lattice
for intermediate  $J_3/J_1 \sim 0.6$,\cite{ll96} in the
$J_1$--$J_2$ model   on  the hexagonal  lattice  for intermediate
$J_2/J_1\sim  0.4$,\cite{fsl01}  and in the  Heisenberg  model  on
the checker-board
lattice.\cite{canals02,fsl03,sfl02,bh02,baa03,tsma03} In the two
first cases  the  ground states are   dressed  columnar VBC   of
dimers. Translation and $C_4$  (resp.   $C_3$ only) symmetries of
the lattice are spontaneously  broken.  The ground state is   4
(resp.  3) times  degenerate.  Spin-spin correlations decrease
exponentially with the  system size.  All  excitations   are
gapped.  Contrary to    the $J_1$--$J_2$    model     on     the
square     lattice,     exact
diagonalizations\cite{ll96,fsl01,fsl03}\index{exact
diagonalizations} give a rather straightforward information on
these systems where the correlation lengths are small enough (far
enough from  the critical points which limit  the boundary of the
VBC phases).

\begin{figure}
\begin{center}
\includegraphics[height=4cm]{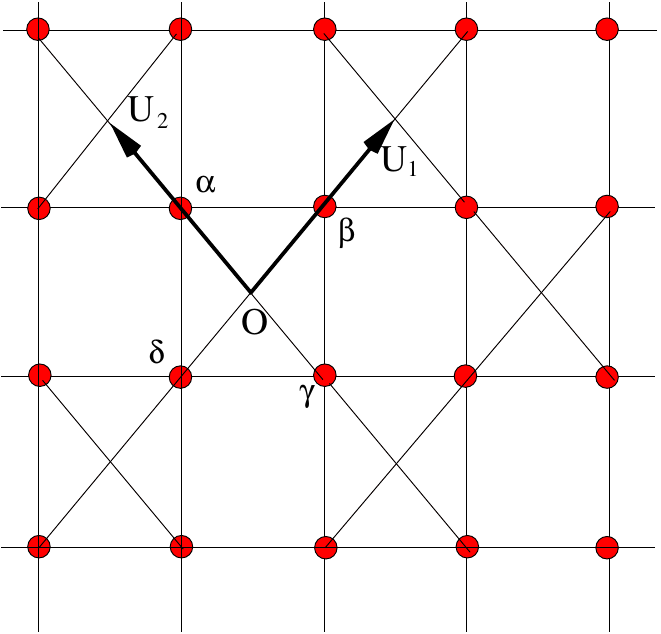}
\hspace*{0.5cm}
\includegraphics[height=4cm]{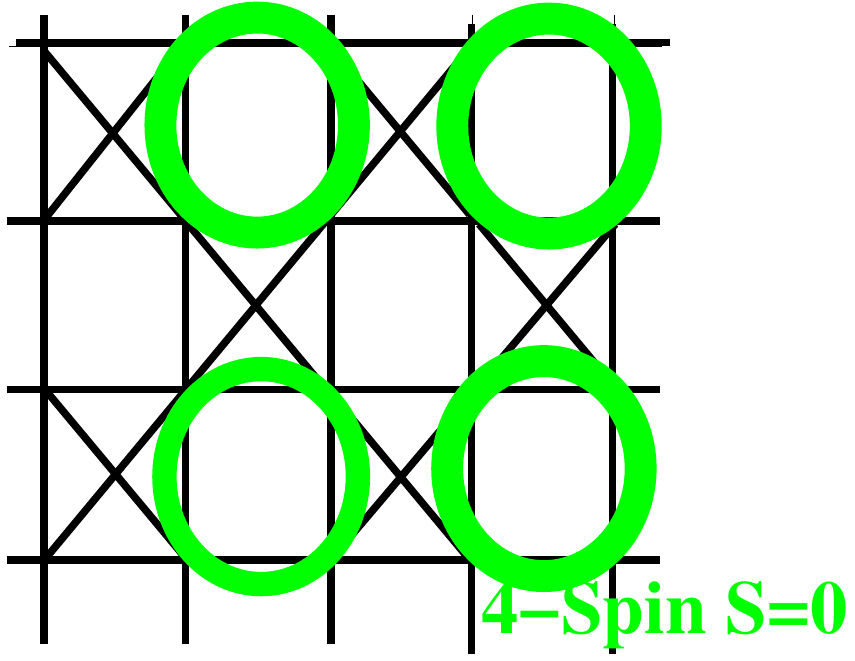}
\end{center}
\caption[99]{Left:  the checkerboard  lattice.  The spins  sit at  the
vertices shown  by bullets, all couplings are  identical, ${\bf u}_1$,
${\bf u}_2$  are the unit vectors  of the Bravais  lattice. Right: the
ground states of the Heisenberg model on the checker board lattice are
translational  symmetry  breaking  VBC  of 4-spin  plaquettes  on  the
uncrossed squares.}
\label{ckb}
\end{figure}

The spin-$\frac{1}{2}$ Heisenberg model  on the checker board lattice,
which  can  also  be  seen  as  a planar  lattice  of  corner sharing
tetrahedrons (see Fig.~\ref{ckb}),  has received the largest attention
for different reasons.\cite{mts01,pc02,canals02,bh02,sfl02,fsl03,baa03,tsma03}
The problem  has   classically  a  continuous  local degeneracy:   the
Hamiltonian can  be rewritten as the sum  of the squares of  the total
spin of each tetrahedron, and every  configuration with a zero spin on
each tetrahedron is a  ground state.  Classically this problem  shares
this  property      with   the Heisenberg   model     on   the kagome,
pyrochlore\cite{mc98,mc98a},    garnet\cite{pp01}   and     pyrochlore
slab\cite{ka02}  lattices (these lattices    made  of corner   sharing
``simplexes'' with  2, 3  or 4  spins  each were  dubbed ``bisimplex''
lattices by Henley\cite{henley2000}).

The quantum spin-$\frac{1}{2}$
antiferromagnet  on the  kagome  lattice has been  found  to  be quite
specific with a  small gap (if any) toward  triplet excitations and an
anomalous density  of  gapless  low-lying singlets   excitations  (see
\S\ref{sec:kagome}).

The quantum  scenario on the  checkerboard lattice is quite different.
The ground state is a (dressed) product  of 4-spin $S=0$ plaquettes on
uncrossed squares: this  state  breaks translational symmetry  but not
$C_4$  (the  point group being  defined  at  the  center of  an  empty
plaquette).  It has a two fold degeneracy  in the thermodynamic limit:
this is easily seen in the  symmetries and finite  size scaling of the
low  lying   levels of  exact spectra.\cite{fsl03}  The  choice of the
4-spin $S=0$  states on the  uncrossed squares corresponds to the most
stable local configuration.   The product of  $S=0$ quadrumers is only
weakly renormalized in the exact ground state.\cite{fsl03}

Excitations of this model  have been studied in different
approaches, exact diagonalizations\cite{f03}, series
expansions,\cite{bh02} \index{series  expansions} real space
renormalization transformation.\cite{baa03} All the  excitations
(singlet and triplet) are gapped.   The triplet excitations
originate from the triplet excitation of an uncrossed plaquette,
they are weakly dispersive with  a  large  gap.   The singlet
excitations cannot be described so simply: from exact
diagonalizations  data one can suspect that some  of these
excitations come  from the reorganization of two adjacent triplet
on crossed squares.   The Contractor Renormalization (CORE)
method  of Berg  {\it et al.}\cite{baa03} on   the other hand
suggests that  these excitations are  domain  walls between   the
two translated plaquette-VBC ground states.


\subsection{Methods}

Spin    waves\cite{uksl96} and  Schwinger bosons\cite{am96a,am96b}
\index{Schwinger bosons} are simple approximations to study the
phase  diagram  of  a quantum frustrated  magnet. But the first
method only gives an approximate knowledge  on the range of
existence of the N\'eel phases, and it is rather difficult  to
include the effect  of fluctuations  beyond the mean-field
approximation within  the Schwinger-Boson formulation. As
discussed   in   the   section devoted to large-$N$
approaches\index{large-$N$ approach} (\S\ref{sec:largeN}),
spontaneous lattice symmetry breaking are very likely  to be due
to topological effects (Berry phase of instantons) which cannot be
captured by the mean-field state. This probably explains why no
spontaneous VBC has ever been found (to our knowledge) in
Schwinger-Boson calculations.\footnote{It seems however  that
spontaneous VBC naturally arise in large-$N$
approaches\index{large-$N$ approach} with fermionic representation
of $SU(N)$ when $1/N$ corrections are considered (see section
\ref{ssec:spin2QDM}).} For unfrustrated problems, as the
Heisenberg model on    the $\frac{1}{5}$-depleted square
lattice,\cite{tku96} QMC  is considered  as  the   method which
can  give benchmarks for other approaches.

Although  VBC  are   naturally  described     with   the help       of
spin-$\frac{1}{2}$ valence-bonds, the appearance of such states in the
low-spin  limit can  sometimes   be anticipated  from  an  appropriate
large-$S$ approach.   An example is given in  the work of Tchernyshyov
{\it   et     al.}\cite{tsma03} (see  also   a    previous    paper by
Henley\cite{henley2000})  on the  checkerboard  Heisenberg  model.  In
this  model, when both   couplings (square lattice bonds and  diagonal
bonds)  are    equal,     ground state  has    a   continuous    local
degeneracy. However  leading $1/S$ corrections select collinear states
out  of  this    huge  manifold.\cite{henley2000} There   remains   an
exponential number of such  states and they  do not have  any magnetic
order.  However    they exhibit long-range  {\em   bond}  order and a
spontaneous symmetry  breaking\cite{tsma03} which is  analogous to the
one observed numerically in the spin-$\frac{1}{2}$ case.

For frustrated problems, exact diagonalizations\index{exact
diagonalizations} can be useful tools in situations where the
system  is not too close from a critical  point, that is when  the
correlations length is not too large.   Successful applications of
exact diagonalizations methods  to   2D Heisenberg magnets
realizing a VBC include studies of  the $J_1$--$J_3$ model on the
square lattice,\cite{ll96}    $J_1$--$J_2$  model   on    the
hexagonal lattice,\cite{fsl01}     Heisenberg model       on
the checkerboard lattice.\cite{fsl03}    In such  situations
analysis of  the  quasi degeneracy of the  low-lying levels    of
the spectra and  of   their finite-size scaling gives an unbiased
and direct informations on  the symmetry breakings in the
thermodynamic limit.\cite{fsl03} However the boundaries of  the
phases  and  the quantum critical  points cannot be
 accurately determined  with  this method.   The  series expansions
described in the previous section appears to be a powerful
approach to determine  phase  boundaries.\cite{gwsoh96,wgsoh97} If
the  spin-spin correlation length is   not very  short, as it   is
the case in   the $J_1$--$J_2$    model       on        the square
lattice      for $J_2/J_1\sim0.5$,\cite{cbps01}  it is very
difficult  to decide from exact diagonalizations between a VBC, a
critical  phase or an RVB \index{RVB} SL.

Concerning excitations, exact  diagonalizations give the gaps in  each
sector and provide a crude approximation of the dispersion laws of the
first  excitations. The large-scale nature  of the excitations (as for
examples  domain  walls  excitations) can  escape   this method.   The
semi-analytical  approaches which  can be  used  for the study  of the
excitations of the    VBC, all use  as  a  basic departure point   the
excitations of a local cluster  of spins conveniently renormalized  by
perturbation\cite{szksu96,bh02}       or    effective  renormalization
technique.\cite{baa03}    Contrary  to   exact  diagonalizations these
methods are not limited by finite-size effects  but the results can be
biased by the departure point.\cite{bh02,baa03}


\subsection{Summary of the properties of VBC phases}\index{VBC}
The   generic  features of  VBC   (whatever the dimensionality  of
the lattice) are:

\begin{itemize}
    \item  A    spin gap,   no   $SU(2)$  symmetry   breaking  and
    short-range spin-spin correlations,

    \item Long-range order   in dimer-dimer  and/or larger  $S=0$
    plaquettes.  The coupling of this order to lattice distortions
    is probable in experimental realizations of spontaneous VBC.

    \item In {\em spontaneous}   VBC  phases the  ground state  is
    degenerate.  From  the theoretical point  of view the discrete
    symmetry of the order parameter of the VBC which spontaneously
    breaks  a   lattice symmetry  may   give  birth   to a  finite
    temperature Ising-like  transition.\cite{ccl90a}  Simultaneity
    between this transition  and a  possible structural transition
    is  likely when  the  couplings of  the  spins  to the lattice
    degrees of freedom (phonons) is considered.\cite{bm02}

    \item VBC have gapped excitations, in the $S=0$ sector as well
    as  in other $S$  sectors.  A wide zoology  of  modes is to be
    expected  as well  as  continua  associated to  multi-particle
    excitations or  scattering of domain walls (in   the case of a
    spontaneous symmetry  breaking of  the ground state).   In two
    dimensions  all   these excitations  have  integer  spins (the
    ordered   back-ground   inducing     a  confinement   of   the
    spin-$\frac{1}{2}$ excitations)
\end{itemize}

Frustration on   the square  lattice  or more  generally  on bipartite
lattices is often  overcome by VBC  phases.  The appearance of  VBC in
triangular geometries is possible in principle but  there is up to now
no examples of  such   phases in pure spin-$\frac{1}{2}$    models (in
Sec.~\ref{sec:QDM} examples  will be given    within the framework  of
quantum dimer models).

It     has  been advocated     in    the  large-$N$    approaches
(see section~\ref{sec:largeN}) that, at least  in two dimensions,
collinear spin-spin correlations  generically lead  to    VBC or
VBS \index{VBS} and    only non-collinear spin-spin correlations
can give birth   to RVB SL  with unconfined spin-$\frac{1}{2}$
excitations. The present  knowledge of $SU(2)$ phase diagrams
supports this prediction.  The VBC found so far numerically in
$SU(2)$   spin models appear   to  be in  regions   of parameter
space where the spin-spin correlations are characterized by some
short-range collinear order in     the large-$S$ limit.  The
$J_1$--$J_2$ model on  the  honeycomb    lattice   has a classical
incommensurate phase  in the regime of  high frustration and there
are some evidences that  in the quantum phase  diagram the
collinear phase is separated from the columnar VBC phase  by a RVB
SL.\cite{fsl01} The multiple-spin exchange  (MSE) model on the
triangular lattice is also believed to be  a RVB SL\cite{mlbw99}
and the corresponding classical ground states generically  have
non-coplanar spin   configurations. Capriotti {\it et
al.}\cite{cbps01} argued that the spin-$\frac{1}{2}$ square
lattice $J_1$-$J_2$ model could be   a RVB SL.  If confirmed, this
would be the first counter-example to  the general rule explained
above (The  Heisenberg  model on  the pyrochlore lattice  might be
an other counter-example\cite{baa03}).


\section{Large-$N$ methods}\label{sec:largeN}
\setcounter{footnote}{0}

Introduced  by   Affleck,\cite{a85}  Affleck  and
Marston,\cite{am88} Arovas       and      Auerbach\cite{aa88} and
Read      and Sachdev\cite{rs89,rs90,rs91} in the context of spin
models, large-$N$ approaches \index{large-$N$ approach} are
powerful methods to investigate    quantum antiferromagnets. When
$N$ is taken  to infinity many of these models can be solved by
saddle point methods and  finite-$N$ corrections can be, at least
in principle, explored in a controlled way.  A success of these
approaches is that they  can describe  the phenomenology  of a
large variety of phases encountered  in quantum magnets : Magnetic
LRO (possibly with  order  by  disorder \index{order by disorder}
selection)  as well  as phases dominated  by short-range
valence-bonds: VBC, VBS and RVB liquids. One       crucial
result       (due to Read and
Sachdev\cite{rs89,rs90,rs91,sr91,s93})  concerning these three
later phases  is  that  the  analysis  of  finite-$N$ corrections
to  some large-$N$ formulations  ($Sp(N)$ for instance, see below)
provides a general criterion to  decide which of these three phase
should appear in a given model.

This criterion is the following in 2D: if the (large-$N$ equivalent of
the) ``spin'' $S$ at each site matches the lattice coordination number
$z$ by $2S=0$~mod~$z$ a VBS is to be expected.  If  it is not the case
(as for a spin-$\frac{1}{2}$  model on the  square lattice) one should
look at the local spin-spin correlations.  If  they are reminiscent of
a   collinear   order, a VBC   with  spontaneous  translation symmetry
breaking is  expected whereas  non-collinear short-range  correlation
generically  give rise to a RVB  phase without any broken symmetry and
deconfined spinon excitations.  These results are of course based on a
large-$N$  generalization of the original  spin model and  there is no
guaranty at all that these rules should always  apply to $SU(2)$ models.
To our knowledge they have however not  been manifestly found in error
up  to now.  In
the following we will  present some of  the important reasoning  steps
leading to this result.

\subsection{Bond variables}
The $SU(2)$  algebra of a spin $S$ at one site
can be  represented by $2$ species of  particles $a^\dag_\sigma$ (with
$\sigma=\uparrow,\downarrow$),  provided  that  the  total  number  of
particles on one site is constrained to be $a^\dag_\uparrow a_\uparrow
+  a^\dag_\downarrow  a_\downarrow=2S$.   The raising  operator  $S^+$
(resp.  $S^-$) is simply represented by $a^\dag_\uparrow a_\downarrow$
(resp.  $a^\dag_\downarrow a_\uparrow$). These particles can be chosen
to   be   fermions   (Abrikosov   fermions)   or   bosons   (Schwinger
bosons). These particles  carry a magnetization $\pm\frac{1}{2}$ since
$S^z=\frac{1}{2}(a^\dag_\uparrow           a_\uparrow-a^\dag_\downarrow
a_\downarrow)$.  For this reason  they are often called {\em spinons}.
The  Heisenberg  interaction  is   a  quartic  interaction  for  these
particles:
\begin{equation}
    {\vec S}_i\cdot{\vec S}_j =
    S^2 - \frac{1}{2} A_{ij}^\dagger A_{ij}
    \label{eq:SSAij}
\end{equation}
with the bond operator $A_{ij}$ defined by:
\begin{equation}
    A_{ij} = a_{j\downarrow} a_{i\uparrow}
    - a_{j\uparrow} a_{i\downarrow}
    \label{eq:ASU2}
\end{equation}
Acting on the  vacuum, $A_{ij}^\dagger$ creates a spin  singlet on the
bond $(ij)$. Physically $A_{ij}^\dagger A_{ij}$ measures the number of
singlets  on   that  bond   and  Eq.~\ref{eq:SSAij}  shows   that  the
antiferromagnetic Heisenberg  interaction just tries  to maximize that
number.

The idea of large-$N$ methods is to generalize the $SU(2)$ symmetry of
the spin$-S$ algebra to a larger group $SU(N)$ (or $Sp(N)$) by letting
the index  $\sigma$ go from  $1$ to  $N$  (or $2N$).  The $SU(N)$  (or
$Sp(N)$) generalization of the Heisenberg  model is solved by a saddle
point calculation of the action.  The $N=\infty$ limit is very similar
to a mean-field decoupling    of  the four-body interaction   of   the
physical    $SU(2)$       model:       $A_{ij}^\dagger    A_{ij}\simeq
A_{ij}^\dagger\left<A_{ij}\right>+\left<A_{ij}^\dagger\right>A_{ij}-\left|\left<A_{ij}\right>\right|^2$.

\subsection{$SU(N)$}
The generalization of the  Heisenberg model to  such a  symmetry group
depends  only  on the choice  of   the  irreducible representation  of
$SU(N)$  according to which the  ``spin'' operators transform (and not
on    the  choice    of fermions   or      bosons  to implement    the
representation). For $SU(2)$ this amounts to specify the magnitude $S$
of  the spin. For $SU(N)$  irreducible  representations are labeled by
Young tableaux.  The case of a general  rectangular tableau with $n_c$
columns and $m$ rows was discussed by Read and Sachdev\cite{rs89b} and
$n_c$ appears  to  continue  to play a   role similar  to $2S$  in the
large-$N$ phase  diagrams.\footnote{Taking the limit $N\to\infty$ with
$m$ fixed  of order 1 and $n_c\sim  N$ is most conveniently  done with
bosons $b^\dagger_{\alpha p}$  where $\alpha=1\cdots N$ is the $SU(N)$
index, while $p=1\cdots m$ label the  different ``colors''.  There are
therefore $N m$ kinds of bosons. On the other hand it is convenient to
use $n_c$ ``colors'' of fermions (still with an $SU(N)$ index) to deal
with   the case $n_c\sim\mathcal{O}(1)$   and    $m\sim N$.    Bosonic
representations   with   $n_c\sim  N$  are   appropriate  to  describe
magnetically ordered  phases\cite{aa88}    but cannot  be  used    for
frustrated models (such  a representation is  not self-conjugate).  On
the other  hand fermionic  representations,  such as the  $m=N/2$  and
$n_c=1$ one,\cite{am88,rokhsar90} can be used on  any lattice but they
do not display magnetically ordered phases and tend to favor dimerized
states.\cite{am88,rs89b,rokhsar90}} In this  review we will focus on a
slightly different large-$N$ generalization of the $SU(2)$ model which
is both able to deal with frustration and magnetic states.

\subsection{$Sp(N)$}
To perform a large-$N$ extension of {\em frustrated} Heisenberg models
one has  to use either fermions\cite{am88} or  bosons with an $Sp(N)$
symmetry.  The  latter  seems to  produce phase diagrams  that closely
resemble the $SU(2)$ problems and we will focus on this representation
which was introduced by  Read and Sachdev.\cite{rs91} The presentation
below           is       largely      inspired        from       their
papers.\cite{rs89,rs90,rs91,sr91,s93}

We now have  $2N$ flavors of bosons  at each site: $b_{i,\sigma}$ with
$\sigma=1..2N$ and we define an $Sp(N)$-invariant bond operator:
\begin{equation}
    A_{ij} = \sum_{\sigma,\sigma'=1..2N}
    \mathcal{J}_{\sigma,\sigma'} \;b_i^\sigma \;b_j^{\sigma'}
\end{equation}
where the  $2N\times2N$ antisymmetric   tensor $\mathcal{J}$ is  block
diagonal
\begin{equation}
    \mathcal{J}=\left[
    \begin{array}{ccc}
    \begin{array}{cc}
    0&1\\-1&0
    \end{array}&    &\\
        &\ddots &\\
        &   &\begin{array}{cc}0&1\\-1&0\end{array}
    \end{array}
    \right]
\end{equation}
and   generalizes the  $SU(2)$   antisymmetric tensor  $\epsilon_{ij}$
($SU(2)$ is identical to $Sp(1)$).    Up to some constant the  $Sp(N)$
Hamiltonian is
\begin{equation}
    \mathcal{H}=-\frac{1}{N} \sum_{ij} J_{ij} A_{ij}^\dagger A_{ij}
\end{equation}
with the constraints
\begin{equation}
    \forall i\;\;
    \sum_{\sigma=1}^{2N} b^{\dag}_{i\sigma}\;b_{i\sigma}=n_c
    \label{eq:constraint}
\end{equation}
$n_c=2S$ in the $SU(2)$ case and  $n_c/N=\kappa$ will be kept constant
when  taking  the  large-$N$  limit.  The  partition function  can be
represented by an imaginary time functional integral:
\begin{equation}
    Z = \int \mathcal{D}[\lambda_i,b_i^\sigma,b_i^{\dag\sigma}]
    \exp{\left(-\int_0^\beta (L_0+\mathcal{H})d\tau\right)}
    \label{eq:Z}
\end{equation}
\begin{eqnarray}
    L_0&=&\sum_{i \sigma}
        b^{\dag\sigma}_i ( \partial_\tau+i\lambda_i) b_{i\sigma}
        -iN\kappa\sum_i \lambda_i
    \label{eq:lL0}
\end{eqnarray}
The $\lambda_i(\tau)$   are  Lagrange multipliers  that   enforce  the
constraint  Eq.~\ref{eq:constraint} at  every  site.   Bond degrees of
freedom  $Q_{ij}$  are introduced   in order  to  decouple the  bosons
(Hubbard-Stratonovitch).  The partition function is now
\begin{equation}
    Z = \int \mathcal{D}[Q_{ij},\bar{Q}_{ij},\lambda_i,b_i^\sigma,b_i^{\dag\sigma}]
    \exp{\left(-\int_0^\beta ( L_0 + L_1) d\tau\right)}
    \label{eq:ZQ}
\end{equation}
with
\begin{equation}
    L_1 = \sum_{ij} \left[
        N \frac{ | Q_{ij} |^2}{J_{ij}}
        - (A_{ij}^\dagger Q_{ij} + {\rm h.c})
        \right]\;,
\end{equation}
so that   a   Gaussian  integration  of   the    $Q_{ij}$  gives  back
Eq.~\ref{eq:Z}. The bond variable  are $Sp(N)$ invariant and they  can
take  non-zero  expectation  values  at   a mean-field level   without
breaking the $Sp(N)$ symmetry.  As we explain  below, they are however
not {\em gauge-invariant}.

\subsubsection{Gauge invariance}
An   important property   of  Eq.~\ref{eq:ZQ}   is  the $U(1)$   gauge
invariance associated to the following transformations:
\begin{eqnarray}
    b_{i\sigma}&\to&b_{i\sigma}e^{i\phi_i} \label{eq:phi}\\
    Q_{ij} &\to& Q_{ij} e^{i(\phi_i+\phi_j)} \label{eq:Qtheta}\\
    \lambda_i &\to& \lambda_i -\partial_\tau \phi_i \label{eq:lambdatheta}
\end{eqnarray}
where  $\phi_i(\tau)$ are arbitrary site- and
imaginary-time-dependent angles.  This gauge invariance comes from
the  the conservation of the local boson number and reflects the
fact the magnitude of the spin is constant at each site. If we
focus on the  phase degrees of freedom of  the bond variables, the
Eq.~\ref{eq:ZQ} describes a system   of charge-1    bosons
coupled    to    a    $U(1)$     lattice  gauge
theory.\cite{wilson-74,bdi74}  These gauge  degrees  of freedom
play a crucial  role in  the analysis  of  the fluctuations  about
mean-field solutions.

{\em Effective action for the bond variables} ---The boson degrees of
freedom can be integrated out to give an effective action for the bond
variables:
\begin{equation}
    Z = \int \mathcal{D}[Q_{ij},\bar{Q}_{ij},\lambda_i]
    \exp{\left(-S^{\rm eff} \right)}
\end{equation}
\begin{equation}
    S^{\rm eff}=N \int_0^\beta \left[
        \sum_{ij} \frac{ | Q_{ij} |^2}{J_{ij}}
        -i\kappa \sum_i \lambda_i
    \right]
    - N  {\rm Tr} \log G
    \label{eq:Seff}
\end{equation}
where $G^{-1}$  is  the  quadratic form which  couples   the bosons in
Eq.~\ref{eq:ZQ} (propagator).  It depends on the bond variables and on
$\lambda_i$.             We     may          write            formally
$G^{-1}=\partial_\tau-i\lambda-Q$. The term $N {\rm Tr} \log G$ is the
free energy    of the  bosons  in  presence  of   the bond  fields. By
construction the action $S^{\rm  eff}$ is gauge-invariant with respect
to  the transformations of  Eqs.~\ref{eq:phi}--\ref{eq:lambdatheta}.
So far  this   is an exact   formulation of  the  original  model  for
arbitrary $N$.

\subsubsection{Mean-field ($N=\infty$  limit)  } Since $N$  factorizes  (no
flavor index  is left in Eq.~\ref{eq:Seff}),  $Z$  is dominated by the
saddle  point of $S^{\rm eff}$ when  $N$ is  large.  For simple models
such as the  first-neighbor antiferromagnet on  the cubic lattice (any
space dimension), the saddle point can be determined analytically. The
$N=\infty$  limit    is almost   equivalent   to   the Schwinger-boson
mean-field theory.\cite{schwingerboson,ceccatto} This can otherwise be
done numerically.\footnote{To our   knowledge, all  the saddle  points
considered  so far are static  (expectation values of the $Q_{ij}$ are
time-independent) and  the  corresponding $\left<Q_{ij}\right>$  could
all be made  real with an  appropriate gauge transformation.  There is
no  chiral order  and  the  time-reversal  symmetry is  unbroken.  The
(oriented) sum of  the complex phases of the  bond variables  around a
plaquette defines a  $U(1)$ flux.  This flux is  related to  the solid
angle formed by the spins  and it vanishes  in collinear as well as in
coplanar states.  In such cases the phases  can be therefore be gauged
away and the $\left<Q_{ij}\right>$ can be  made real.  For this reason
complex bond variables are  usually not observed.\cite{ot03} } In this
large-$N$  limit, two kinds of mean-field   solutions can appear.  For
large  enough  $\kappa$ the  bosons condense  at some wave vector, the
spectrum of the  mean-field Hamiltonian is gapless.  This  corresponds
physically to N\'eel long-range order.  On the other hand, for smaller
$\kappa$  (smaller ``spin'') the  mean field Hamiltonian is gapped and
the ground state preserves  the $Sp(N)$ symmetry.  Fluctuations around
the saddle  point are not   expected to change drastically  the N\'eel
ordered phases  but  they play  an important  role in   the physics of
$Sp(N)$ symmetric phases.  The following  is a brief discussion of the
effects of fluctuations in these non-magnetic phases.

\subsubsection{Fluctuations about the mean-field solution} At the mean-field
level described  above some  $Q_{ij}$ acquire a  (static in  all known
cases) non-zero expectation value: $\left<Q_{ij}\right>=\bar{Q}_{ij}$.
For  this reason  such a  state spontaneously  breaks the  local gauge
invariance        of       Eqs.~\ref{eq:phi},\ref{eq:Qtheta}       and
\ref{eq:lambdatheta}.   However  this does  not  mean  that the  gauge
degrees are  all gapped and  do not play  any role at low  energy.  In
fact, as remarked by Read and Sachdev, depending on the {\em geometry}
of  the  lattice  defined   by  the  bonds  where  $Q_{ij}\ne0$,  some
long-wavelength gapless  gauge excitations survive  and the associated
fluctuations  must  be  taken   into  account.   More  precisely,  the
fluctuations  of  the  bond  variables  about  the  saddle  point  are
decomposed into an amplitude and a phase
\begin{eqnarray}
    Q_{ij}=\left(\bar{Q}_{ij} + q_{ij}\right)\exp(i\theta_{ij})
    \label{eq:qtheta}
\end{eqnarray}
and we  expand $S_{\rm eff}$ with these  new variables. Two cases must
then be considered:
\begin{itemize}
\item[i)]  The  lattice  made  of  the  sites  connected  by
non-zero $\bar{Q}_{ij}$ bonds is bipartite.  This is automatically
the case if the original lattice defined  by bonds where the
exchange $J_{ij}\ne0$ is  bipartite.  This can   also be true  if
the  original lattice is a non-bipartite lattice but some bonds
have $\bar{Q}_{ij}=0$ so that the remaining lattice  is bipartite.
This  is the case,  for instance, in the  $J_1$--$J_2$ model on
the   square lattice,\cite{rs91}  in some regions    of the phase
diagram   of   the   Shastry-Sutherland model\cite{cms01} and  in
on      the   checkerboard Heisenberg model.\cite{mts01} Such
configurations of the bond variables give {\em collinear} spin
structures: spin-spin correlations  can  either be long-ranged
(large $\kappa$, N\'eel phase) or short-ranged but in both cases
the magnetic structure factor is  peaked at a simple wave vector
${\bf k}_0$   such that $2{\bf k}_0$ is    a reciprocal lattice
vector ($k_0=(\pi,0)$, $k_0=(0,\pi)$    or $k_0=(\pi,\pi)$ in
square geometries). \item[ii)] The  lattice made of the sites
connected by non-zero $\bar{Q}_{ij}$ bonds  is not  bipartite.
This    happens  in some   phases  of the $J_1$--$J_2$--$J_3$
model  on the square lattice,\cite{rs91}  on the triangular  or
kagome lattices\cite{sachdev92},\index{Kagom\'e lattice}   in the
Shastry-Sutherland model for  some    values of     the exchange
parameters\cite{cms01} and    on    an    anisotropic triangular
lattice.\cite{cmm01} Such mean-field  states generically have
planar but non-collinear spin-spin correlations.
\end{itemize}
It is simple to check that case i) preserve a {\em global} continuous
symmetry while such a symmetry is absent in ii). Consider the
following {\em global} gauge transformation in case i) :
\begin{eqnarray}
    b_{i\sigma}\to b_{i\sigma}e^{i\phi} \;&&\;
    b_{j\sigma}\to b_{j\sigma}e^{-i\phi} \label{eq:phiij} \\
    Q_{ii'} \to Q_{ii'} e^{2i\phi}  \;&&\;
    Q_{jj'} \to Q_{jj'} e^{-2i\phi}\label{eq:Q2}\\
    Q_{ij} &\to& Q_{ij} \label{eq:Qij}
\end{eqnarray}
where  $i,i'$  belongs  to   sublattice  $A$  and  $j,j'$  belongs  to
sublattice $B$.   This transformation  does not change  the mean-field
parameters   $\bar{Q}_{ij}$.   The  only   fields  affected   by  this
transformation   are  those   connecting   two  sites   on  the   same
sublattice.\footnote{In  a gauge theory  language the  fields $Q_{ii'}$
(resp.  $Q_{jj'}$) of Eq.~\ref{eq:Q2} transform like a charge-2 scalar
for  the $U(1)$  gauge field.   Instead, from  Eq.~\ref{eq:phiij}, the
bosons (spinons)  carry a  charge 1.}  They  have a  zero expectation
values in  case i) (or  do not even  exist if the physical  lattice is
itself bipartite).  For this reason  it is possible to make low-energy
(and long-wavelength) gapless gauge excitations about the saddle-point
by   replacing   the   global   staggered  phase   shift   $\phi$   of
Eq.~\ref{eq:phiij}  by  a  slowly  varying  (staggered)  $\theta_{ij}$
(Eq.~\ref{eq:qtheta}).  A  gradient expansion of  the effective action
performed  at the  appropriate points  in the  Brillouin zone  for the
phase   fluctuations  only  involves   gradients  of   $\theta$.   The
corresponding action is that of $U(1)$ lattice gauge theory coupled to
charge-1 boson (spinons).

\subsubsection{Topological effects - instantons and spontaneous dimerization}
So far only small fluctuations around the saddle point were considered
and   the  contribution   of  topologically   non-trivial  gauge-field
configurations  were  ignored.   Consequently,  the magnitude  of  the
``spin'' $n_c/2$  was a continuous parameter and  {\em the information
about the integer  or half-integer (for instance) character  of $S$ as
disappeared}.   From  Haldane's  work   on  quantum  spin  chains  and
non-linear sigma  models\cite{haldane83} it  is well known  that Berry
phases  in  spin  systems  give  rise  to  topological  terms  in  the
low-energy effective action which can play a crucial role depending on
the parity of $2S$.

In non-linear sigma models in  2+1 dimensions the Berry phase vanishes
for  configurations which  are  smooth  on the  scale  of the  lattice
spacing\cite{haldane88,noHopfTerm2D88}  (unlike  the 1+1  dimensional
case).  However  ``hedgehog'' space-time singularities\cite{haldane88}
give  non-trivial  Berry phases.   Read  and  Sachdev  found that  the
closely  related  instantons  of  the effective  $U(1)$  gauge  theory
described  before\footnote{In  that  gauge  theory associated  to  the
phases of the  link variables an instanton corresponds  to a tunneling
event during  which the total  magnetic field piercing the  lattice is
changed by $\pm2\pi$.}  also play a crucial role in the physics of the
$Sp(N)$ (as well as $SU(N)$) spin models.

The Berry phase associated to such a singular configuration depends on
details   of  the  lattice  geometry.    In  the   short-range ordered
$(\pi,\pi)$ phase of the   square lattice antiferromagnet  this  Berry
phase is  a multiple of $in_c\pi/2$.   Although dealing  with a gas of
interacting ($1/r$ Coulomb-like  potential) instantons is a  difficult
problem (see  Ref.~\cite{rs90}  and references therein),  we can guess
that  the  physics will   depend on  $n_c\;{\rm mod}\;4$.  A  detailed
analysis\cite{rs90} shows  that   when   $n_c\ne0\;{\rm mod}\;4$   the
instantons  condense and   {\em   spontaneously  break  the    lattice
translation symmetry}.  This generates a static electric field for the
$U(1)$   gauge field.   Since  the  electric field  is  coupled to the
difference  of amplitudes of the  bond  variables, such state acquires
spatially inhomogeneous expectation values of the bond variables, {\em
it  is a  VBC  and spinons are confined   in pairs}.  In the $J_1$--$J_2$
model around $J_2/J_1\simeq0.5$ the   mean-field state is  short-range
ordered with correlations peaked at $(\pi,\pi)$.  A columnar dimerized
state is  predicted by Read  and Sachdev   from this analysis  of  the
fluctuations, in agreement  with  a number  of numerical works  on the
$SU(2)$ $J_1$--$J_2$ spin-$\frac{1}{2}$ model.

In a recent work by Harada, Kawashima and Troyer\cite{hkt03} the
phase diagram of the (first neighbor - unfrustrated) $SU(N)$
antiferromagnet on  the  square  lattice  with $n_c=1$ was  found
to be in  complete agreement with Read and Sachdev's predictions.
They showed by quantum Monte Carlo \index{quantum Monte Carlo}
simulations   that for $N\leq4$   the  systems  is N\'eel ordered
whereas  it  is a columnar   VBC for $N>5$.   This provides an
additional support to the field theory  arguments described above.
It also underlines that   the mechanism of spontaneous  symmetry
breaking and formation of a VBC  may  come from  quantum
fluctuations only  and that frustration is not  always required
(although it clearly enhances quantum fluctuations).

On the other hand when $n_c=0\;{\rm mod}\;4$ the  analysis of Read and
Sachdev shows that fluctuations   should not bring any broken  lattice
symmetry.  Spinons are also  confined and this state closely resembles
the     valence-bond   solid (VBS)  proposed     by   Affleck {\it  et
al.}\cite{aklt87} as a possible ground state when the spin $S$ matches
the  coordination  number  $z$   according  to $2S=0$   mod  $z$  (see
\S\ref{ssec:VBS}).

\subsubsection{Deconfined phases}\index{deconfined phases}
 Now  we suppose    that,  starting from   a
mean-field  solution  with collinear correlations (case i),   a
parameter of the  original  spin model is varied  so that the mean
field  solution is changed  and some bonds $Q_{ii'}$ ($i$ and $i'$
belong to the same sublattice) acquire a non zero expectation
value (case ii).  In  the  framework of  square lattice
antiferromagnets,  a finite  third-neighbor coupling  ($J_3$)
would  be needed.\cite{rs91}  From   the   point   of  view  of
the long-wavelength gauge fluctuations (related  to the continuum
limit of the  phases  $\theta_{ij}$)   discussed   above,  the
appearance    of $\bar{Q}_{ii'}\ne0$ is  equivalent  to the
condensation   of a (Higgs) boson of charge 2.   This is a
spontaneous  break  down of  the global $U(1)$ staggered symmetry
of Eqs.~\ref{eq:phiij}--\ref{eq:Qij} down to a    $\Z$ one   since
the   field   $Q_{ii}$ is   not invariant  under Eq.~\ref{eq:Q2}
except if  $\phi=0$  or $\pi$.   Based on  results of Fradkin  and
Shenker\cite{fs79}   concerning  confinement  in  compact lattice
gauge theories coupled to matter, Read and Sachdev argued that
this Higgs mechanism  suppresses the low-energy gauge fluctuations
and liberate the spinons.\index{spinons}  This  confinement
transition is described by  a $\Z$ gauge theory.  The suppression
of the $U(1)$ gauge  fluctuations also forbids  the condensation
of instantons discussed above and the ground state remains
unique\footnote{Except for a discrete topological degeneracy.} and
{\em   bond  variables    have  uniform  expectation values.  It
is a genuine SL  without any broken symmetry and deconfined
spinons.} \index{deconfined spinons}


\section{Quantum Dimer Models}\index{quantum dimer models, QDM}\label{sec:QDM}
\setcounter{footnote}{0}

In  a previous  section   (\S\ref{sec:VBC})  we  showed  that  pairing
spins-$\frac{1}{2}$   into singlets at  short   distances is a  rather
natural    way     to     overcome   frustration     in     Heisenberg
antiferromagnets.  QDM   are    defined  in the     Hilbert   space of
nearest-neighbor valence-bond (or dimer) coverings of the lattice.  By
construction these models focus  on the dynamics  in the singlet space
and     ignore magnetic   (gapped      magnons   or gapped    spinons)
excitations. For  this reason they  are (a priori)  not appropriate to
describe  the physics of spin  systems  where magnetic excitations are
gapless.

The Hamiltonian of a QDM usually contains kinetic as well as
potential energy terms for these dimers.  Such Hamiltonians can
often be simpler than   their   spin parents  and  are   amenable
to  several analytic treatments  because of their    close
relations to  classical   dimer problems,\cite{k61,f61,k63} Ising
models   and     $\Z$   gauge theory.\cite{msf02,msp02,ms02} These
models  can   offer simple descriptions    of VBC\cite{rk88}   as
well as     RVB\index{RVB} liquids.\cite{ms01,msp02} It is  in
particular possible to write down some QDM    that  have a simple
and exact VBC ground state with spontaneous broken symmetries
(such as Rokhsar and Kivelson's model on the square
lattice\cite{rk88} with attractive potential energy only - in
which case the exact ground state is very simple).  Simple
solvable QDM    which have a dimer-liquid   ground state   can
also  be constructed.\cite{msp02}

\subsection{Hamiltonian}

\newcommand{\hh}{\begin{picture}(13,9)(-2,2)
	\put (0,0) {\line (1,0) {8}}
	\put (8,8) {\line (-1,0) {8}}
	\put (0,0) {\circle*{3}}
	\put (0,8) {\circle*{3}}
	\put (8,0) {\circle*{3}}
	\put (8,8) {\circle*{3}}
	\end{picture}
}
\newcommand{\vv}{\begin{picture}(13,9)(-2,2)
	\put (0,0) {\line (0,1) {8}}
	\put (8,8) {\line (0,-1) {8}}
	\put (0,0) {\circle*{3}}
	\put (0,8) {\circle*{3}}
	\put (8,0) {\circle*{3}}
	\put (8,8) {\circle*{3}}
	\end{picture}
}

The first QDM \index{QDM} was introduced by Rokhsar and
Kivelson.\cite{rk88} It is defined   by  an   Hamiltonian  acting
in  the   Hilbert space  of first-neighbor dimer  (valence-bonds)
coverings of  the square lattice and reads:

\begin{equation}
\mathcal{H}=\sum_{\rm Plaquette} \left[
-J\left(\left|\vv\right>\left<\hh\right| +{\rm H.c.}\right)
+V\left(\left|\vv\right>\left<\vv\right|+\left|\hh\right>\left<\hh\right|\right)
\right] \label{eq:sQDM}
\end{equation}

Flipping two parallel dimers around a square plaquette is the simplest
dimer move on the square lattice and the $J$ terms precisely represent
such dynamics.  The $V$   terms are diagonal in   the dimer basis  and
account for an    attraction or  repulsion between    nearest-neighbor
dimers.   These are the two  {\em  most local}  terms (respecting  all
lattice  symmetries) which   can  be   considered.\footnote{They  were
originally  derived\cite{rk88} as the lowest  order  terms of a formal
overlap     expansion  (see \S\ref{ssec:spin2QDM}     below)   of  the
Heisenberg. In that  calculation $J\sim x^4$  and $V\sim x^8$.  Notice
that  a  three-dimer kinetic term     (extending over two  neighboring
plaquettes)  is generated  at order    $x^6$ and  is not included   in
Eq.~\ref{eq:sQDM}.}

\subsection{Relation with spin-$\frac{1}{2}$ models}
\label{ssec:spin2QDM}

There exists  different interesting  mappings between frustrated  Ising
models  and QDM.\cite{msc00,ms02} Here,     however, we focus   on the
relations between  QDM and $SU(N)$ (or $Sp(N)$)  spin models  in which
dimers are related to singlet valence-bonds.

{\em Overlap expansion}. --- A valence-bond state (product of two-spin
singlets -  belongs  to the spin-$\frac{1}{2}$  Hilbert space)  can be
associated  to any dimer   covering.\footnote{There is however  a sign
ambiguity (a valence-bond is antisymmetric under  the exchange of both
spins)  that can be fixed by  choosing  an orientation on every bond.}
Two such valence-bond states $\left|a\right>$ and $\left|b\right>$ are
not orthogonal but,  as first discussed by  Sutherland\cite{s88} their
overlap decays exponentially with the length $L$ of the loops of their
transition    graphs     (defined    in     \S\ref{sssec:trgr})     as
$|\left<a|b\right>|=2^{1-L/2}$.   Rokhsar   and    Kivelson\cite{rk88}
introduced  a    formal    expansion  parameter   $x$    and  replaced
$|\left<a|b\right>|$  by   $2x^{L}$.   Their  idea  is that   although
$x=\frac{1}{\sqrt{2}}$ for physical $SU(2)$ spins, the physics of some
models may     be captured  by  the   first  orders  of  a  small  $x$
expansion. Truncating this expansion to order $x^n$ gives an effective
Hamiltonian which contains {\em  local}  terms involving at most   $n$
dimers.\footnote{The {\em signs}  of the non-diagonal  (kinetic) terms
of the effective QDM  obtained by such small  $x$ expansion depends on
the   sign  convention which  was  chosen   to  map   valence-bonds to
dimers. An important  question  is to know  whether,  at least at  the
lowest non-trivial  order, a sign convention  giving the same sign for
all kinetic terms exists (as in  Eq.~\ref{eq:sQDM}).  This is the case
on the square\cite{rk88}  and triangular lattices.\cite{ms01}} In this
approach the dimer states of the QDM  are in one-to-one correspondence
with    {\em  orthogonalized}   valence-bonds   in   the spin  Hilbert
space.\footnote{This implicitly assumes  that the  valence-bond states
are  linearly  independent.  This  can be  demonstrated on  the square
lattice and appears   to be the   case on  the triangular  and  kagome
lattices for large enough sizes. The operator which orthogonalizes the
valence-bond basis  into  the  dimer  basis is   $\Omega^{-1/2}$ where
$\Omega_{a,b}=\left<a|b\right>$ is the overlap matrix.}

{\em Fluctuations about large-$N$ saddle points}.--- From the argument
above it could   seem that the connection between   spin-$\frac{1}{2}$
models and QDM relies on a variational approximation: the spin Hilbert
space  is   restricted     to   the   nearest-neighbor    valence-bond
subspace. This connection is in fact probably deeper, as some theories
describing  fluctuations  about  some  large-$N$ saddle points  ($1/N$
corrections) are equivalent to  (generalized)  QDM.  This mapping  was
discussed by  Read   and Sachdev\cite{rs89b} for  representation  with
$m=1$ (number  of   rows  in  the  Young    tableau of  the    $SU(N)$
representation), $N\to\infty$ and $n_c\sim\mathcal{O}(1)$, it leads to
a {\em generalized} QDM where $n_c$ dimers emanate  from each site.  A
QDM  also    describes $1/N$  corrections      in  the case  of    the
fermionic\footnote{Young  tableau with    $n_c=1$ column  and  $m=N/2$
rows.}   $SU(N)$ generalization\cite{am88,rokhsar90} of the Heisenberg
model:
\begin{eqnarray}
  \mathcal{H}&=&\frac{1}{N}\sum_{ij} J_{ij}:B_{ij}^\dagger B_{ij}: \nonumber \\
  &=& -\frac{1}{N}\sum_{ij} J_{ij}\; B_{ij}^\dagger B_{ij} +{\rm cst}
  \label{eq:HSUNf} \\
    {\rm where}\;\;B_{ij}&=& \sum_{\sigma=1}^N c_{j\sigma}^\dagger c_{i\sigma}
\end{eqnarray}
and where  the $c_{i\sigma}$ are  $N$ flavors of  fermions satisfying
a  constraint similar to Eq.~\ref{eq:constraint} :
\begin{equation}
  \sum_{\sigma=1}^N c_{i\sigma}^\dagger c_{i\sigma}=N/2
  \label{eq:constraintF}
\end{equation}
Rokhsar   showed\cite{rokhsar90}  that   in  the   $N\to\infty$  limit
``dimerized states'' (or  Peierls states) becomes exact  ground states
of Eq.~\ref{eq:HSUNf} for a large class of models.\footnote{ Let $J_0$
be  the largest value of  the $J_{ij}$. Each  dimerization (no site is
left   empty) where only bonds  where  $J_{ij}=J_0$ are  occupied is a
ground state.  Here a  ``dimer''   between to neighbors  $i$  and  $j$
consists of a  $SU(N)$  singlet made  with $N$  fermions (one of  each
flavor) hoping back and forth between $i$ and  $j$.  It is constructed
from            $\prod_{\sigma=1}^N          \left(c_{i\sigma}^\dagger
+c_{j\sigma}^\dagger\right)  \left|0\right>$.   by projecting  out the
components  which do  not  satisfy  Eq.~\ref{eq:constraintF}.   Notice
however that in  the $N\to\infty$ limit  the relative  fluctuations of
the total number of fermion on each site are of order $1/\sqrt{N}$ and
can  be neglected.}  Quite  naturally, $1/N$ corrections will induce a
dynamics into this subspace of  dimerized states; it can be  described
by a  QDM (with kinetic  energy terms only  at this order).  At lowest
order, on the square lattice, a kinetic term identical to the $J$ term
in Eq.~\ref{eq:sQDM} is generated and  naturally favors a columnar  or
resonating-plaquette  crystal (in agreement with  a number of works on
the spin-$\frac{1}{2}$ model).  The  same arguments were discussed for
the kagome    lattice.\cite{mz91}  In  that  case the    leading $1/N$
corrections to  the fermionic saddle  point generate three-dimer moves
around     hexagons and      stabilize  a   crystal     of  resonating
hexagons.\footnote{To our knowledge however there is no clear evidence
of such ordering in  the spin-$\frac{1}{2}$ case.} This  formalism was
also applied to the checkerboard  model\cite{canals02} to conclude  to
the presence of a VBC phase.

\subsection{Square lattice}

\begin{figure}
  \begin{center}
    \includegraphics[height=3cm]{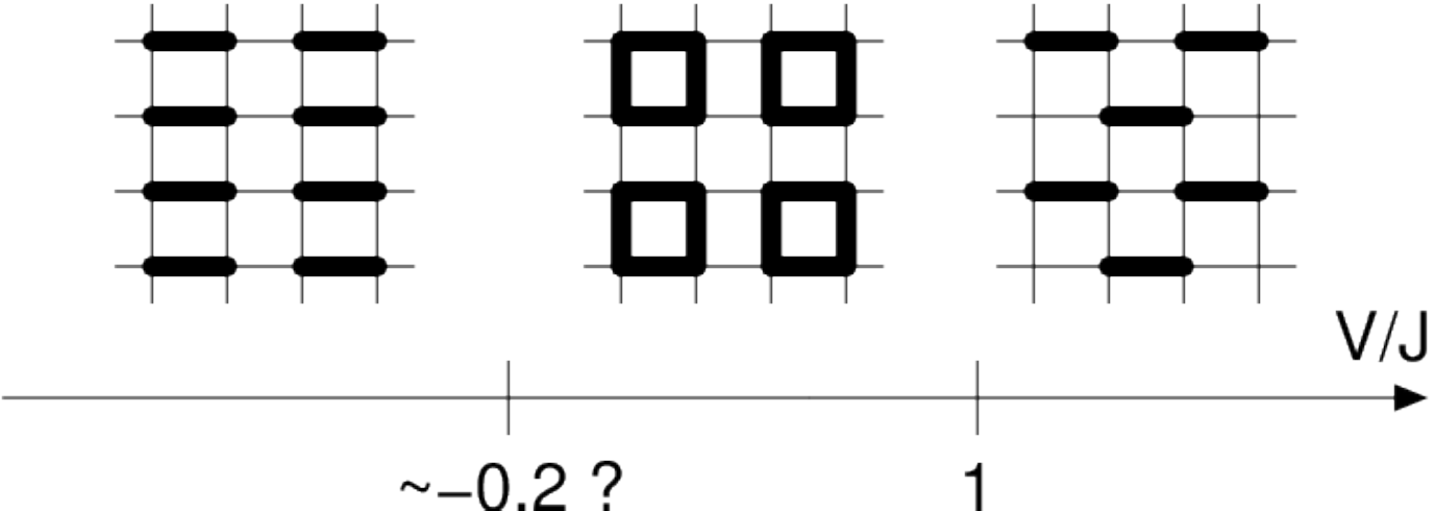}
    \caption[99]{Schematic  phase diagram  of the   square lattice
      QDM.  The  possible location of an  intermediate plaquette phase
      is taken from the work of Leung {\it et al}.\cite{lcr96}}
    \label{fig:sQDM}
  \end{center}
\end{figure}

The phase diagram of the  Rokhsar and Kivelson's square lattice
QDM is shown Fig.~\ref{fig:sQDM}.  Since  a change in the signs of
the basis dimer   configurations   can   change  $J$   into $-J$
(see Ref.~\cite{rk88})   we  will  choose   $J>0$  without loss
of generality.

\subsubsection{Transition graphs  and topological sectors}
\label{sssec:trgr}

In     order to understand the  particularities    of QDM on bipartite
lattices it  is useful to   describe how the  set of  dimer  coverings
splits into {\em  topological  sectors}. To do   so we first   have to
define transition graphs: the transition  graph of two dimer coverings
$c$ and $c'$ is obtained  by superimposing $c$ and  $c'$ on the top of
each other; it  defines a set  of non-intersecting loops covering  the
lattice.  On  each bond  where the dimers  of $c$   and  $c'$ match, a
trivial loop of  length 2 is obtained.   When the lattice is bipartite
(two sublattices $A$ and $B$) these loops can be {\em oriented} in the
following way: any dimer belonging to $c$ is  oriented from $A$ to $B$
and dimers  of $c'$ are  oriented $B\to A$.   The  transition graph is
then  made of  loops of  the   type $A\to  B\to  A\to  B\cdots$.  With
periodic  boundary     conditions    two   winding  numbers\cite{rk88}
$-L/2\leq\Omega_x,\Omega_y\leq    L/2$   are  associated  to    such a
transition graph ($L\times  L$ sites).  $\Omega_x$ (resp.  $\Omega_y$)
is the net number of  topologically non-trivial loops (clockwise minus
counterclockwise) encircling  the     torus in the    $x$ (resp.  $y$)
direction.

Dimer coverings can   be   grouped into different   {\em   topological
sectors}. By definition two dimer coverings belong  to the same sector
if they can be transformed into each other by repeated actions of {\em
local} dimer moves (the  transition graph associated to each  movement
does not  wind  around the whole  system  if it has  periodic boundary
conditions).  On the square lattice  two-dimer moves are sufficient to
connect any two  states in the  same sector;  that  is the Hamiltonian
Eq.~\ref{eq:sQDM}  is ergodic  within  each topological  sector.  In a
torus geometry, $c$ and $c'$ belongs to the same sector if and only if
their transition graph has winding numbers $\Omega_x=\Omega_y=0$.  The
different topological sectors can  be labeled by their winding numbers
with respect to some  reference columnar configuration.  Their  number
is of order $\mathcal{O}(L^2)$ for a system of linear size $L$.

\subsubsection{Staggered VBC for $V/J> 1$}
When $V$ is sufficiently large the system tries to minimize the number
of    parallel   dimers.     The    staggered   configuration    shown
Fig.~\ref{fig:sQDM}  has  no such  {\em  flippable  plaquette}. It  is
always  a zero-energy  eigenstate  of Eq.~\ref{eq:sQDM}  and becomes  a
ground state  for $J\geq  V$.   It breaks  several lattice  symmetries
(four-fold degenerate) and is a VBC.

\begin{figure}
  \begin{center}
    \includegraphics[height=3cm]{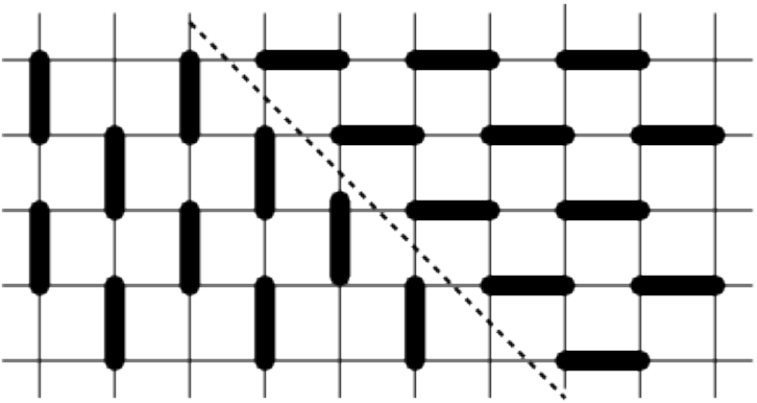}
    \caption[99]{Domain wall in a staggered VBC.}
    \label{fig:stagg2}
  \end{center}
\end{figure}

The  expectation value  of the  energy per  plaquette  satisfies ${\rm
min}(0,V-J)\leq E_0/N_p\leq {\rm max}(0,V+J)$.  For $V/J>1$ this gives
$0 \leq E_0/N_p$ and any  zero-energy state saturates this lower bound
and is a therefore a  ground state.  One should however notice that it
is possible  to make  zero-energy domain walls  in this VBC  since the
state   shown  Fig.~\ref{fig:stagg2}  is   also  annihilated   by  the
Hamiltonian. No local  dimer movement can take place  in the staggered
VBC  (with or  without domain  walls).  Each  of these  states  form a
topological   sector   with   a   single   configuration   which   has
$\left|\Omega_x\right|+\left|\Omega_y\right|=L/2$  with  respect to  a
columnar state.

\subsubsection{Columnar crystal for $V<0$}

When parallel dimers  attract each other the system  tries to maximize
the number  of flippable plaquettes. Columnar  configurations as shown
on the left side of Fig.~\ref{fig:sQDM} do maximize this number.  Such
a VBC is exactly realized  for $V<0$ and $J=0$. Elementary excitations
consist of a pair of (say) vertical dimers in a background of vertical
columns  of horizontal  dimers. Such  excitations are  gapped ($\Delta
E=2|V|$) and this VBC phase will  survive to the inclusion of a finite
$J$  term.  Notice  that unlike  the  staggered VBC  presented in  the
previous paragraph  the columnar dimer  configuration is not  an exact
eigenstate  when $J\ne0$.   The exact  diagonalizations data  of Leung
{\it et al}\cite{lcr96} suggest  that the columnar phase may disappear around
a critical value $V/J\simeq -0.2$.

\subsubsection{Plaquette phase}

When the kinetic energy dominates  ($|V|\ll J$) the system will try to
maximize   the   number    of   resonating   plaquettes   $\left|   \|
\right>+\left|=\right>$. This  can be achieved  through the resonating
plaquette  crystal shown   Fig.~\ref{fig:sQDM} and  the numerical work
(exact  diagonalizations  up  to  $8\times8$ sites)  of  Leung {\it et
al}\cite{lcr96} suggest  that this phase   is realized in  an interval
$-0.2\leq V/J \leq 1$. Although  this model would  not suffer from the
fermion sign  problem we are  not aware of  any QMC simulation on this
model.

\subsubsection{Rokhsar-Kivelson point}

The point $J=V$ (Rokhsar-Kivelson (RK) point) plays a special role. As
remarked  by  Rokhsar   and  Kivelson\cite{rk88}  the  equal-amplitude
superposition of all dimerizations in a given topological sector is an
exact ground state. When $J=V$ the Hamiltonian can be written as a sum
of projectors:
\begin{eqnarray}
  \mathcal{H}_{J=V=1}&=&\sum_{p} \left|\Psi_p\right>\left<\Psi_p\right| \label{eq:sRK}\\
  \left|\Psi_p\right>&=&\left|\vv\right> - \left|\hh\right>
\end{eqnarray}
The linear superposition  of all dimer coverings belonging  to a given
sector $\Omega$
\begin{equation}
  \left|0\right>=\sum_{c\in\Omega}\left| c\right>
\label{eq:RK}
\end{equation}
is annihilated  by Eq.~\ref{eq:sRK}  and is therefore  a ground state.
The  argument  is the  following.   Consider  a  plaquette $p$  and  a
configuration $\left|  c\right>$. If $\left|  c\right>$ has one  or no
dimer at all  on the edges of $p$  we have $\left<\Psi_p |c\right>=0$.
If two dimers  are present, then there exists  a configuration $\left|
c'\right>$ in the same sector which only differ from $\left| c\right>$
by a  two-dimer flip on $p$.   In such a case  the combination $\left|
c\right>+\left|c'\right>$       is      again       orthogonal      to
$\left|\Psi_p\right>$. This shows that $H\left| 0\right>=0$.

When open boundary    conditions are considered    (this restricts
the topological  sector to   $\Omega_x=\Omega_y=0$)  the RK  state
is  the linear     combination      of      an    exponential
number     of configurations.\footnote{This   is also   true for
periodic boundary conditions provided the two   winding numbers do
not scale like  the linear size $L$.}  This is  very different
from the crystalline states considered so far where some  periodic
configurations were favored and it closely resembles Anderson's
RVB picture\index{RVB}.   As we shall see this RK state is not a
``true'' liquid on the square lattice since dimer-dimer
correlations are not  short-range  but   algebraically decaying
($\sim1/r^2$) with distance.     The  calculation    of
dimer-dimer correlations in  the  RK  state (Eq.~\ref{eq:RK}) maps
onto   a {\em classical    }   dimer  problem   solved  by
Kasteleyn,\cite{k61,k63} Fisher\cite{f61}    and Fisher and
Stephenson.\cite{fs63}  From this Rokhsar  and Kivelson\cite{rk88}
constructed gapless excitations  (in single-mode approximation)
which dispersion relation vanishes as ${\bf k}^2$ at small
momentum (the origin is taken at $(\pi,\pi)$). Quoting Rokhsar and
Kivelson,\cite{rk88}  these excitations  (dubbed ``resonons'') are
the {\em Goldstone mode\index{Goldstone modes}  of the gauge
symmetry which allows the phases of the different   topological
sectors to be varied without changing the energy}.  Another   mode
of gapless excitations (around $(\pi,0)$ and  $(0,\pi)$  in the
Brillouin zone),  specific to the fact that  the  ground state has
critical (algebraic)  dimer-dimer correlations, was recently
discovered.\cite{ms03}

The QDM \index{QDM} on the square  lattice is thus  believed to be
ordered (VBC) everywhere  except  at  the  RK  point  ($J=V$)
where  it has  quasi long-range (critical) dimer-dimer
correlations.

\subsection{Hexagonal lattice}\index{quantum hexagonal lattice}

\begin{figure}
  \begin{center}
    \includegraphics[height=3cm]{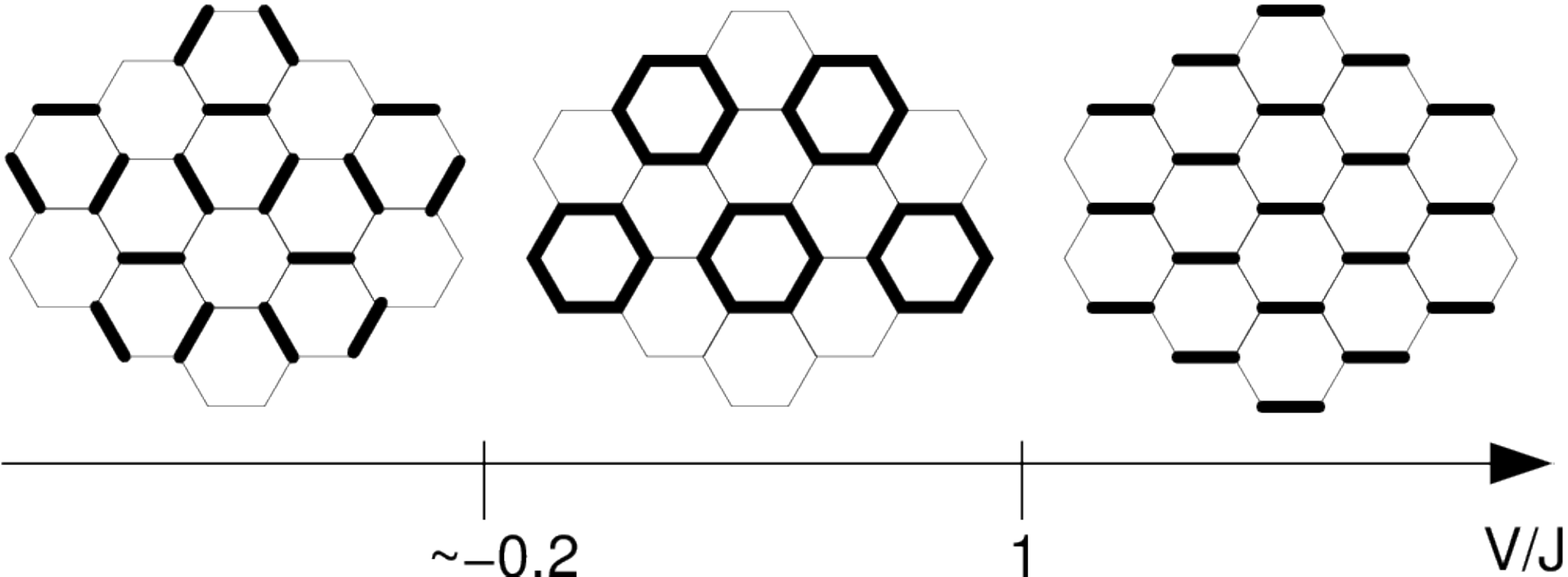}
    \caption[99]{  Phase  diagram of  the  hexagonal  QDM obtained  by
      Moessner  {\it et  al.}\cite{msc01} Although  the VBC  shown for
      $V<0$ do not  have all dimers parallel to  the same direction it
      is equivalent  to the columnar  VBC found in the  square lattice
      QDM\index{QDM} because it maximizes  the number of flippable plaquettes. It
      also  corresponds  to  the  ordering pattern  predicted  in  the
      large-$N$ theory of Ref.~\cite{rs90} As for the VBC obtained for
      $V/J>1$, it is the  hexagonal counterpart of the {\em staggered}
      VBC  of  the  square  lattice  (no  flippable  plaquette,  exact
      eigenstate and maximum tilt in a height representation).}
    \label{fig:hQDM}
  \end{center}
\end{figure}

The QDM \index{QDM} on  the honeycomb lattice was  studied by
Moessner, Sondhi and Chandra,\cite{msc01} in particular  with
Monte Carlo simulations.  The phase   diagram is very similar  to
the square lattice-case discussed above.  It  possesses three
crystalline phases  (Fig.~\ref{fig:hQDM}) and has algebraically
decaying dimer-dimer correlations at the Rokhsar Kivelson  point
(where the ground state  in  each sector has the  same form  as
Eq.~\ref{eq:RK}).
 The  absence of liquid phase  (with
exponentially   decaying 2$n$-mer-2$n$-mer  correlations)   is  believed  to
prevail in  bipartite lattices.
This relation  between the absence of a
deconfined dimer  liquid  phase
\footnote{
Bipartiteness  seems to  forbid deconfinement but  not short-range
dimer-dimer correlations. The 4-8 lattice (squares and octogons) is an
example  where dimer-dimer correlations   are short-range.   We thank
R.~Moessner  for   pointing   this  to    us.  On this   lattice   the
equal-amplitude superposition of all coverings  would be similar to an
explicit VBC wave function (thus  confining). Such situations are only
possible when the number of sites is even in the  unit cell.}
and the  bipartite character  of   the lattice  as been discussed   by
several  authors\cite{hkms03} and is related  to the  possibility of a
{\em height   representation}\cite{height_representation}    of  dimer
coverings when the lattice is bipartite.\footnote{Consider a bipartite
lattice with coordination number $z$.  For  each dimer covering we can
associate integers  (heights) on  the  dual  lattice by  the following
rule.  Set the  height to be zero  on a plaquette  at the origin.  The
height  is  then defined  on the  whole  lattice  by turning clockwise
(resp.      counterclockwise)  around  sites    of  the $A$-sublattice
(resp. $B$-sublattice) and changing the  height by $z-1$ when crossing
a  dimer and by  $-1$ when crossing  an empty bond.    It is simple to
check the difference  of  heights $\delta h(x)=h_1(x)-h_2(x)$  between
two dimerizations is constant   inside each loop of their   transition
graph.  Notice that  the loops of a  transition graph can be naturally
oriented on a bipartite lattice.   Then, $\delta h(x)$ changes by $+z$
(resp.  $-z$) when crossing a clockwise (resp.  counterclockwise) loop
of the  transition  graph.   Columnar dimerizations  have  an averaged
height  which is flat  and staggered ones have  the maximum tilt.  The
winding numbers $(\Omega_x,\Omega_y)$ correspond to the average height
difference between both sides of the sample.  The kinetic energy terms
of  Eqs.~\ref{eq:sQDM} and   \ref{eq:hQDM}  change the height   of the
corresponding  plaquette  by $\pm z$  and  the potential terms ($V>0$)
favor tilted configurations.}

\newcommand{\ha}{
  \begin{picture}(32,14)(-16,-2)
    \Hex\pA{\La}\pC{\Lc}\pE{\Le}
  \end{picture}
}
\newcommand{\hb}{
  \begin{picture}(32,14)(-16,-2)
    \Hex
    \pB{\Lb}\pD{\Ld}\pF{\Lf}
  \end{picture}
}

\begin{eqnarray}
  \mathcal{H}=&-J&\sum_h \left(
  \left|\ha\right>\left<\hb\right| +{\rm H.c.}
  \right) \nonumber \\
  &+V&\sum_h\left(
  \left|\ha\right>\left<\ha\right|+\left|\hb\right>\left<\hb\right|
  \right)
\label{eq:hQDM}
\end{eqnarray}

Fouet   {\it  et   al.}\cite{fsl01}   studied  the
spin-$\frac{1}{2}$ $J_1$--$J_2$--$J_3$    model   on the hexagonal
lattice  by  exact diagonalizations\index{exact diagonalizations}
and  found evidences of a staggered VBC of  the type predicted for
$V/J>1$ by Moessner {\it et al.}\cite{msc01}  in  the QDM. Other
phases (N\'eel ordered phase  and a possible short-range RVB SL)
are also present in the spin-$\frac{1}{2}$ model.\cite{fsl01}

\subsection{Triangular lattice}\index{triangular lattice}
\label{ssec:QDMTri}

\newcommand{\rhomb}{
  \pA{\C}\pB{\C}\pZ{\C}\pC{\C}
 }
\newcommand{\rhombF}{
  \begin{picture}(18,10)(-5,-6)
    \rhomb
    \pA{\La}\pB{\Lb}\pZ{\Le}\pC{\Ld}
  \end{picture}
}
\newcommand{\rhombH}{
  \begin{picture}(22,10)(-8,-6)
    \rhomb
    \pA{\La}\pC{\Ld}
  \end{picture}
}
\newcommand{\rhombV}{
  \begin{picture}(22,10)(-8,-6)
    \rhomb
    \pB{\Lb}\pZ{\Le}
  \end{picture}
}

\begin{figure}
\begin{center}
\includegraphics[width=8.5cm]{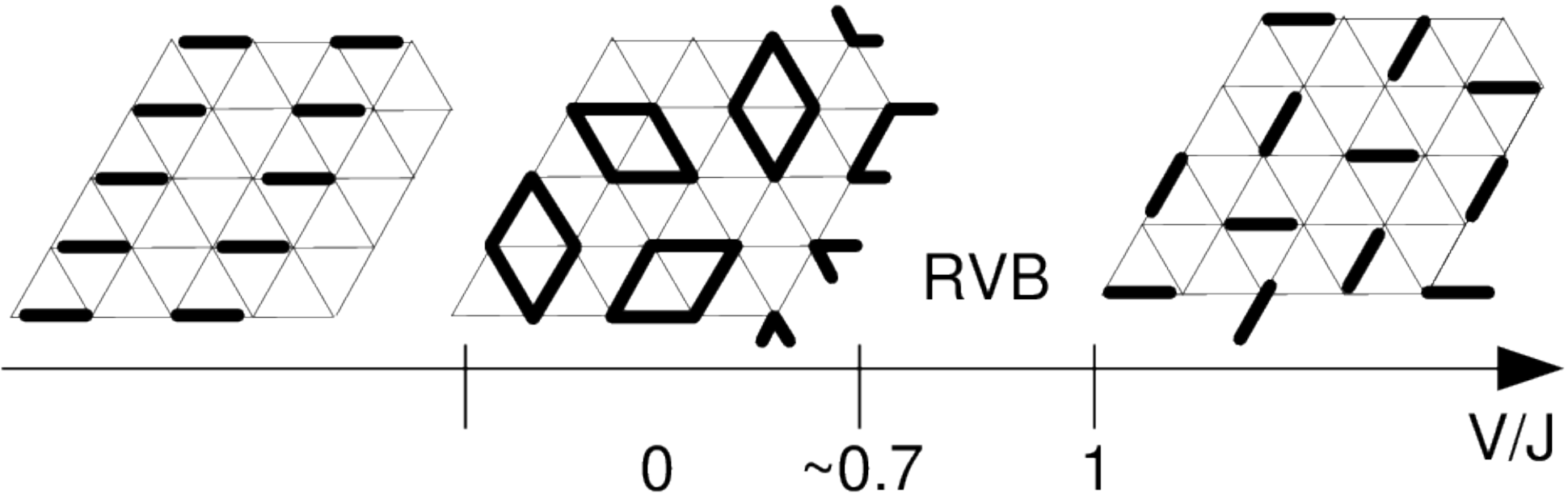}
\caption[99]{Phase diagram of the triangular lattice QDM
obtained by Moessner and Sondhi.\cite{ms01}}
\label{fig:QDMtri}
\end{center}
\end{figure}

The most  local dimer Hamiltonian  on the triangular  lattice contains
kinetic and potential two-dimer terms  on each rhombus; it was studied
by Moessner and Sondhi:\cite{ms01}
\begin{eqnarray}
  \mathcal{H}=&-J&\sum_r \left(
  \left|\rhombV\right>\left<\rhombH\right| +{\rm H.c.}
  \right) \nonumber \\
  &+V&\sum_r\left(
  \left|\rhombV\right>\left<\rhombV\right|+\left|\rhombH\right>\left<\rhombH\right|
  \right)
\label{eq:tQDM}
\end{eqnarray}
where  the sums run over  all  rhombi $r$  of the  lattice (with three
possible orientations).   This model was  shown to  possess (at least)
three crystalline phases,  including a   columnar and staggered    one
(Fig.~\ref{fig:QDMtri}) as in the two previous examples. An additional
VBC  (with resonating diamonds plaquettes)  with a large unit cell (12
sites) was also   predicted around $V=0$.   When $V<0$  and $J=0$  the
ground state  is highly  degenerate  since it  is   possible, from  an
ordered columnar configuration,   to shift all  the dimers  along  any
straight line   without changing the  number  of  flippable plaquettes
(contrary to the square lattice case).   However, an infinitesimal $J$
is expected to lift this degeneracy and to order the ground state in a
columnar way.

The phenomenology of these ordered phases is that  of usual VBC and we
refer to the original paper\cite{ms01} for details.

\subsubsection{RVB liquid at the RK point}\index{RVB}

The new physics of this model appears through  the existence of a {\em
liquid} phase in the interval $0.7\lesssim V/J \leq 1$. As for the two
previous  QDM  the ground states are  exactly  known at   the RK point
$J=V$. As before dimer-dimer correlations are obtained exactly at this
point by  a Pfaffian calculation\cite{ms01,fms02,iif02} but the result
shows a   {\em  finite} correlation length.    From  their Monte Carlo
simulations Moessner {\it et al.}\cite{ms01}  argued that the spectrum
is   gapped  at the  RK   point and  that  this   gap persists down to
$V/J\simeq 2/3$,  that is  over  a  finite range   of  coupling.  This
picture   is consistent with the   exact diagonalizations performed on
this model.\cite{ifiitb02,mlms02}

\subsubsection{Topological order}
\label{subsec:topologicalorder}

When the lattice is not bipartite the  loops of a transition graph can
no longer be oriented.   The winding numbers $\Omega_x$ and $\Omega_y$
are now  defined as the (positive)  number  of non-trivial loop around
$x$  and $y$  in the transition  graph  with a reference configuration
(say columnar).   These two integers are not  conserved by local dimer
moves. They are however {\em conserved modulo  two}, which leaves only
four   sectors.\footnote{The        two-dimer moves     included    in
Eq.~\ref{eq:tQDM} are not  sufficient to guaranty ergodicity with each
of the  four  sectors.  Staggered   states  (12-fold degenerate  - not
$\mathcal{O}(L)$ like on   the  square  lattice) have   no   flippable
plaquette  but  can be    connected  to other states with   four-dimer
moves.\cite{ms01}} Consequently {\em the  dimer liquid ground state is
four-fold degenerate} at the RK  point.  This degeneracy holds exactly
at the RK point even on finite-size samples but it is expected to hold
in the thermodynamic limit in the whole liquid phase ($0.7\lesssim V/J
\leq 1$).

Conventional  orders are  often associated  to a  spontaneously broken
symmetry and  lead to  ground state degeneracies in  the thermodynamic
limit.  The  four-fold degeneracy discussed above is  the signature of
some kind of order, called topological
order.\cite{wen91}  The peculiarity  of this  order  is that it is not
associated  to any local  order  parameter: a local observable  cannot
decide whether a given dimerization is in one  sector or another.  The
existence of topological   order   is intimately associated   to   the
fractionalized nature of the  elementary excitations (see  below).  In
the case of a RVB dimer liquid these excitations have been known to be
{\em  Ising vortices}  for   a long  time\cite{k89,rc89}  (dubbed {\em
visons}  in the   recent  literature\cite{sf00,sf01}).   We  will  now
discuss these excitations in more  details in the  framework of a  QDM
which realizes the same dimer liquid phase but for  which not only the
ground state but all the eigenstates are known exactly.

\subsection{Solvable QDM on the kagome lattice}\index{Kagom\'e lattice}
\label{ssec:QDMKag}

An   exactly  solvable QDM on  the  kagome  lattice was  introduced by
D.~Serban,  V.~Pasquier  and one of us.\cite{msp02}   It offers a very
simple and explicit realization of the  ideas discussed above (visons,
topological order etc.).

\subsubsection{Hamiltonian}
The kagome lattice  QDM introduced in  Ref.~\cite{msp02} contains only
kinetic terms and has no external parameter. The Hamiltonian reads:
\begin{eqnarray}
  \mathcal{H}&=&-\sum_h \sigma^x(h) \label{eq:kQDM}\\
{\rm where }\;\;
  \sigma^x(h)&=&\sum_{\alpha=1}^{32}
  \left|d_\alpha(h) \right>\left< \bar{d}_\alpha(h)\right|
  +
  {\rm H.c}
  \label{eq:sigmax}
\end{eqnarray}
The sum runs over the  32 loops on  the lattice which enclose a single
hexagon    and around      which     dimers can   be     moved    (see
Table~\ref{tab:kloops}  for  the 8 inequivalent loops).   The shortest
loop is the  hexagon itself, it involves  3 dimers.  4, 5 and 6-dimers
moves are also possible  by including 2, 4  and 6 additional triangles
(the  loop length must be  even).  The largest loop is  the star.  For
each loop $\alpha$   we associate the  two  ways dimers can be  placed
along                that      loop:  $\left|d_\alpha(h)\right>$   and
$\left|\bar{d}_\alpha(h)\right>$.    Notice that $\sigma^x(h)$ measures
the relative  phases  of dimer configurations displaying  respectively
the   $d_\alpha(h)$    and $\bar{d}_\alpha(h)$   patterns   in the wave
function.

\begin{table}
   \tbl{
The 8 different classes of loops which can  surround an hexagon of the
kagome lattice.  Including all possible symmetries we find 32 possible
loops. The first column indicates the number of dimers involved in the
coherent motion around the hexagon.  \label{tab:kloops}}
{ \begin{tabular}{|c|c|c|c|}
      \hline
      3&
      \begin{picture}(50,34)(-24,-15)
    \pA{\La}\pB{\Lb}\pC{\Lc}\pD{\Ld}\pE{\Le}\pF{\Lf}\KagHex
      \end{picture} &  & \\
      \hline 
      4&
      \begin{picture}(50,34)(-24,-13)
    \pA{\La}\pB{\La}\pG{\Lc}\pC{\Lc}\pD{\Ld}\pE{\Ld}\pJ{\Lf}\pF{\Lf}
    \KagHex\pG{\C}\pJ{\C}
      \end{picture} &
      \begin{picture}(50,34)(-24,-13)
    \pA{\La}\pB{\La}\pG{\Lc}\pC{\Lc}\pD{\Ld}\pE{\Le}\pF{\Le}\pK{\La}
    \KagHex\pG{\C}\pK{\C}
      \end{picture} &
      \begin{picture}(50,34)(-24,-13)
    \pA{\La}\pB{\La}\pG{\Lc}\pC{\Lb}\pH{\Ld}\pD{\Ld}\pE{\Le}\pF{\Lf}
    \KagHex\pG{\C}\pH{\C}
      \end{picture}\\
      \hline 
      5&
      \begin{picture}(50,34)(-24,-13)
    \pA{\La}\pB{\La}\pG{\Lc}\pC{\Lb}\pH{\Ld}\pD{\Ld}
    \pE{\Ld}\pJ{\Lf}\pF{\Le}\pK{\La}
    \KagHex\pG{\C}\pJ{\C}\pK{\C}\pH{\C}
      \end{picture} &
      \begin{picture}(50,44)(-24,-23)
    \pA{\Lf}\pB{\Lb}\pL{\Lb}\pC{\Lb}\pH{\Ld}\pD{\Ld}
    \pE{\Ld}\pJ{\Lf}\pF{\Le}\pK{\La}
    \KagHex\pL{\C}\pJ{\C}\pK{\C}\pH{\C}
      \end{picture} &
      \begin{picture}(50,44)(-24,-23)
    \pA{\Lf}\pB{\La}\pL{\Lb}\pC{\Lc}\pG{\Lc}\pD{\Ld}
    \pE{\Ld}\pJ{\Lf}\pF{\Le}\pK{\La}
    \KagHex\pL{\C}\pJ{\C}\pK{\C}\pG{\C}
      \end{picture}\\
      \hline 
      6&
      \begin{picture}(50,50)(-24,-23)
    \pA{\Lf}\pB{\La}\pL{\Lb}\pC{\Lb}\pG{\Lc}\pD{\Lc}
    \pE{\Ld}\pJ{\Lf}\pF{\Le}\pK{\La}\pH{\Ld}\pI{\Le}
    \KagStar
      \end{picture}&  & \\
      \hline 
    \end{tabular}
}
\end{table}

\subsubsection{RK ground state}
As  for   the  QDM discussed   previously  the  ground state   of this
Hamiltonian   is the  equal  amplitude   superposition  of  all  dimer
coverings belonging    to   a given  topological  sector   (as  on the
triangular  lattice there  are four sectors).   This can be
readily shown by writing  $\mathcal{H}$ as a sum  of projectors:
\begin{eqnarray}
  \mathcal{H}&=&-N_h+\sum_h \sum_{\alpha=1}^{32}
  \left[\;\left|d_\alpha(h) \right>-\left| \bar{d}_\alpha(h)\right>\;\right]
  \left[\;\left<d_\alpha(h) \right|-\left< \bar{d}_\alpha(h)\right|\;\right]
\end{eqnarray}
where $N_h$ is the number of hexagons  on the lattice.  When expanding
the products the diagonal terms give a simple constant since
\begin{equation}
  \sum_{\alpha=1}^{32} \left|d_\alpha \right>\left<d_\alpha\right|
  +
  \left| \bar{d}_\alpha\right>\left<\bar{d}_\alpha\right| = 1
\end{equation}
This  reflects  the fact that,  for  any  dimerization, the  dimers on
hexagon $h$  match {\em one and only  one} of the $2\times32$ patterns
$\left\{d_\alpha,\bar{d}_\alpha\right\}$.

Unlike  the  square    or  triangular case,    the   RK  ground states
$|0\rangle=\sum_{c\in\Omega}|c\rangle$ are   not  degenerate with some
staggered VBC.\footnote{Because resonances loops  of  length up to  12
are present the  dynamics is ergodic in each  of  the four topological
sectors.\cite{msp02}}   This    means  that    the    Hamiltonian   of
Eq.~\ref{eq:kQDM} is not at  a phase transition to a  VBC.  As we will
explain it  is {\em inside} a  liquid RVB phase.

The RK wave function can be viewed as dimer condensate.  It is
similar to  the ground state  of liquid $^4$He  which   has the
same positive amplitude      for  any       configuration     and
its    permuted images.\cite{feynman52-53} An important
difference, however,  is that the  QDM  \index{QDM} state is
incompressible and cannot sustain acoustic phonons. This can be
related to  the fact that the $U(1)$  symmetry of the Bose liquid
is absent in the QDM on non-bipartite lattices.  It is replaced
instead by   a discrete  $\Z$  gauge  symmetry  (see
\S\ref{sssec:Z2} below).

\subsubsection{Ising pseudo-spin variables}

\begin{figure}
  \begin{center}
    \includegraphics[height=4.5cm,width=6cm]{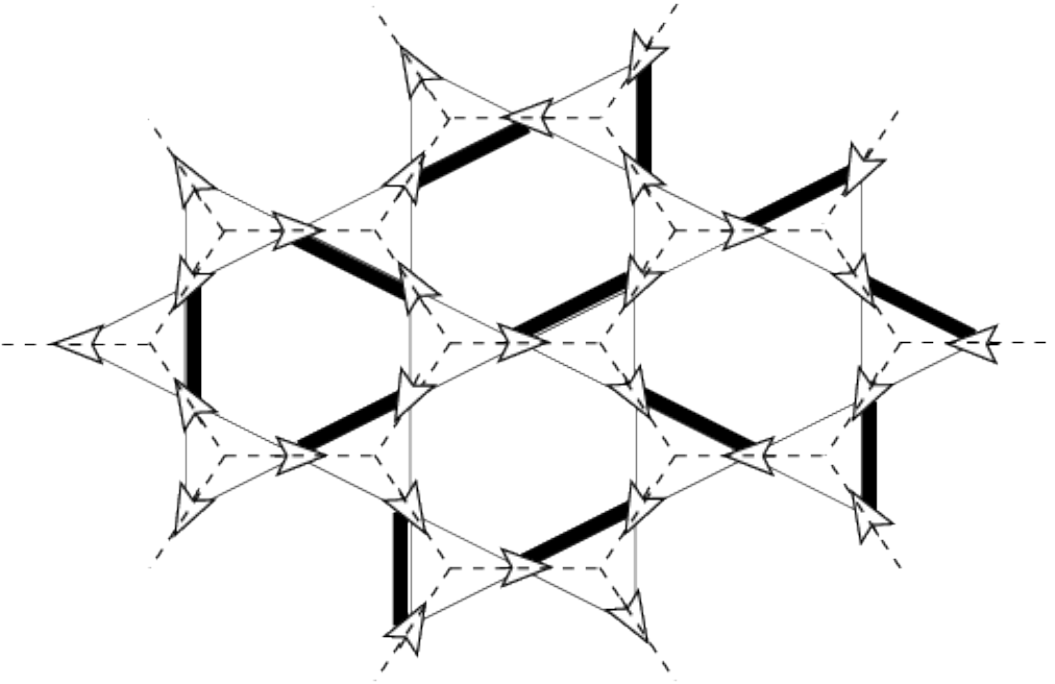}
    \caption[99]{A  dimer  covering  on  the kagome  lattice  and  the
    corresponding arrows. Dashed lines: honeycomb lattice.}
    \label{fig:arrow}
  \end{center}
\end{figure}

The kinetic energy operators $\sigma^x$ defined in Eq.~\ref{eq:sigmax}
commute with each  other. This is obvious when  two such operators act
on remote  hexagons but it   also  holds for  neighboring ones.   This
property can   easily be  demonstrated  with the   help of  the  arrow
representation   of  dimer     coverings  introduced    by    Zeng and
Elser.\cite{ez93} This mapping of  kagome  dimerizations to arrows  on
the   bonds     of   the  honeycomb       lattice  is      illustrated
Fig.~\ref{fig:arrow}.  Each  arrow has   two possible  directions:  it
points toward the interior of  one of  the two neighboring  triangles.
If site $i$ belongs to a dimer $(i,j)$ its arrow must point toward the
triangle the site $j$ belongs to.  A dimer covering can be constructed
from any   arrow configuration provided   that the number  of outgoing
arrows is odd (1 or 3) on every triangle.

The operators $\sigma^x$ have a  particularly simple meaning in  terms
of the arrow  degrees  of freedom: $\sigma^x(h)$  flips  the  6 arrows
sitting  on $h$.\footnote{Flipping all   the arrows around  any closed
loop (such  as around an   hexagon)  preserves  the local   constraint
imposed on  arrow configurations.    Flipping   the arrows around    a
topologically non-trivial loop changes the topological sector.}  It is
then clear that the $\sigma^x$ commute and that $\sigma^x(h)^2=1$.  In
fact these operators can  be used as  Ising pseudo-spin  variables and
the  Hamiltonian now   describes  non-interacting pseudo-spins  in   a
uniform   magnetic field  pointing  in  the   $x$ direction.   In  the
ground state we have $\sigma^x(h)=1$ on every hexagon.

\subsubsection{Dimer-dimer correlations}
The  ground state is the most possible  disordered dimer liquid as the
dimer-dimer  correlations strictly  vanish    beyond  a few    lattice
spacings.  Such  correlations  can    be  computed by    the  Pfaffian
method. On the kagome lattice the determinant  of the Kasteleyn matrix
(which is directly related to the partition  function of the classical
dimers problem) is exactly constant in Fourier space.\cite{hw88} Since
dimer-dimer correlations  are obtained from  the  Fourier transform of
the inverse  of  this determinant, they  turn out  to be strictly zero
beyond a few lattice spacings (as soon as the two bonds  do not touch a
common  triangle).\cite{msp02} This result can also   be obtained by a
simpler  argument\cite{msp02,msp03b} using  the  $\sigma^x$ operators.
This result is related  to the kagome geometry.\footnote{The model  of
Eq.~\ref{eq:kQDM}   can  be  generalized   to any    lattice   made of
corner-sharing triangles.\cite{msp02}}  This absence   of  long-range
dimer-dimer correlations  demonstrates that the   RK state is  a dimer
liquid and that it breaks no lattice symmetry.

On the  triangular lattice,   even at high    temperature, dimer-dimer
correlations decay exponentially  with distance but these correlations
remain {\em finite}  at  any distance.  On   the  square lattice  such
correlations are  even  larger  because they  decay  only as  a  power
law. This means that the  infinite hard-core dimer repulsion makes QDM
non-trivial  even at infinite temperature; dimers  cannot be free when
they are fully-packed.  From this point of view we see that the kagome
lattice is particular: it is as close as possible to a free dimer gas,
except for non-trivial correlations over  a few lattice spacings. This
is a reason why  dimer coverings on the  kagome lattice can be handled
with independent pseudo-spin  variables and why  the RK state  on this
lattice is the most possible disordered RVB liquid.

\subsubsection{Visons excitations}
\label{sssec:vison}

\begin{figure}
  \begin{center}
    \includegraphics[height=2.5cm]{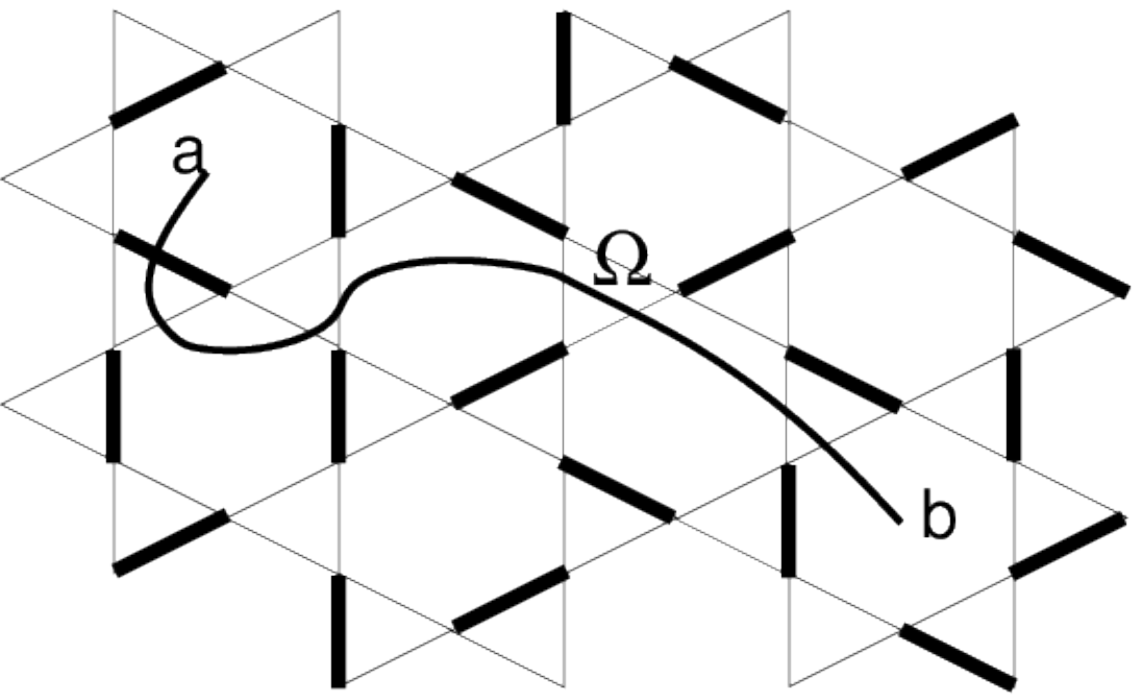}
    \includegraphics[height=2.5cm]{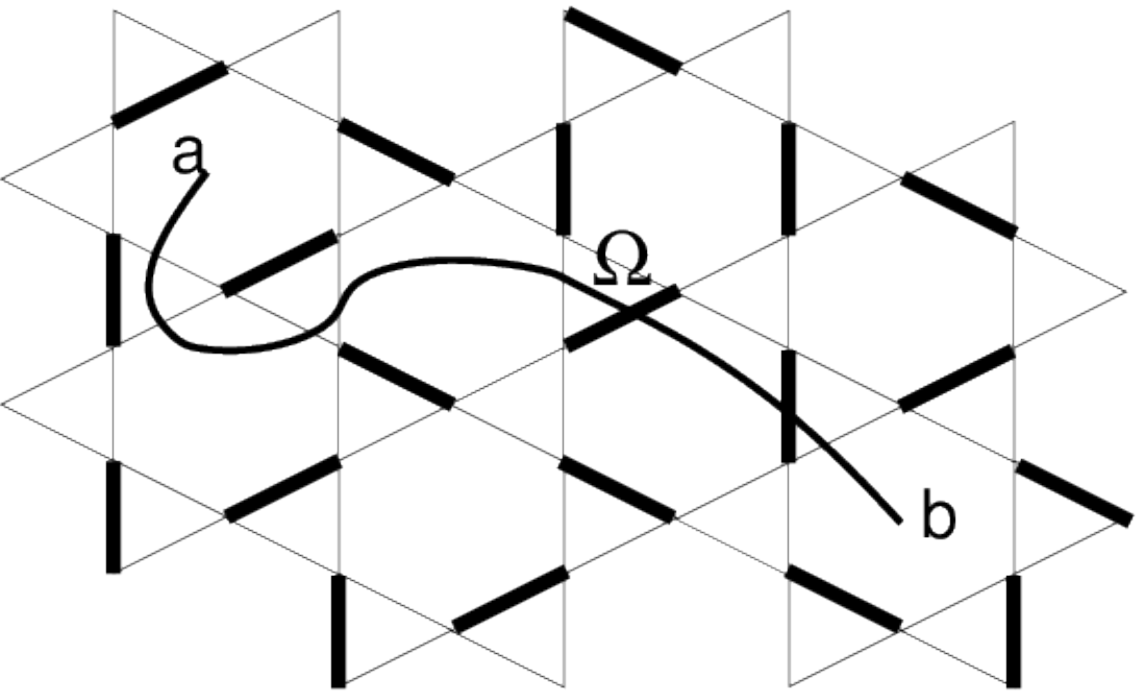}

  \caption[99]{A pair of visons (located in $a$ and $b$) is created by
  applying  to the  RK wave function  a factor  $(-1)$  for each dimer
  crossing   the string $\Omega$. The  dimerization  shown there on the left
   appears  in the  linear superposition  of  the two-vison state
  with the sign $-1$ whereas the one on the right has the sign $+1$.}

  \label{fig:Vison} \end{center}
\end{figure}

The $\sigma^x$ operators  can be simultaneously diagonalized but  they
must satisfy the global constraint  $\prod_h \sigma^x(h)=1$ since this
product flips every  arrow {\em twice}.  It  must  therefore leave all
dimerizations unchanged.   The  lowest  excitations  have therefore an
energy $4$ above the ground state and they are made of a {\em pair} of
hexagons $a$ and $b$  in a $\sigma^x(a)=\sigma^x(b)=-1$ state. $a$ and
$b$  are the    locations     of  two   Ising   vortices   (or    {\em
visons}\cite{sf00,sf01}).   As  remarked before  this   means that the
relative  phases   of  the  configurations   with   $d_\alpha(h)$  and
$\bar{d}_\alpha(h)$ patterns  have now changed sign. The corresponding
wave function  is  obtained in the   following way.  Consider a string
$\Omega$ which goes from $a$ to $b$ (see Fig.~\ref{fig:Vison}) and let
$\Omega(a,b)$ be the operator which measures  the parity $\pm1$ of the
number of  dimers crossing that   string.  $\Omega(a,b)$ commutes with
all  $\sigma^x(h)$,    except  for   the     ends   of the     string:
$\sigma^x(a)\Omega(a,b)=-\Omega(a,b)\sigma^x(a)$. A dimer move changes
the sign of  $\Omega(a,b)$ if and only if  the associated loop crosses
the  string an   odd  number of  times,  which  can  only  be done  by
surrounding one  end of  the  string.  This shows  that  $\Omega(a,b)$
flips the   $\sigma^x$ in $a$ and  $b$.\footnote{Up   to a global sign
(reference         dependent)   $\Omega(a,b)$       is  equal       to
$\sigma^z(a)\sigma^z(b)$  where  the $\sigma^z$   operators are  those
introduced by  Zeng and Elser.}  As  the  RK ground state $|0\rangle$,
$\Omega(a,b)|0\rangle$  is   a  linear   combination  of   all   dimer
configurations belonging to one sector. However the amplitudes are now
$1$ and $-1$ depending on the number of dimers crossing $\Omega$. This
wave function  therefore has nodes, it is  an excited  state of energy
$4$ with  two vortices  in $a$ and   $b$.  It is   easy to see that  a
different choice $\Omega'$ for the string connecting $a$ and $b$ gives
the same state up to a global sign which depends  on the parity of the
number of kagome sites enclosed by $\Omega\cup\Omega'$.

These vortex excitations carry a $\Z$ charge  since attempting to put
two  vortices on  the same  hexagon does  not change  the state.  Such
excitations are not local  in terms of  the dimer degrees  of freedom.
Indeed,  determining the sign of a  given dimerization in a state with
two visons which  are far apart requires  the  knowledge of the  dimer
locations along the whole string connecting the  two vortex cores.  In
this model the visons appear to be static and non-interacting. This is
a particularity of  this  solvable model  but the  existence of gapped
vison excitations  is   believed  to be  a   robust  property   of RVB
liquids. In more  realistic models the  visons will acquire a dynamics
and  a  dispersion  relation  but will   remain gapped.\footnote{It is
possible to add potential energy  terms to Eq.~\ref{eq:kQDM} to  drive
the system outside of the liquid phase and this transition corresponds
to a vison condensation.}  They will also  have some interactions with
each  other but  should remain {\em    deconfined}.  This property  is
particularly clear in the  kagome QDM: visons are  necessarily created
by pairs but the energy is independent of their relative distances.

The  Ising vortices  also offer a  simple picture  of the  topological
degeneracy.  Consider  a  ground state $|+\rangle$ of  the model which
lives in the sector where the  winding number $\Omega_y$ (with respect
to  some  arbitrary  but  fixed    dimerization) is  even.     Another
ground state $|-\rangle$ is obtained in the odd-$\Omega_y$ sector. Now
consider the combination $|0\rangle=|+\rangle+|-\rangle$ and apply the
operator $\Omega(0,L_x)$ corresponding  to  a closed loop  surrounding
the torus  in the $x$ direction. This   amounts to creating  a pair of
nearby visons at  the origin, taking  one of them  around the torus in
the $x$  direction and annihilating them.  This  can also be viewed as
the creation  of a  vison in  one hole  of  the torus (with  no energy
cost).         It       is      simple      to       check        that
$\Omega(0,L_x)|0\rangle=|+\rangle-|-\rangle$   (up to an    irrelevant
global sign).  This provides  a simple relation between the vison-pair
creation  operator and the  existence   of two topologically  distinct
ground states $|+\rangle+|-\rangle$ and $|+\rangle-|-\rangle$.

\subsubsection{Spinons deconfinement}
\label{sssec:QDMdeconf}

We assume that dimers represent  ``dressed'' singlet
valence-bonds, as in  the overlap expansion
(\S\ref{ssec:spin2QDM}).  Since the Hilbert space is   made   of
fully-packed dimer  coverings  the     model of Eq.~\ref{eq:kQDM}
only describes spin-singlet states.  However, as any QDM,   it can
be extended  to   include  static  holes or
spinons.\index{spinons} Configurations  with unpaired sites
(spinon or holon) are now allowed but the kinetic terms of the
original Hamiltonian which loop passes on an empty site  gives
zero. Consider a  system with two static spinons in $x$ and   $y$.
As on  the square\cite{rk88} and triangular lattices\cite{ms01} at
the  RK point the exact ground state $\left|x,y\right>$ remains
the sum of all dimer coverings and   the ground state energy is
independent of the distance between the  two spinons (except at
very short distance if they belong  to a  common hexagon).  This
is  a first indication that RVB  {\em spin} liquid has deconfined
spin-$\frac{1}{2}$ excitations   (spinons).   In  the  QDM
language these excitations are simply unpaired sites in a dimer
liquid background. Such  unpaired sites are necessarily  created
by pairs but they can then propagate freely (no attractive
potential) when they are sufficiently far apart.

Another calculation allows  to test the  deconfinement properties of a
dimer liquid.  We   consider the state   $\left|\psi\right>=\sum_{{\bf
r}\ne0}\left|0,{\bf r}\right>$ where  $\left|0,{\bf r}\right>$ is  the
(un-normalized) ground state with  two spinons  in $0$ and  ${\bf  r}$.   The
probability  to find a spinon  in ${\bf r}$ in the $\left|\psi\right>$
can be obtained by the relatively  involved calculation of the monomer
correlation\footnote{Ratio of number of dimer coverings with two holes
in  $0$ and ${\bf r}$  to the number  without  hole.}  with Pfaffians.
One   the  square   lattice   this  probability   goes   to   zero  as
$1/\sqrt{r}$.\cite{fs63} This shows that the second spinon is (quasi-)
confined  in the vicinity  of  the first  one  on  the square  lattice
because escaping  far away represents a large  ``entropy'' cost in the
dimer background.  On the triangular  lattice it goes exponentially to
a constant.\cite{fms02} This result  is a signature of  deconfinement.
In fact the  same  signature  can be  obtained on  the  kagome lattice
without  any technical  calculation since the   monomer correlation is
exactly $1/4$ at any distance.\cite{msp03c}

If unpaired  sites are  allowed  one can describe spinons   or holons.
Unfortunately in the presence of simple kinetic energy terms for these
objects the model can no longer be solved.  However one can consider a
static spinon  and its  interaction with  visons: when the   spinon is
adiabatically taken around a vison the dimers are shifted along a path
encircling the vison.  Because the  vison wave function is  particularly
simple in this model it   is easy to  check  that this multiplies  the
wave function  by  a  factor  $-1$.    This is   the signature  of   a
long-range statistical  interaction\cite{rc89,k89} between visons and
spinons (or holons).  In more realistic models,  as long as the visons
are gapped  excitations the spinons  are expected to be deconfined. On
the  other hand if  the  visons condense their long-range statistical
interaction with spinons frustrates  their  motion. This is no  longer
true if they propagate   in {\em pairs},   in which case they are  not
sensitive  any more to  visons (see Ref.~\cite{msp02} for an extension
of the present QDM with  a  vison condensation). This simple  physical
picture illustrates the relation between vison condensation and spinon
confinement.

\subsubsection{$\Z$ gauge theory}
\label{sssec:Z2}

The forces responsible for confinement are usually associated to gauge
fields and their fluctuations.   Whereas $U(1)$ compact gauge theories
are  generically confining in  $2+1$ dimensions,\cite{polyakov87,fs79}
$\Z$    gauge    theories   are    known    to   possess    deconfined
phases.\cite{kogut79} For this reason  some attention has been paid to
the  connections between  $\Z$  theories and  fractionalized phases  in
2D electronic systems.\cite{sf00}

It is known\cite{msf02} that QDM can be obtained as special limits of
$\Z$ gauge  theories, the gauge variable  being the dimer  number on a
bond.  However, on the kagome lattice this connection can be made exact
and completely  explicit since  there is a  one to  one correspondence
between   dimer   coverings    and   physical   states   ({\it   i.e.}
gauge-invariant) of  a $\Z$ gauge theory.\cite{msp02}  In this mapping
the  gauge  fields are  Ising  variables living  on  the  link of  the
honeycomb lattice  ({\it i.e.} kagome sites) and  are constructed from
the  arrows described  previously.  As  for the  constraints  of gauge
invariance they correspond to the odd parity of the number of outgoing
arrows on  every triangles. The  $\sigma^x$ operator used to  define a
solvable QDM  translate into a gauge-invariant  plaquette operator for
the gauge  degrees of  freedom (product of  the Ising  gauge variables
around an hexagon). With this mapping the visons appear to be vortices
in the  $\Z$ gauge field  and the solvable model  of Eq.~\ref{eq:kQDM}
maps to the deconfined phase
of the gauge theory (precisely at infinite temperature).

\subsection{A QDM with an extensive ground state entropy}
\label{ssec:QDMmu}

So far we   have discussed QDM\index{QDM} that  realize  either
spontaneous VBC, critical states  or RVB liquids.   We wish to
mention here that these three  scenarios may  not  be   the  only
possible ground states  for QDM. In particular,  a  QDM on the
kagome lattice  with an extensive ground state  entropy was
recently discussed.\cite{msp03} This model was  introduced from
the observation  that  the  dimer kinetic energy terms  arising
from an  overlap  expansion (\S\ref{ssec:spin2QDM}) generally have
non trivial {\em signs}  as soon as resonance loops of {\em
different lengths} are considered.  It was then realized that such
signs (which make  the QDM  no longer  appropriate  for QMC
simulations) can lead to  qualitatively  new phases, different
from      VBC or RVB   liquids.     The    Hamiltonian introduced
in Ref.~\cite{msp03}   is           similar      to       that of
Eqs.~\ref{eq:kQDM}-\ref{eq:sigmax} except that each resonance loop
$\alpha$   is now included with  a  {\em sign} $(-1)^{n_\alpha}$
where $n_\alpha=3,\cdots,6$ is the number of dimers involved:
\begin{equation}
    \mathcal{H}=\sum_h (-1)^{n_\alpha} \left[
    \left|d_\alpha(h) \right>\left< \bar{d}_\alpha(h)\right|
    +
    \left| \bar{d}_\alpha(h)\right>\left<d_\alpha(h)\right|
    \right]
    \label{eq:muQDM}
\end{equation}
These signs are precisely those appearing in the overlap expansion (at
the  order of one   hexagon)  of the  Heisenberg  model on  the kagome
lattice.  This expansion was  carried out by Zeng and Elser\cite{ze95}
in an  insightful   paper which laid  the  basis of the analysis  of the
kagome antiferromagnet in the first neighbor valence-bonds subspace.

Although  not exactly solvable,  the Hamiltonian of
Eq.~\ref{eq:muQDM} was shown to be a dimer  liquid (short-range
dimer-dimer correlation) and to have a huge ground state
degeneracy $\sim 2^{N/6}=1.122^N$ ($N$ is  the   number of  kagome
sites).    In addition,  several numerical indications pointed to
a critical behavior of this system,\cite{msp03} with    a
possible     algebraic     decay    of     energy-energy
correlations.\footnote{Notice   that  a   one-dimensional   analog
of Eq.~\ref{eq:muQDM} can be defined and exactly maps onto the
Ising chain in  transverse field at its  critical point.}  It  was
argued that the effective   QDM    describing      the singlet
dynamics   of   the spin-$\frac{1}{2}$ Heisenberg  antiferromagnet
on  the kagome  lattice\index{Kagom\'e lattice} could be {\em
close} (in parameter  space) to Eq.~\ref{eq:muQDM}.  If correct,
this sheds light  on the very large  density of singlet states
observed at low energy in  the numerical  spectra  of that spin
model (see \S\ref{sec:kagome}).

\section{Multiple-spin exchange
models}\label{sec:MSE}\index{multiple-spin exchange, MSE}
\setcounter{footnote}{0}


\subsection{Physical realizations of multiple-spin interactions}

\subsubsection{Nuclear magnetism of solid $^3$He}

Solid $^3$He was the first magnetic system in  which the importance of
MSE interactions   was recognized.\cite{thouless65,rhd83,cf85}  Due to
the large zero-point motion  of the atoms  about their mean  positions,
tunneling events during which 2, 3 or 4 atoms exchange their positions
in a cyclic way  are frequent.  These  processes generate an effective
interaction between the (nuclear) spins which can be written
\begin{equation}
  \mathcal{H}=\sum_P -J_P (-1)^P\left( P_{spin} + P_{spin}^{-1}\right)
\end{equation}
where the  sum runs over  permutations  $P$,  $J_P>0$ is the  exchange
frequency  of the associated tunneling   process (in  real space)  and
$P_{spin}$ acts on the  Hilbert space of spin-$\frac{1}{2}$ located on
the site of  the crystal. The sign $-(-1)^P$  depends of the signature
of the permutation  $P$ and is a consequence  of  the Pauli principle.
For a {\em cyclic}  permutation involving $n$ spins  this sign is just
$(-1)^n$  and  is  responsible   for  the ferromagnetic  character  of
processes involving  an odd number   of spins.  For spin-$\frac{1}{2}$
particles, two  and three-spin exchange  terms  reduce to the familiar
Heisenberg interaction:
\begin{eqnarray}
  P_{12}=2\vec{S}_1\cdot\vec{S}_2+\frac{1}{2} \label{eq:P2} \\
  P_{123}+P_{321}=P_{12}+P_{23}+P_{31}-1 \label{eq:P3}
\end{eqnarray}
but this is no longer true for $n\geq4$:
\begin{equation}
P_{1234}+P_{4321}=P_{12}P_{34}+P_{14}P_{23}-P_{13}P_{24}+P_{13}+P_{24}-1
\end{equation}
which can be  expressed (thanks to Eq.~\ref{eq:P2}) as  a sum of terms
with two and four Pauli matrices.

$^3$He can form solid  atomic mono-layers  with a triangular  geometry
when adsorbed on a graphite substrate at ultra low temperatures (milli
Kelvin range).  This 2D  magnet has been  studied for a long time (see
Refs.~\cite{greywall,gr95}  and references therein) and the importance
of MSE   interactions  involving up    to  six  atoms has  now    been
recognized.\cite{roger84,rbbcg98} The exchange frequencies of the most
important processes have been   computed by Path Integral  Monte Carlo
(PIMC)\cite{cj87,bcl92,ceperley95,bc99}  (analytic  WKB   calculations
have  also been carried  out\cite{roger84,ah00})  as a function of the
density.    The proposed  MSE   Hamiltonian  describing the   magnetic
properties of this 2D quantum crystal reads
\begin{eqnarray}\label{Hmulti}
\mathcal{H}=(J_2-2J_3) \sum_{
        \begin{picture}(17,10)(-2,-2)
                \put (0,0) {\line (1,0) {12}}
                \put (0,0) {\circle*{5}}
                \put (12,0) {\circle*{5}}
        \end{picture}
} P_{12}
+J_4 \sum_{
\begin{picture}(26,15)(-2,-2)
        \put (0,0) {\line (1,0) {12}}
        \put (6,10) {\line (1,0) {12}}
        \put (0,0) {\line (3,5) {6}}
        \put (12,0) {\line (3,5) {6}}
        \put (6,10) {\circle*{5}}
        \put (18,10) {\circle*{5}}
        \put (0,0) {\circle*{5}}
        \put (12,0) {\circle*{5}}
        \end{picture}
} \left( P_{1\ldots 4}+{\rm H.c}\right)\\
-J_5 \sum_{
\begin{picture}(26,15)(-2,-2)
        \put (0,0) {\line (1,0) {24}}
        \put (6,10) {\line (1,0) {12}}
        \put (0,0) {\line (3,5) {6}}
        \put (18,10) {\line (3,-5) {6}}
        \put (6,10) {\circle*{5}}
        \put (18,10) {\circle*{5}}
        \put (0,0) {\circle*{5}}
        \put (12,0) {\circle*{5}}
        \put (24,0) {\circle*{5}}
        \end{picture}
} \left( P_{1\ldots 5}+{\rm H.c}\right) \nonumber
+J_6 \sum_{
\begin{picture}(26,30)(-2,-15)
        \put (6,10) {\line (1,0) {12}}
        \put (6,-10) {\line (1,0) {12}}
        \put (0,0) {\line (3,5) {6}}
        \put (0,0) {\line (3,-5) {6}}
        \put (18,10) {\line (3,-5) {6}}
        \put (18,-10) {\line (3,5) {6}}
        \put (6,10) {\circle*{5}}
        \put (6,-10) {\circle*{5}}
        \put (18,10) {\circle*{5}}
        \put (18,-10) {\circle*{5}}
        \put (0,0) {\circle*{5}}
        \put (12,0) {\circle*{5}}
        \put (24,0) {\circle*{5}}
        \end{picture}
} \left( P_{1\ldots 6}+{\rm H.c}\right)
\label{eq:H26}
\end{eqnarray}
where Eq.~\ref{eq:P3} was used to absorb the  three-spin terms
into an effective      first-neighbor      Heisenberg     exchange
$J_2^{\rm eff}=J_2-2J_3$. At high density the hard-core potential
between Helium atoms only leaves three-body  exchanges possible
($J_3\gg J_{n\ne3}$) and Eq.~\ref{eq:H26} reduces    to a first
neighbor   Heisenberg  {\em ferromagnet},\cite{drh80} as observed
experimentally for  the  first time by Franco {\it et
al.}\cite{frg86} in high-density layers. On the other hand the
second layer solidifies  at  lower density and  higher order
exchange terms  cannot be  ignored.\footnote{The first layer  is
then so  dense that exchange is strongly  suppressed.  The first
layer can  also   be replaced  by   an   $^4$He  or  HD
mono-layer.}   PIMC simulations\cite{bc99} and  high-temperature
fits of the experimental data\cite{rbbcg98}  showed that   the
relative strength   of two- and four-spin terms   if  roughly
$J_2^{\rm    eff}/J_4\sim -2$  in   the low-density second layer
solid.  

The $J_2$--$J_4$ model was studied by exact
diagonalizations\index{exact diagonalizations}. It exhibits many distinct phases. There are evidence of  
 a   short-range RVB SL   phase
with no  broken symmetry,\cite{mblw98,mlbw99,lmsl00} and different gapless exotic phases,
with  $SU(2)$ symmetry breaking but non on-site magnetization. Exotic order parameters have been evidenced in different part of the
phase diagram, as a chiral current\cite{Lauchli2005} and an octupolar moment.\cite{Momoi2006}

The most recent ultra-low temperature measurements      of   specific
heat\cite{kmyf97} and uniform susceptibility\cite{cthrbbg01} are
not incompatible with  a gapless phase, but the signature of the octupolar moment remains to be observed.\cite{FukuyamaKITP2012}

\subsubsection{Wigner crystal}

The   Wigner crystal \index{Wigner crystal} is  another fermionic
solid with a triangular geometry where MSE interactions can play
an important role.  At very low density the Coulomb energy
dominates, the crystal   is  almost classical and MSE interactions
are very small.   Exchange frequencies $J_P$ can be computed in
this regime by  a  semi-classical (WKB)
approximation\cite{roger84,kh00,hk01} and,  as  for  the high
density solid of  $^3$He,  three-body exchanges dominate  and give
rise  to ferromagnetism. However, at higher  density and close to
melting, PIMC calculations  of the exchange frequency\cite{bcc01}
showed  that the magnetism  may be described by  a MSE model with
parameters ($J_2^{\rm eff}$  and $J_4$) close to those   where the
triangular  MSE model is expected to be a RVB \index{RVB} SL.
Unlike the $^3$He case, the  particles (electrons) are charged and
an external magnetic  field has  also an orbital effect, it
introduces complex phases in the exchange energies: $P+P^{-1}\to
e^{i\alpha}P+e^{-i\alpha}P^{-1}$  where the angle $\alpha=2\pi
\phi/\phi_0$  is  proportional to  the magnetic flux $\phi$
passing through the area  enclosed by the exchange trajectory  and
$\phi_0$ is the  unit  flux  quantum. This  can give  rise  to
very  rich  phase diagrams\cite{bcc01,hk01}  where complex  MSE
terms  compete  with the Zeeman  effect  (see Ref.~\cite{ok98} for
some  early experimental attempts to explore this physics).

\subsubsection{Cuprates}

The possibility  of significant  four-spin exchange  around  square Cu
plaquettes in copper oxide compounds  was first suggested by Roger and
Delrieu.\cite{rd89} They interpreted  the  anomalously large width  of
Raman scattering spectra as a signature of  four-spin exchange in this
copper oxide superconductor.  The importance of these MSE interactions
in CuO$_2$ planes  ($J_4\sim 0.25 J_2$) has  then been emphasized by a
number   of  groups and   in   different  materials  and  by different
experimental                      and                      theoretical
approaches.\cite{sugai90,coldea01,mr02,kk03} Four-spin  plaquette ring
exchange    also   plays      a     significant   role     in   ladder
compounds.\cite{mkebm00,bmmnu99,sku01,gkt03}   For  instance, exchange
parameters    with  values    $J_{rung}=J_{leg}=110$~meV           and
$J_{ring}=16.5$~meV  were  proposed  for La$_6$Ca$_8$Cu$_{24}$O$_{41}$
based     on    the       dispersion      relation    of      magnetic
excitations.\cite{mkebm00,bmmnu99}

\subsection{Two-leg ladders}\index{two-leg ladders}
\label{ssec:ladderMSE}

Numerous  works   were  devoted   to  ladder  models   with  four-spin
interactions.  These include  general bi-quadratic interactions as well
as models  with ring-exchange  terms.  We will  only discuss  here the
simplest of these MSE models:
\begin{eqnarray}
  \mathcal{H}=&J&\sum_n \left(
    \vec{S}_{n,1}\cdot \vec{S}_{n,2}
    +\vec{S}_{n,1}\cdot \vec{S}_{n+1,1}
    +\vec{S}_{n,2}\cdot \vec{S}_{n+1,2}
    \right) \nonumber \\
    +&K&\sum_\Box \left(
    P_{1234} +H.c
    \right)
    \label{eq:ladder}
\end{eqnarray}
Thanks  to several  studies\cite{bmmnu99,mvm02,hmh03,lst03}  the phase
diagram of  this Hamiltonian  is now rather  well understood  and five
different phases were identified.
\begin{itemize}

\item   {\bf   Ferromagnetic  phase}.   The   ground state  is   fully
polarized. This phase includes  the $(J=-1,K=0)$ and the  $(J=0,K=-1)$
points.

\item {\bf  Rung-singlet phase}.  This phase  includes the ground state of
the ladder without  MSE term $(J=1,K=0)$.  The  spectrum is gapped and
the   ground state  is  unique.    A moderate  $K/J\gtrsim0.23\pm0.03$
destroys this  phase\cite{bmmnu99,honda01,hn02,lst03} in favor of the VBC
below.

\item {\bf Staggered VBC} with dimers  on the legs.  In one of the two
degenerate ground states the dimerized bonds are $(2n,1)-(2n+1,1)$ and
$(2n+1,2)-(2n+2,2)$.       The      VBC     disappears             for
$K/J\gtrsim0.5$.\cite{lst03} Such  a staggered VBC was first predicted
in  the framework   of   a ladder  with  bi-quadratic interaction   by
Nersesyan   and   Tsvelik.\cite{nt97} Using     Matrix-Product Ansatz,
Kolezhuk    and  Mikeska\cite{km98}    constructed  models which   are
generalizations    of  Eq.~\ref{eq:ladder}  and   which    have  exact
ground state with long-range  staggered dimer correlations.   In this
phase the   magnetic excitations are  very  different from  the magnon
excitations of the rung-singlet  phase above. Here the  excitations do
not form well-defined quasi-particles but a  continuum made of pairs of
domain walls connecting two dimerized ground states.\cite{nt97,km98}

\item  {\bf   Scalar  chirality   phase}.   The  order   parameter  is
$\langle\vec{S}_{n,1}\cdot(\vec{S}_{n,2}\times\vec{S}_{n+1,2})\rangle$
and it spontaneously breaks the time-reversal symmetry and translation
invariance.  The  ground state is two-fold  degenerate up to  the next
transition at $K/J\simeq2.8\pm0.3$.\cite{lst03} There exists a duality
transformation\cite{hmh03,mhnh03} which  maps the  scalar    chirality
order parameter onto   the   dimer   order   parameter of   the    VBC
above.\footnote{The Hamiltonian of Eq.~\ref{eq:ladder} is self-dual at
$2K=J$.}    Applying    such a  transformation     to  the  exact  VBC
ground states   mentioned   above,  models   with  an  exactly   known
ground state       and   scalar      chirality     LRO   can        be
constructed.\cite{hmh03,mhnh03} Although  chiral  SL  have  been  much
discussed  in  the  literature, this is   to  our knowledge  the first
realization of such a phase in a SU(2) symmetric spin-$\frac{1}{2}$ model.

\item   {\bf  Short-range   ordered   phase  with   vector-chirality}
correlations.        The         strongest        correlations     are
$\langle(\vec{S}_{n,1}\times\vec{S}_{n,2})\cdot(\vec{S}_{n',1}\times\vec{S}_{n',2})\rangle$
but they  remain short-range.    The   spectrum is gapped  and    the
ground state  is unique.  This   phase includes  the pure $K=1$  model
where  $J=0$.  This phase   is  related by the duality  transformation
discussed    above to   the   rung-singlet   phase.\cite{mhnh03}  This
transformation    indeed relates   the  N\'eel  correlations  $\langle
(\vec{S}_{n,1}-\vec{S}_{n,2})\cdot(\vec{S}_{n',1}-\vec{S}_{n',2})\rangle$
(which  are  the strongest  ones  in the   rung-singlet phase)  to the
vector-chirality correlations.  Close    to the ferromagnetic    phase
($J<0$) one observes   a crossover to  a  region where the   strongest
correlations are  ferromagnetic spin-spin correlations along  the legs
and antiferromagnetic along the rungs.\cite{lst03}
\end{itemize}
\subsection{MSE model on the square lattice}
The phase  diagram   of the Hamiltonian  \ref{eq:ladder}  on  the {\em
2D   square-lattice}   has   been  recently studied    by
L\"auchli\cite{lauchli03}  by      exact   diagonalizations.   N\'eel,
ferromagnetic, columnar VBC  and staggered VBC phases were identified,
as  in   the   ladder model   above.   In  addition,   a nematic phase
characterized     by    long-range   vector   chirality  correlations
(alternating spin currents) was found  around the $K=1$, $J=0$  point.
To our knowledge this could be the first  microscopic realization of a
nematic order in a 2D spin-$\frac{1}{2}$ model.

\subsection{RVB phase of the triangular $J_2$--$J_4$
MSE}\index{RVB}
\label{ssec:RVBMSE}

\begin{figure}
\begin{center}
\includegraphics[width=8cm]{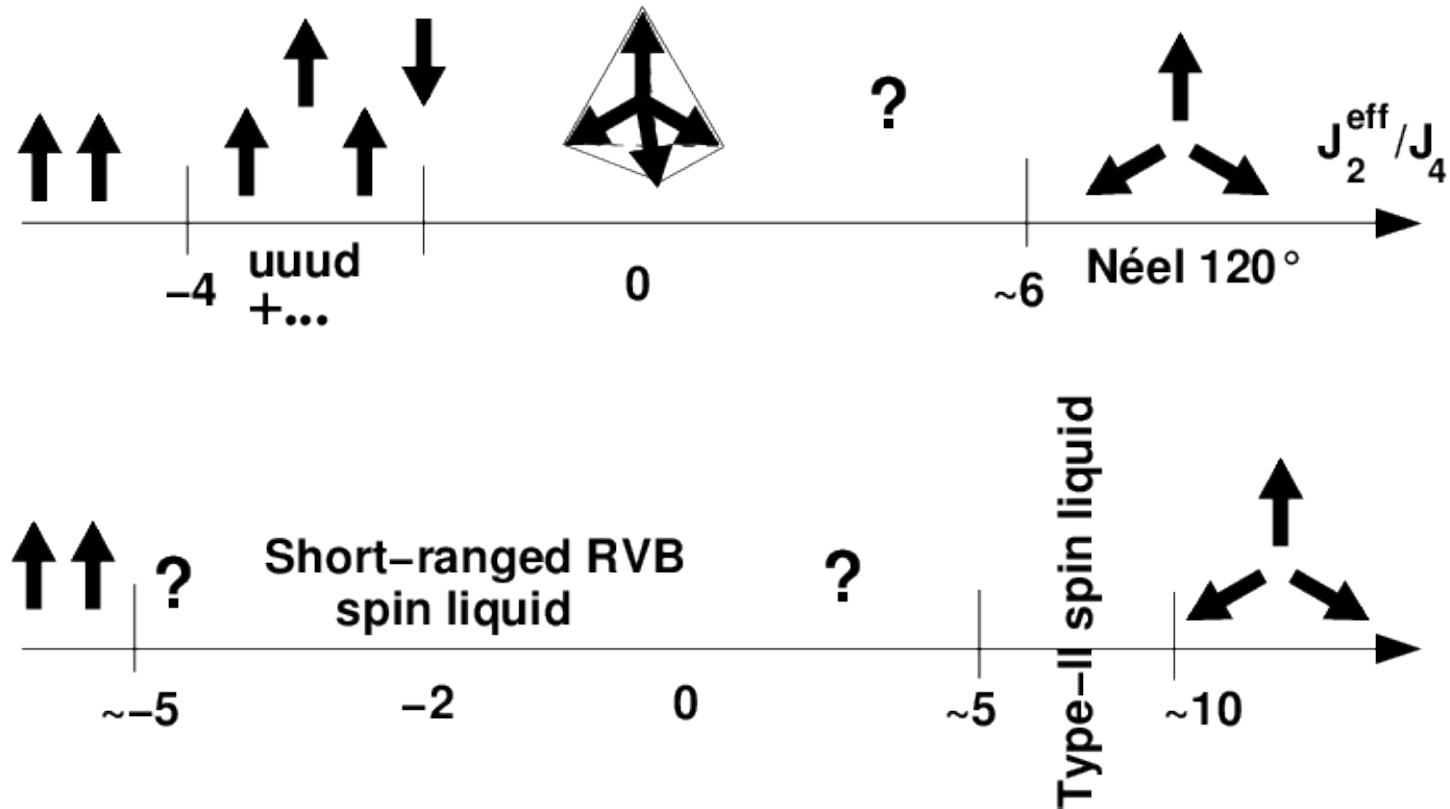}
\end{center}
\caption[99]{
Classical (top)  and  quantum (bottom)  phases   of the $J_2-J_4$  MSE
Hamiltonian.      The classical  model   was    studied  by  Kubo  and
Momoi\cite{km97} and is based on a variational approach.  The quantum
phase diagram  is  the  simplest scenario  compatible with  the  exact
diagonalizations data of Refs.~\cite{mblw98,mlbw99,lmsl00}.  While not
completely understood, in  the type-II spin-liquid  region the spectra
are characterized by a very  large number of singlet excitations below
the first triplet state. This is not the case in the RVB phase.}
\label{fig:phasediagMSE}
\end{figure}

Because  of its relevance  to solid $^3$He  films and Wigner crystals,
the MSE model on the  triangular lattice has been  the subject of many
studies.\cite{km97,mkn97,mbl98,ksmn98,mblw98,mlbw99,lmsl00}  We   will
discuss here  some properties  of the  simplest MSE  model with up  to
four-spin cyclic exchange interactions ($J_2  -2J_3$ and $J_4$ only in
Eq.~\ref{eq:H26}).         The   classical           phase     diagram
(Fig.  \ref{fig:phasediagMSE}) of this  model has been studied by Kubo
and   collaborators\cite{km97,mkn97}  and the  quantum   one has  been
roughly scanned  in  Ref.\cite{lmsl00}: we  will mainly focus   on the
short-range RVB  spin liquid (see Fig.~\ref{fig:phasediagMSE}), which
might be {\it the first RVB  SL encountered in  an SU(2)-symmetric
spin model}.

\subsubsection{Non-planar classical ground states}

It     is  well-known that      an  Heisenberg  model  (with  possible
second-neighbors,  third-neighbors, ...   interactions)  on a  Bravais
lattice  always admits a planar  helical ground state at the classical
level. This is no longer  true when  MSE  are present and finding  the
classical  ground state for arbitrary  $J_2$  and $J_4$ is an unsolved
problem.  A  mean-field phase diagram was  obtained  for the classical
model\cite{km97} but   very   few exact  results  are  known.   In the
neighborhood of $J_4=1$ $J_2=0$ the classical ground state is known to
be  a  four-sublattice  configurations   with  magnetizations pointing
toward the  vertices of  a  regular tetrahedron.\cite{km97}  This is a
quite interesting model where the  ground state spontaneously breaks a
discrete Ising  symmetry associated to the sign  of the triple product
$\vec{S}_1\cdot(\vec{S}_2\times\vec{S}_3)$   around a triangle.   This
broken symmetry  gives rise to  a finite-temperature  phase transition
which has  been observed in  Monte Carlo simulations.\cite{mkn97} This
phenomena is similar to    the transition predicted in  the  $(\pi,0)$
phase of the $J_1$--$J_2$ model on the square lattice.\cite{ccl90a}

\subsubsection{Absence of N\'eel LRO}

The   classical  ground states   at   $J_4=1,J_2=0$  are   tetrahedral
configurations.  Although  this phase appears to be  stable within the
framework    of   linear    spin-wave    calculations\cite{mkn97}   or
Schwinger-Boson mean-field  theory,\cite{mbl98} exact diagonalizations
indicate   that  the   magnetic   LRO  is   washed   out  by   quantum
fluctuations.\cite{mkn97,mlbw99}   The   chiral  order  predicted  to
survive at  long distances and finite  temperatures\cite{mkn97} in the
classical  system for  $J_2=0$  is also  likely  to be  washed out  by
quantum fluctuations.\cite{mlbw99}

When $J_2=1$ a relatively small  amount of $J_4\sim 0.1$ is sufficient
to   destroy  the   three-sublattice  N\'eel   LRO  realized   by  the
first-neighbor Heisenberg model.\cite{lmsl00}  The nature of the phase
on  the  other  side  of  this  transition  is  not  settled  but  the
finite-size spectra display a  large density of singlet excitations at
low   energy    which   could   be   reminiscent    of   the   kagome
situation.\cite{lmsl00}

From  exact diagonalizations\index{exact diagonalizations} (up  to
36 sites) no  sign  of N\'eel LRO could be found at
$J_4=1,J_2=-2$.\cite{mblw98,mlbw99} In addition, the finite-size
analysis showed  that the spin-spin correlation length is quite
short  at $J_2=-2$ and  $J_4=1$ and a spin  gap of  the order of
$\Delta\sim 0.8$ exists at this  point. Much of the numerical
effort to  elucidate the nature of the  MSE ground state was
concentrated on this   point  because it    is close to   the
parameters  realized in low-density $^3$He films (when higher
order exchanges are neglected).

\subsubsection{Local singlet-singlet correlations  -
  absence of  lattice   symmetry breaking}

\begin{figure}
\begin{center}
\includegraphics[width=8cm]{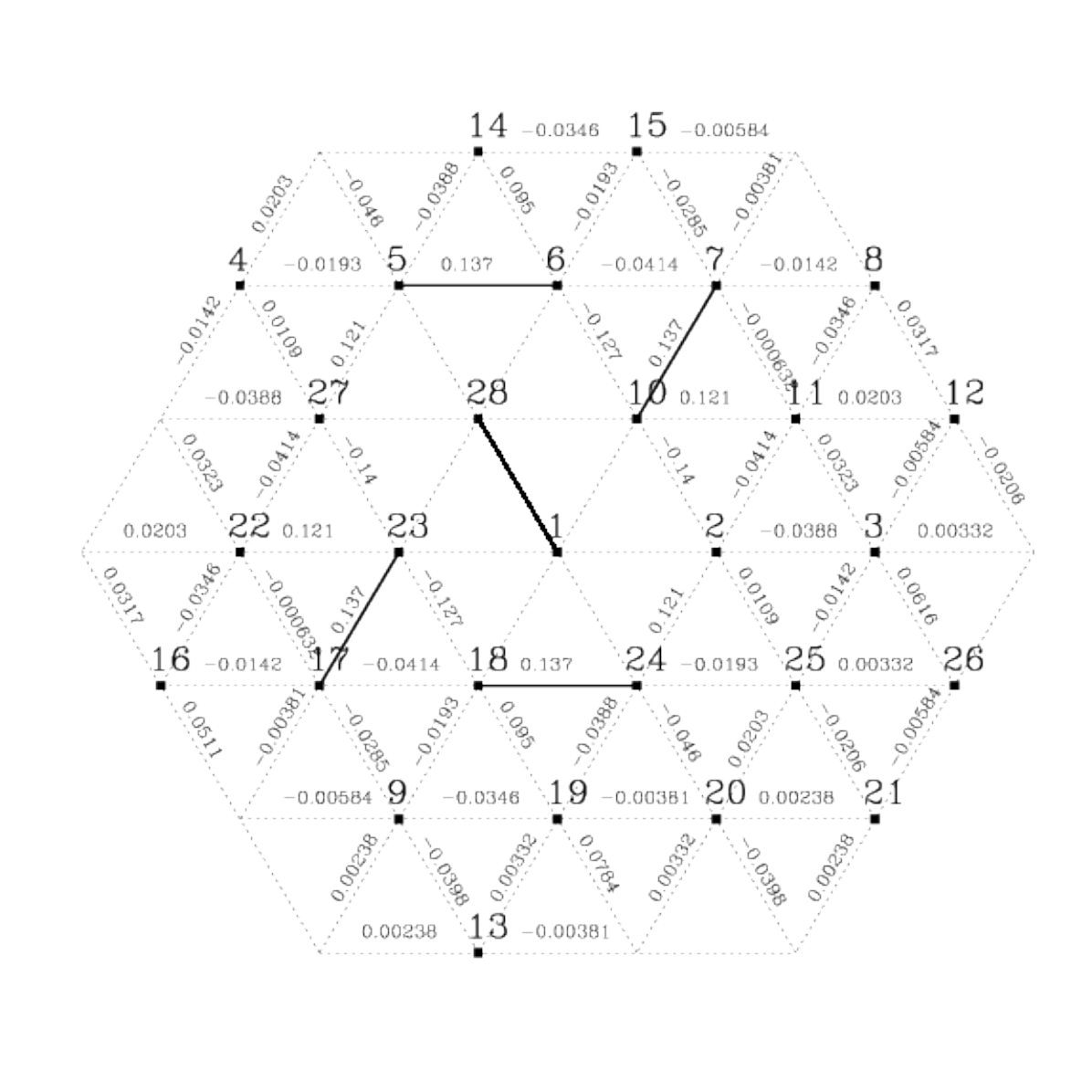}
\end{center}
\caption[99]{Dimer-dimer correlations in the ground state
of  the $J_2$--$J_4$ MSE  model on the triangular  lattice  (28 sites) at
$J_2=-2$, $J_4=1$ (from  Ref.~\cite{mlbw99}). Numbers are proportional
to          $\langle\hat{d}_0\hat{d}_x\rangle        -\langle\hat{d}_0
\rangle\langle\hat{d}_x\rangle$  where the  operator       $\hat{d}_x$
projects onto the singlet space of the bond $x$ and $\hat{d}_0$ refers
to the reference  bond $(1,28)$. These  results shows a clear tendency
for repulsion between parallel dimers.}
\label{fig:ddcorrel}
\end{figure}

Having  excluded the possibility of  a  N\'eel ordered
ground state at $J_4=1,J_2=-2$ it is  natural to look for  a
possible VBC.  Because of the complexity of the  MSE Hamiltonian
it   is not clear what kind  of spatial  order should  be favored.
From   the analysis of dimer-dimer correlations (see
Fig.~\ref{fig:ddcorrel})  it  appears that  parallel valence-bonds
repel each-other at  short distance.  This is similar to what is
observed in the staggered phase of the $J_2$--$J_4$ MSE ladder and
square-lattice   models.  For  this   reason  it appears  that  a
plausible VBC would be the staggered VBC encountered in the
triangular QDM \index{QDM} for  $V>J$  (\S\ref{ssec:QDMTri}).
However this scenario seems difficult   to    reconcile   with the
weakness of dimer-dimer correlations.\cite{mlbw99} In addition,
the low-energy singlet states and their quantum
numbers\cite{StagVBC} do  not reflect the   12-fold
quasi-degeneracy   that should  be present    if  the  system  was
to spontaneously break some lattice symmetry according to a
staggered VBC pattern.  Small systems usually  {\em  favor}
ordered  phases  because low-energy and long-wavelength
fluctuations that could destabilize an ordered state are reduced
compared to larger  systems.  From the fact that the finite-size
spectra do not show the signatures of a staggered VBC symmetry
breaking it is unlikely that the  MSE model could develop a VBC of
this kind in the thermodynamic limit.

\subsubsection{Topological degeneracy and Lieb-Schultz-Mattis Theorem}

Because no  VBC   phase could be  identified   in  the  MSE model  at
$J_4=1,J_2=-2$ the numerical  data were compared  with the predictions
of an RVB liquid scenario.

In one  dimension    a  famous   theorem due   to  Lieb,   Schultz  and
Mattis\cite{lsm61} (LSM) states  that in   a one-dimensional spin system
with an half-integer spin in the unit cell there is at least on excited
state collapsing to the ground state in the thermodynamic limit
(periodic boundary conditions).
There are in fact several arguments suggesting that this theorem might,
at least to some extent,
also  apply to higher dimensions.\cite{al86,o00,mlms02,hastings03,o03}
If  that is the case  a gapped system with  an odd integer spin in the
unit cell must have a degenerate  ground state.  The simplest scenario
to  explain this degeneracy is a  translation  symmetry breaking.  One
could   think that   this  would  rule  out  the  possibility of   any
(translation invariant) RVB liquid  in such models.  This is incorrect
because a ground state degeneracy  can have a  topological origin on a
system with  periodic boundary  conditions,  as  we discussed in   the
framework of QDM (\S\ref{sec:QDM}).  Such  a phase is characterized by
a four-fold topological ground state degeneracy  when the system is on
a torus.   That    degeneracy  allows the  system   to  fulfill  LSM's
requirement   without     any spontaneous       translation   symmetry
breaking.\cite{mlms02}

On a finite-size system the topological degeneracy is only approximate
but  some constraints   exist for the  quantum  numbers  (momentum  in
particular)  of the quasi-degenerate  multiplet.\cite{mlms02} A system
with periodic boundary conditions with an even  number of sites but an
{\em odd  number of rows}  is expected to  have two ground states with
differ by a momentum $\pi$ in  the direction parallel  to the rows, in
close analogy to the  LSM theorem in  dimension one. The numerical
spectra of the MSE model exhibit a set  of three singlet energy levels
collapsing  onto     the  ground state  when  the   system     size is
increased\cite{mlbw99} and their    quantum number  turn  out  to   be
consistent with  the   constraints   derived from the      general RVB
picture.\cite{mlms02}

\subsubsection{Deconfined spinons}\index{deconfined spinons}

\begin{figure}
\begin{center}
\includegraphics[height=4cm,clip=true]{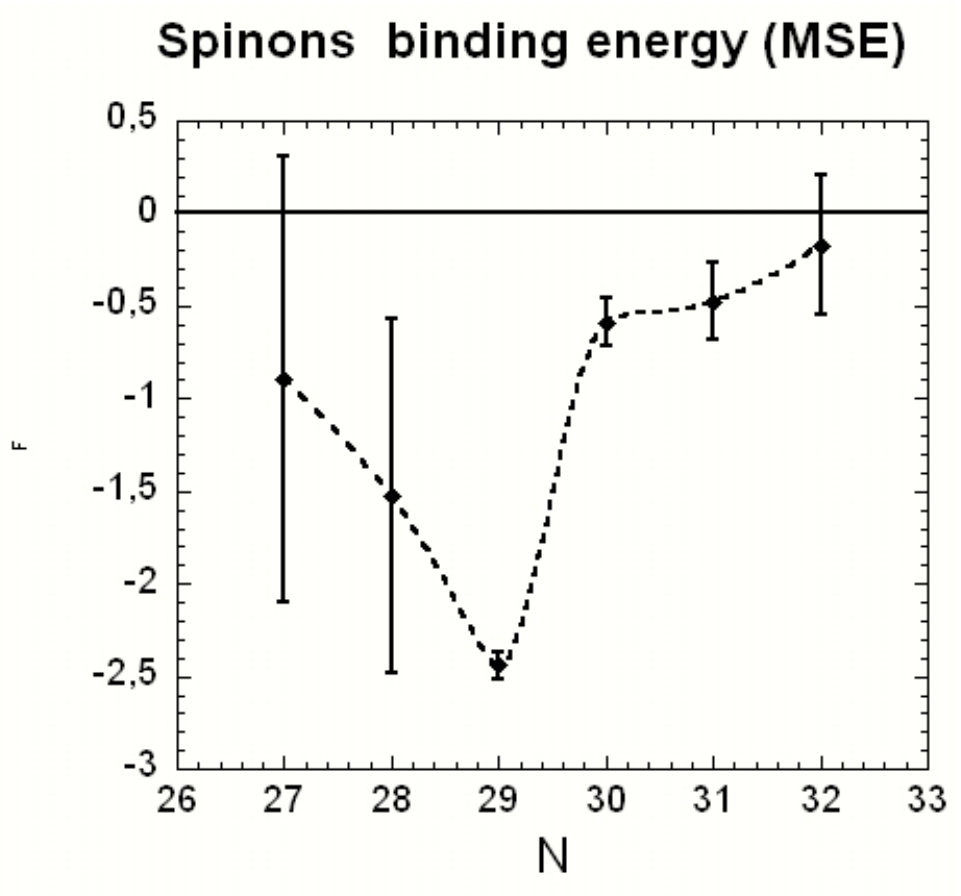}
\caption[99]{Spinons binding energy as a function of the system size in
the MSE model at $J_2=-2$ and  $J_4=1$. The vertical bars correspond to
the  range    of    values found     for  different  sample    shapes.
}\label{spinon}
\end{center}
\end{figure}

The  SL phase described above is   expected to have deconfined spinons
($S=\frac{1}{2}$ excitations).  These excitations should show up as an
incoherent   continuum    in   the   spin-spin    dynamical  structure
factor.  However such a feature would  probably be rather difficult to
observe on small 2D lattices, in particular due to the small number of
inequivalent ${\bf k}$-vectors  in  the Brillouin zone.  On  the other
hand, the binding energy of two spinons can  be evaluated by comparing
the ground state  energy and the first   magnetic excitation energy on
even and odd samples.  In the case  of the MSE model at $J_4=1,J_2=-2$
the results show the existence of  a bound-state (it is more favorable
to put  two spinons in the  same  small sample than  in separate ones,
which is not surprising) but this {\em does  not mean that the spinons
are confined} (contrary  to  the conclusions  of  Ref.~\cite{mlbw99}).
Interestingly this binding energy seems to  go to zero for the largest
available sizes~(Fig.~\ref{spinon}):  this might  indicate the absence
of attraction between   spinons for  large  enough  separation and  an
asymptotic deconfinement.

It is important to  stress here that  the RVB SL phase discussed  here
(and    its   QDM    counterparts    of     \S\ref{ssec:QDMTri}    and
\S\ref{ssec:QDMKag})  is not the only  way to  spinon deconfinement in
2D. There is  at least another scenario, inherited from  one
dimension, which is the  {\em sliding Luttinger liquid}.  Indeed,  the
Luttinger liquid behavior and the one-dimensional critical behavior of
magnetic  chains  seem  to  be  robust  to  small (or  moderate)  {\em
frustrating transverse couplings between chains},  as observed both in
theoretical\cite{nge98,ahllt98,efkl00,be01,vc01,sp02,ssl02}        and
numerical  approaches.\cite{sfl02}  This  regime between  one  and two
dimensions which may have been observed in Cs$_2$CuCl$_4$\cite{cttt01}
is the subject of a number of recent studies.\cite{be01,vc01}

\subsection{Other models with MSE interactions}

Multiple-spin interactions are present in a number of models that were
found   to exhibit fractionalization   or  an RVB liquid ground state.
Well known examples of MSE interactions with an Ising symmetry are $\Z$
gauge theories, where the gauge  invariant plaquette term is a product
of    Pauli  matrices $\prod_i   \sigma^z_i$.    Such  theories have a
deconfined phase in 2+1 dimension\cite{kogut79} and their relevance to
fractionalized phases of  2D  electronic systems has  been
pointed  out by  Senthil and Fisher.\cite{sf00}  The connexion between
$\Z$ gauge theories and QDM was mentioned in
\S\ref{sssec:Z2}. Some MSE  spin models with  an Ising  symmetry and a
fractionalized  ground state were  discussed by Kitaev,\cite{kitaev97}
Nayak and   Shtengel.\cite{ns01} In the other  limit  of a  $U(1)$ (or
$XY$) symmetry several models  have been studied.  Recent examples are
based  upon  the   spin-$\frac{1}{2}$  four-spin   XY  ring   exchange
interaction\cite{bfg02,pbf02,sdss02,sm02}
\begin{equation}
  \mathcal{H}=-K\sum_{\langle ijkl\rangle} \left(
  S_i^+S_j^-S_k^+S_l^- + {\rm H.c}
\right)
\label{eq:4XY}
\end{equation}
which is the $XY$ analog of the $SU(2)$ MSE interaction $P_{1234}+{\rm
H.c}$.


\section{Antiferromagnets on the kagome  lattice}
\label{sec:kagome}\index{Kagom\'e lattice}
\setcounter{footnote}{0}

The   spin-$\frac{1}{2}$ nearest-neighbor Heisenberg model on the kagome
lattice   has  attracted much attention in the long quest of Spin Liquids.
The first researches  go back to the end of the eighties~\cite{e89,ce92,s92,le93,ze95} and more than 300 papers appeared in regular journals since this period.
This model possesses the two ingredients that are considered important to obtain a liquid ground state: frustrated geometry (non bipartiteness)
and low coordination number which enhance the quantum effects.
Moreover the classical model is at
a high-degeneracy point in parameter space, where several phases meet.
The first breakthrough in the study of this tough quantum problem has been done thanks to large scale exact diagonalisations:\cite{le93,ze95,lblps97,web98,smlbpwe00}
These studies could not detect any form of LRO and the model was therefore an interesting candidate for a Spin Liquid.
But the subject remained controversial as the sizes of the clusters in exact diagonalisations  were very limited (36 spins during nearly two decades,
48 very recently\cite{LauchliKITP2012}  and a large number of  competing hypotheses were considered
(\S\ref{subsec:competing phases}).
A new impetus was  recently  provided by the discovery of  various  Cu compounds    exhibiting  spin liquid behaviors (\S\ref{subsec:experiments}).
In the following we will first describe the properties of simpler ``parent'' models (classical model \S\ref{subsec:classical}, Ising model \S\ref{subsec:Ising},
quantum dimer model \S\ref{subsec:qdm}). We will then give the 2012 results of  state of the art DMRG numerical approaches (\S\ref{subsec:numerics})
for the pure nearest-neighbor spin-1/2 Heisenberg  model,
describe various proposals  of nearby phases   (in parameter space) and a brief account of the experimental situation.

\subsection{Ising model}
\label{subsec:Ising}
The classical  model   remains disordered at     all
temperatures.\cite{kn53,hr92}
The system fails to order even at $T=0$ and has a large finite entropy per site:
$S_{\rm   kag}^{\rm  Ising}= 0.502$,  more   than half  the
independent spin value, much  larger than the triangular lattice value
$S_{\rm tri}^{\rm Ising}=  0.323$   and   of  the order of     Pauling
approximation      for    independent       triangles          $S_{\rm
Pauling}=0.501$.\cite{p38} Moessner  and Sondhi have  studied  this
Ising model   in  a transverse magnetic   field  (the simplest way  to
include some quantum fluctuations):  the model  fails to
order for any transverse field, at any temperature.\cite{msc00,ms01b}

\subsection{Classical Heisenberg models on the kagome lattice}
\label{subsec:classical}

The nearest-neighbor classical $O(3)$ Heisenberg model on the kagome lattice
also has   a  huge ground state degeneracy.    This property
holds on different lattices  with corner sharing units
such as the checkerboard  lattice or the three  dimensional pyrochlore
lattice (Moessner and Chalker\cite{mc98,mc98a}). On all these lattices
the nearest-neighbor Heisenberg Hamiltonian  can be written as the sum
of the square of the total spin $\vec{S}_{\alpha}$ of individual units
$\alpha$ (a triangle   for the     kagome  lattice and a   tetrahedron in the  2d checkerboard or 3d pyrochlore cases),   which   share only      one
vertex. Classical ground states are obtained whenever $\forall\alpha\;
\vec{S}_{\alpha}=\vec{0}$.    This     condition fixes   the  relative
positions of the three classical spins of a  triangle at $120$ degrees
from each   other  in a  plane.   But  it does  not  fix  the relative
orientation of the plane of   a triad with  respect  to the planes  of
the triads on neighboring triangles:  the model has a continuous  local
degeneracy\cite{chs92,hr92}    at  $T=0$.\footnote{ Counting  the {\em
planar}  ground states amounts to determine in  how  many ways one can
associate  one of the three  letters $A$, $B$ and  $C$ to each site so
that each triangle  has spins along  the three different orientations.
This  already  represents  an extensive  entropy.\cite{baxter70,hr92}
In such a planar ground state, on may look for a closed loop involving only two
spin directions, say $A-B-A-B-\cdots$. By construction, all the spins connected to this loop
point in direction $C$.
It is then possible to rotate simultaneously all the spins of the loop by some arbitrary angle about the $C$ axis, without any energy cost.
This creates a non-planar ground state.}

The classical model has a large  density of low-lying excitations at  low temperature.\cite{k94} Thermal  fluctuations  select    coplanar configurations because they
have the largest phase space for low-energy fluctuations and are therefore entropically favored.\cite{chs92,hr92,rcc93}
The plausibility of  long-range order in  spin-spin  correlations or other observables at   very low temperature   has also been discussed.\cite{hr92,rb93,zhitomirsky2008,Henley2009}
The recent progresses made in numerical simulations point to a selection of the so-called $\sqrt3\times\sqrt3$ order at ultra-low temperatures.\cite{cepas2011,Chern2012}

The $T=0$ instabilities of the classical model to infinitesimal perturbations has been studied very early.\cite{HarrisKallinBerlinsky92}
Depending on its sign, a second-neighbor Heisenberg  coupling $J_2$ leads to   two different coplanar phases with  unit cells of 3 or 9 spins.
They correspond respectively to the sol-called ``$q=0$'' and the ``$\sqrt3 \times \sqrt3$'' magnetic structures in reference to the order wave vector.
The Dzyaloshinsky-Moriya interaction
(with DM vector perpendicular to the spin plane) favors the ``q=0'' phase.\cite{e02,ecl02}
It has been recently shown that an infinitesimal antiferromagnetic interaction $J_3$ between  third neighbors across hexagons favors a non-planar chiral order with a 12-spin unit cell.\cite{Janson_2008} Under the action of  quantum fluctuations these classical orders can be destabilized to give birth to distinct classes of spin liquids with different fluxes around close contours.\cite{wwz89,messioPSG2012}

\begin{figure}
\begin{center}
\includegraphics[trim= 60 395 100 125,clip,width=6cm]{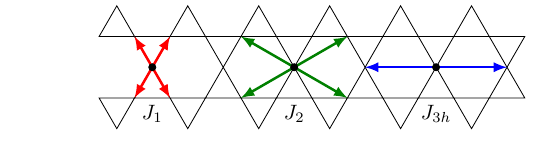}\\
	\includegraphics[trim= 2 0 0 0,clip,width=6cm]{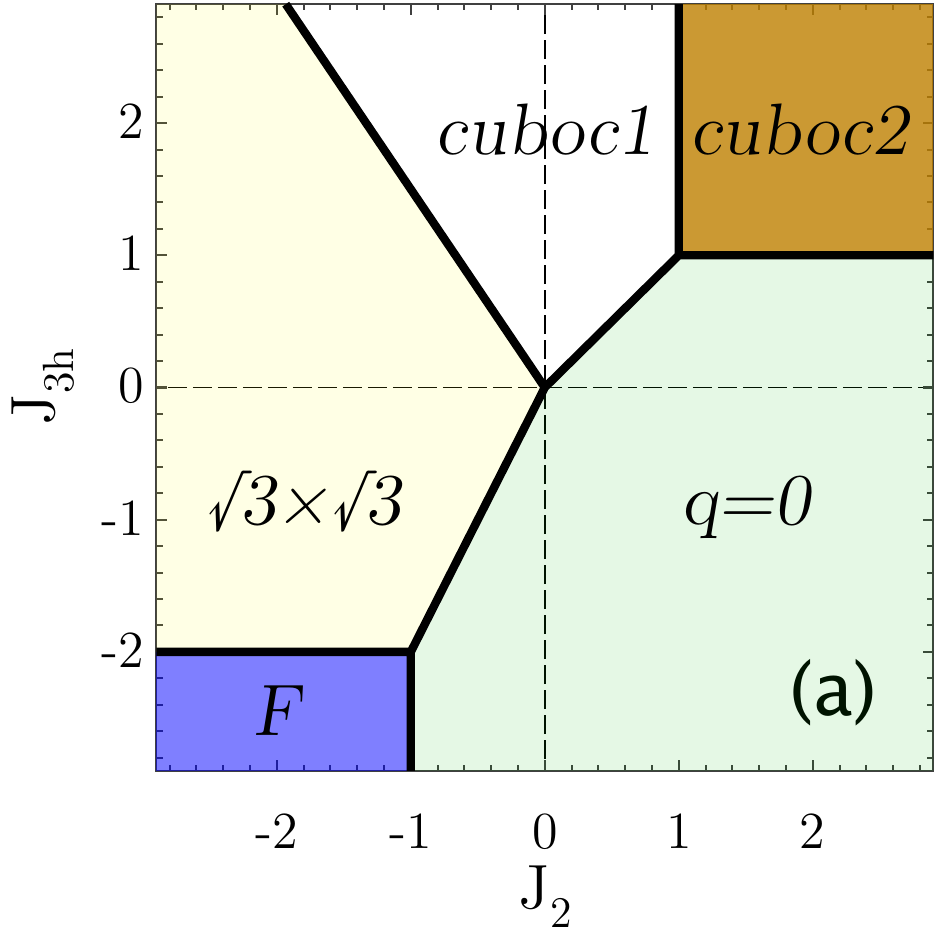}\\
     \includegraphics[trim = 180 180 180 180, clip,width=2.5cm]{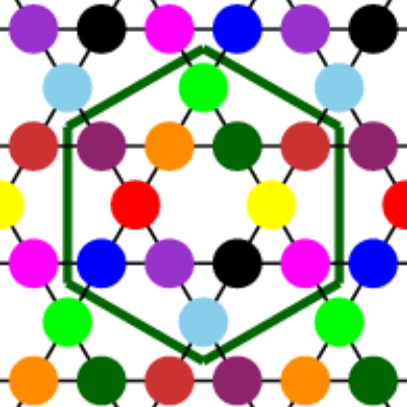}\hspace{0.5cm}
     \includegraphics[trim = 25 140 45 150, clip,width=2.5cm]{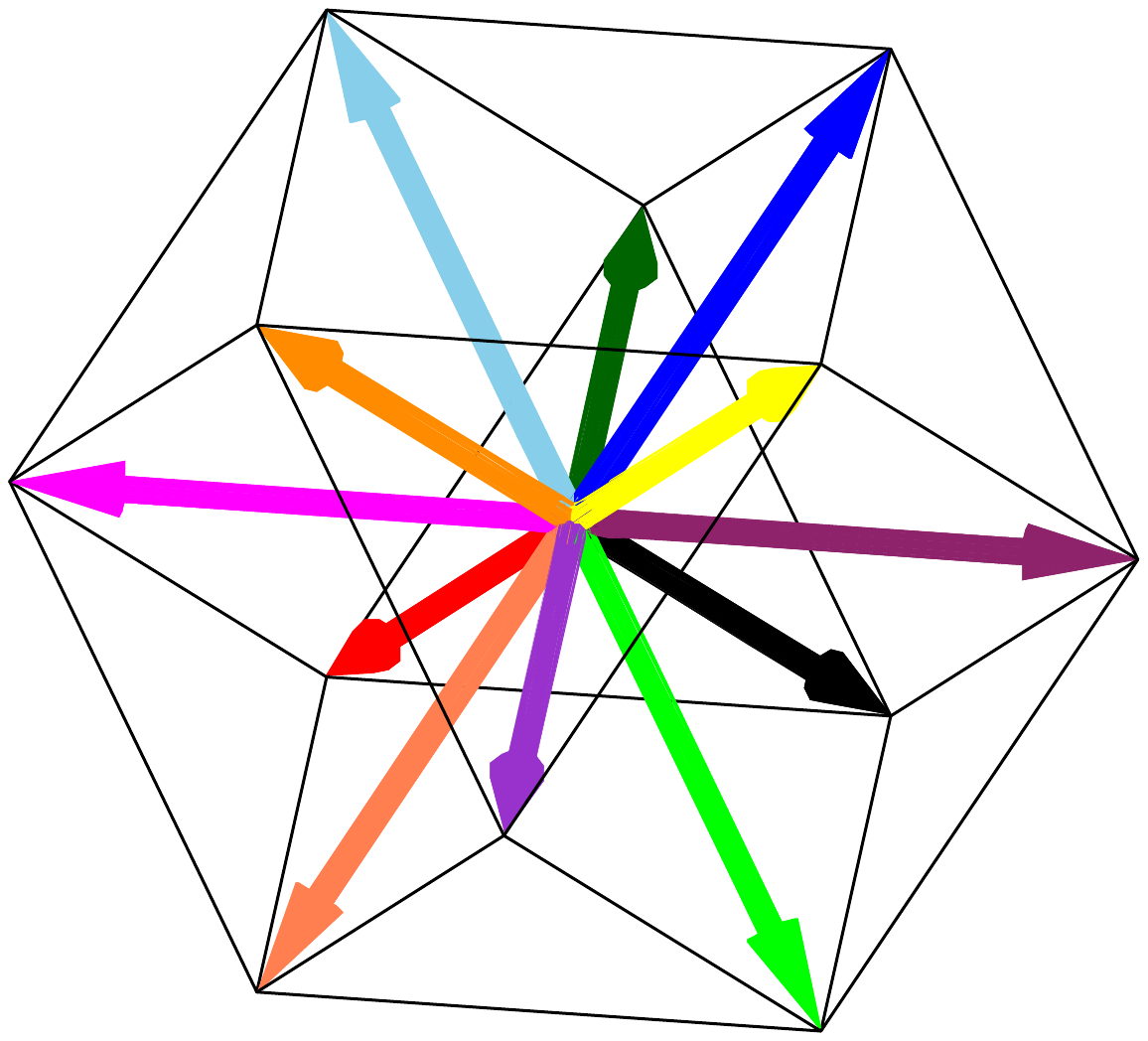}
\end{center}
\caption{Phase diagram of the model with up to third neighbors interactions at $T=0$ and $J_1=1$.
Top: definition of the interactions $J_1$, $J_2$ and $J_{3h}$.
Middle:  exact classical phase diagram
The point $(J_2,J_{3h})=(0,0)$ is a tricritical point.
  Bottom: Description of the {\it cuboc1} order.
  Left: on the kagome lattice each color corresponds to a different magnetic sublattice.
  The thick line indicates the 12 site unit-cell.
  Right: arrows are the spin orientations with the same color coding as in the left figure.
 The black lines connecting the end of the vectors form a cuboctaedron.
 On each triangle, the spins are coplanar at 120 degrees;  for opposite sites on each hexagon, spins are anti-parallel.
 The triple products (determinant) of three spins of the hexagons, either first or second neighbors, are non zero and measure the chirality of the phase.
 They change sign in a mirror symmetry or in a spin flip.
 }\label{fig:diagramJ1J2J3}
\end{figure}

A new algebraic method\cite{Messioregular2011} has been used to enlarge our knowledge of the zero-temperature phase diagram of the classical $J_1$-$J_2$-$J_3$ model, as shown in Fig.~\ref{fig:diagramJ1J2J3}.
This method, which is based on symmetry arguments, allows a classification of all the regular classical magnetic  orders that can be harboured on a lattice whatever the details of the $SU(2)$ invariant
spin Hamiltonian.

 \subsection{ Nearest-neighbor   RVB   description  of  the  spin-$\frac{1}{2}$
kagome antiferromagnet} \label{subsec:qdm}
Assuming that the ground state of the spin-$\frac{1}{2}$ Heisenberg model has a large enough spin-gap,
an hypothesis supported by early exact diagonalisations\cite{le93}, 
 Zeng and Elser\cite{ze95}    proposed a
variational
description of the ground state and low-lying excitations  of  the
kagome antiferromagnet in  the  basis  of  nearest neighbor  valence bonds.
They analyzed in  this  context the  dimer dynamics and showed on
a $N=36$ sample that  the hexagon VBC --favored by  the shortest
(three-)dimer moves--  melts  when introducing higher order
tunneling.  Mila and Mambrini\cite{m98,mm01} confirmed that this
reduced Hilbert space of next neighbor  valence bonds captures
some of the perplexing features  of this   magnet and
specifically  the absence of (measurable) gap in  the singlet
sector of small samples.\cite{web98}


Some recent numerical results\cite{dmnm03} (in  the full spin-$\frac{1}{2}$
Hilbert space as well as  in the RVB subspace) showed that
(static)  non-magnetic impurities  (holes) experience an
unexpected {\em   repulsion} in this   system and that  no
significant magnetic moment is created  in the vicinity of  the
impurities. These non-magnetic
impurities provide a valuable probe of spinon
confinement in    2D antiferromagnets\cite{sv00}
and the absence of magnetic moment formation suggest that they are {\it deconfined}.

\subsection{
Spin-$\frac{1}{2}$  Heisenberg  model on    the   kagome lattice: numerics}
\label{subsec:numerics}
{Understanding the nature of the ground state of the nearest neighbor spin-1/2 Heisenberg  model on the kagome lattice is clearly a difficult problem:
classical degeneracy, numerous competing phases at the quantum level, and sign problem for
quantum Monte Carlo methods.
The first  ``unbiased'' tool used to approach this problem has been  exact diagonalisations.~\cite{lblps97,web98,Sindzingre2009,laeuchli2009,laeuchli2011}
This approach has shown the absence of Néel or VBC LRO,\cite{ms07} but the sizes up to $N\leq 36$ were insufficient  to give a definite answer on the nature of the spin liquid:
a careful analysis has even shown that both a critical gapless spin liquid, as well as a gapped one remained plausible on the basis of the $N=36$ spectra.\cite{Sindzingre2009}
In a numerical ``tour de force'',  L\"auchli {\it et al.} have very recently obtained a part of the low-lying levels of the exact $N=48$ spectrum.\cite{LauchliKITP2012}
A refined analysis of these data will probably give some valuable information in a near future.

In the meanwhile a breakthrough in large scale DMRG   calculations  on very large and long cylinders has  brought more insight on the ground state of  this puzzling system, as we will explain below.~\cite{jiangsheng2008,Yan2011,Jiangwangbalents2012,DepenbrockSchollwock2012}

\subsubsection{Ground-state energy per spin}
All these results agree on the extremely low energy per bond of the ground state ($\langle{2\vec  S}_i\cdot{\vec  S}_j\rangle\simeq  -0.438)$ $  \sim
87\%$ of the energy per bond in  an isolated triangle. On this lattice
the energy per bond  of  the spin-$\frac{1}{2}$  system  is much lower than  the
classical energy, $\frac {E_{qu.}}{E_{cl.}}  \sim  1.74$, a ratio  much
larger than   in any other  2D magnet,  that can  only be
compared to the value obtained for the Bethe chain (1.77).  The kagome
lattice   is  the 2D lattice   which  offers the largest
stabilization due to quantum fluctuations.

\subsubsection{Correlations}

Early exact diagonalisations\cite{le93}
as well as the latest DMRG simulations\cite{DepenbrockSchollwock2012} found spin-spin and dimer-dimer
correlations to be short-ranged, in agreement with the early series expansion.\cite{sh92}
The scalar chirality correlators $\langle \left(\vec S_1\cdot(\vec S_2 \times \vec S_3)\right) \left(\vec S_{1'}\cdot(\vec S_{2'} \times \vec S_{3'})\right)\rangle$
between elementary triangles have   been shown  to be short ranged\cite{ce92,DepenbrockSchollwock2012}.
Depenbrock \textit{et al.} added that:   ``Chiral correlators for other loop types and sizes decay even faster''
and will come back to this point later when discussing a putative chiral topological spin Liquid.\cite{Messiocuboc12012}

\subsubsection{Spin gap}
The spin gap of the symmetric $N=36$ toroidal cluster is $0.14$,
but the  data on such small systems are difficult to extrapolate safely to the thermodynamic limit.
The most recent results of DMRG on  large  and very long cylinders
(with diameter up to $\sim 16$ lattice spacings)   point  to a value of the spin gap $\sim 0.13$.\cite{web98,jiangsheng2008,Yan2011,Jiangwangbalents2012,DepenbrockSchollwock2012}
The 2012 belief is thus that nearest-neighbor Heisenberg model  on the kagome lattice is indeed a gapped spin liquid.

\subsubsection{Singlet gap}

The extension of the Lieb-Shultz Mattis theorem to two dimensions\cite{o00,Hastings04} guaranties
that the spin-$\frac{1}{2}$ kagome antiferromagnet (as any spin model with conserved $S^z$, short-range interactions, and a half-odd integer spin per unit cell)
cannot simultaneously be gapped and have a unique ground state on a closed surface of genius larger or equal to one.
 If we admit that spin excitations are indeed gapped, this implies
that the spectrum is either gapless in the singlet sector, or the ground state is degenerate in the thermodynamic limit.
If the singlet excitations are also gapped, the ground state must be degenerate. If this degeneracy is not the consequence of some
conventional lattice symmetry breaking (valence bond crystal for instance), it is said to be a ``topological degeneracy'' (see \S\ref{subsec:topologicalorder} and \S\ref{sssec:Z2}), and this kind of SL is now called a topological spin liquid.

From the presence of a spin gap and absence of long-range dimer dimer correlations, a topological liquid is thus presently the most natural scenario for the spin-1/2 Heisenberg problem on the kagome lattice.
But the structure of the singlet sector is not completely unveiled at the moment. In a simple $\mathbb{Z}_2$ topological spin liquid we expect a four-fold degeneracy on a two torus, and a two-fold degeneracy on a
cylinder at the thermodynamic limit. In fact,  small systems up to 36 sites exhibit a quasi continuum of $S=0$ excitations and absence of a sizeable gap in the singlet
sector.~\cite{web98,Sindzingre2009}\footnote{This feature, inconsistent with simple pictures of gapped spin liquids,  remained puzzling for a long time and may have recently
received an explanation.~\cite{Messiocuboc12012}} There is an interesting evolution of the spectrum between the sizes $N=36$ and $N=48$\cite{LauchliKITP2012}
but  a coherent picture for the spectrum of low lying excitations obtained in ED remains to be constructed.

On the other hand the first excited singlet state in DMRG is found at a finite energy of about $0.04 $ \textrm{or} $0.05$ above the ground state,\cite{Yan2011} which
seems contradictory with the Lieb-Shultz-Mattis theorem.
There might be different explanations to this: either the  splitting induced by the finite cylinder circumference is larger than expected,
or  the DMRG method is biased toward stabilizing a linear combination of the (quasi-degenerate) ground states with the smallest entanglement entropy.\cite{Jiangwangbalents2012}}

\subsubsection{Entanglement entropy and signature of a $\mathbb{Z}_2$ liquid}

The entanglement entropy (EE) is a powerful tool to detect the presence of ``topological order'' in states without any conventional local order parameter.
In a generic wave function with short-range correlations, the EE $S_A$ of a subsystem $A$ has a
leading  contribution corresponding to the ``area law'' $S_A\simeq a l_A^{d-1}$, where $l_A$ is the typical linear size of $A$ and $d$ the space dimension.
This contribution simply  originates from the short-range correlations taking place  ``across'' the boundary of the subsystem $A$.
Hamma {\it et al.}\cite{hiz05},
Kitaev and Preskill\cite{kp06} and Levin and Wen\cite{lw06} discovered that, in presence of topological order,  the EE may also contain a sub-leading contribution of order $\mathcal{O}(l^0)$:
$S_A\simeq a l_A ^{d-1} - \gamma $. This contribution $\gamma$ indicates the presence of some non-local form of quantum entanglement. What makes this sub-leading contribution conceptually interesting and practically
very useful is the fact that it is ``universal'': it depends on the
topological phase (through the so-called total quantum dimension), but not on the microscopic or short-distance properties of the wave functions.
For a $\mathbb{Z}_2$ liquid,  the simplest topological phase, we have $\gamma=\ln(2)$.

Extracting $\gamma$ from a given wave function is a non-trivial
problem and the original suggestion\cite{kp06,lw06}
was to compute some EE difference between suitably chosen subsystems so that the leading term cancels. This subtraction scheme has been implemented numerically in a few cases.\cite{fm07,ihm11,zgv11}
It is however easier and more accurate to obtain $\gamma$ from the EE in a cylinder geometry. There, the translation invariance of the boundary of $A$ allows
to extract $\gamma$ simply from a fit to (at least) two system size. The cylinder geometry approach to topological EE has been used successfully in several
numerical works,\cite{lbh10,smp12,zgtov12,jyb12} including
some on the kagome antiferromagnet with first-neighbor coupling only,\cite{DepenbrockSchollwock2012} or with additional next-nearest neighbor $J_2$.\cite{Jiangwangbalents2012}
For the first--neighbor model, the DMRG results of Ref.~\cite{DepenbrockSchollwock2012} gave $\gamma$ consistent with $\ln(2)$ with a  $\simeq 10\%$ accuracy.
With a further neighbor interaction (a point  sitting deeper in a possibly distinct liquid phase\cite{Messiocuboc12012}), Ref.~\cite{Jiangwangbalents2012}
obtained $\gamma\simeq \ln(2)$ with an accuracy of the order of $1\%$.

\subsubsection{Spin liquids on the kagome lattice and Projective symmetry groups}

X.-G. Wen was the first to develop a constructive approach of fermionic mean-field theories of spin liquids  on the square lattice.~\cite{Wen_PSG}
The general idea is the following.

In order  to be able to describe liquid states with fractionalized excitations (spin-1/2 excitations), it is natural to start
with a description in terms of Schwinger bosons or fermions (\S\ref{sec:largeN}). Since the microscopic degrees of freedom
are instead spin-1 objects, the spinon are inevitably coupled to a gauge field. This gauge field describes  the quantum fluctuations of the singlet bonds.

If the effective gauge theory is non-confining the system has spin-1/2 fractionalized excitations and is a spin liquid.
This spin liquid can thus be characterized by
the symmetries of the effective  theory: space symmetry group (point group and lattice translations),
the time reversal symmetry, and the  local  gauge transformations.
Whereas a spin liquid does not break any lattice symmetry, its effective theory in terms of spinons  and gauge fields may break the space symmetries
insofar as the effect of each  space symmetry can be compensated by a gauge transformation, thus achieving an invariant physical spin liquid state.
The projective representation of the symmetry group (in short PSG) is defined as the set of all combinations of space and gauge symmetry operations that   leave   an effective  theory
(or mean-field Hamiltonian) invariant.
The above requirements allow the determination of the distinct mean-field theories describing a fermionic or bosonic spin liquid on a given lattice. This number is finite but may be large
(288 fermionic mean-field structures on the square lattice).\cite{Wen_PSG}
This construction does not depend on the details of spin Hamiltonian, except for its symmetries. It gives
a first general classification of the spin liquids, at least valid in the regime of small gauge fluctuations.
It also provides a direct way to determine the low-energy gauge group for a given mean-field state (a particular subset of the PSG,
called invariant gauge group (IGG)).\footnote{
Using the schwinger bosons as a starting point, one generically finds that the low-energy gauge group is $U(1)$
on bipartite lattices, and $\mathbb{Z}_2$ on non-bipartite lattices.
The former situation leads to confinement (monopole proliferation in the $U(1)$ gauge theory) whereas the situation may or may not
-- depending if the $\mathbb{Z}_2$ gauge theory is in its confined or deconfined phase --  correspond to a liquid with deconfined spinons.}
For  a detailed  discussion in the context of the kagome lattice, see Ref.~\cite{Huhvisonskagome2011}.

The PSG approach is a first step toward a classification of spin liquid phases, but many questions remain unanswered.\footnote{
See the oral communications (by M.~Hermele, X.-G.~Wen, Y.~Ran, and others) at the
KITP Program: ``Frustrated Magnetism and Quantum Spin Liquids:
From Theory and Models to Experiments'' (Aug 13 - Nov 9, 2012).
Available online at: http://online.kitp.ucsb.edu/online/fragnets12}
What is the domain of validity of the PSG classification beyond mean field ? How to ``measure'' the PSG in a spin wave function ?\footnote{This question is tackled in Ref.~~\cite{messioPSG2012},
where it is proposed to determine a PSG through the measurement of fluxes which are physical observables, that can be expressed in terms of spin operators.} Can two different different PSG correspond to the same phase once fluctuations are included ?
What is the relation (duality?) between fermionic and bosonic $\mathbb{Z}_2$ liquids ?\footnote{This has been anlyzed in details in one case on the honeycomb lattice.\cite{luranSLhoneycomb2011}}
Is there some spin liquids that cannot be reached by these approaches?

On the  kagome lattice Wang \textit{et al.} studied the PSG of SU(2) invariant hamiltonian in the Schwinger boson representation  and found  eight different Schwinger
boson mean-field Ansaetze of $\mathbb{Z}_2$ SLs  which preserve all space symmetries (4 of them only have non zero amplitude on  nearest neighbor bonds).~\cite{wangVishwanath2006}
Lu \textit{et al.}\cite{SFMFT_kagome_PSG} have done the same analysis in the Schwinger fermion representation and found 20 different spin liquids (with only 5 gapped spin liquids
with non zero gauge field  on nn bonds).~\cite{SFMFT_kagome_PSG} The relation between the
20 $\mathbb{Z}_2$ SLs in Schwinger fermion representation  and the 8 $\mathbb{Z}_2$ SLs in Schwinger boson representation has not been clarified up to now.
The precise gauge structure of the spin liquid exhibited in recent DMRG   approaches on the kagome lattice is not known,
although two proposals  were made in Refs.~\cite{SFMFT_kagome_PSG} and~\cite{Messiocuboc12012}.

\subsection{Competing phases}
\label{subsec:competing phases}
The classical model is known to be at a high degeneracy point in parameter space [see \S\ref{subsec:classical}] and we can expect that perturbations larger than the spin gap
and/or larger than the vison gap
will
destabilize the $\mathbb{Z}_2$ Spin liquid in favor of different competing phases. It has been shown for instance that a large enough Dzyaloshinskii-Moriya (DM) coupling  (larger than about one tenth the Heisenberg coupling constant) drives the quantum spin
liquid towards the $q=0$ semi-classical N\'eel long-range order.\cite{cepas2008,messioDM2010} We expect a large enough anti-ferromagnetic second neighbor
coupling to have the same effect. Other transitions towards the semi-classical $\sqrt3\times\sqrt3$ and cuboc1 phases  upon increasing ferromagnetic 2nd neighbor or anti-ferromagnetic third neighbor interactions
are also expected. But the true nature of these transitions and/or the presence of  spin liquid phases   in between are today  unknown. As explained above  the number of possibilities is considerable.

We  thus  give  below a rapid survey of  the various   states that have been proposed as plausible spin-1/2 ground states of the Heisenberg problem, as they could be interesting for nearby phases.
\subsubsection{Valence Bond Crystals}
The ``simplest''
crystal\cite{sma02} has a unit cell of 12 spins and is made of resonating  {\em stars} with 6
dimers.   The second
VBC, with a unit cell of 36 spins, is  made   of  resonating  (trimerized)  {\em
hexagons}. It was discussed as the most reasonable crystal for the pure $J_1$ model
 by Marston and  Zeng\cite{mz91} and Zeng  and
Elser\cite{ze95}  and rediscussed more recently.\cite{ns03,sh2007,singh2008}
In
both  scenarios it is the energy gained by {\em local
resonances} (involving respectively 6 and 3 valence-bonds) which
drives the system toward a  VBC. From the energy point  of view,
the  star  VBC is   probably  less realistic  since it involves a
much longer  resonance loop. In the pure Heisenberg model
the resonance  loop involving 6 valence-bonds
around  a star  has a vanishing amplitude at  the  lowest
non-trivial  order of the overlap expansion  in   the   RVB
subspace,  as  was    shown  by Zeng    and Elser.\cite{ze95}
In the approximation  where  only the
shortest resonance  loops are present,  the model was  indeed
found to be in the {\em  hexagon} VBC phase. A crucial
(numerical)  result of Zeng and Elser\cite{ze95}  is  however that
at the pure  nearest neighbor Heisenberg  point this VBC {\em   melts} when higher order resonances loops are
included. Extra couplings would be  needed to stabilize this VBC phase.
For further reading concerning possible VBC instabilities as well as
variationnal approches (tensor networks, RVB subspace, projected fermionic wave functions)
see Refs.~\cite{Poilblanc2010,Evenbly2010,Schwandt2010,Iqbal2011,PoilblancMisguich2011}.

\subsubsection{ $U(1)$  Dirac Spin Liquid}

Ran \textit{et al.}\cite{ran2007} have constructed a variational wave function  of a  gapless $U(1)$  spin liquid.
At the mean-field level it corresponds to fermionic spinons with a conical dispersion relation at the Fermi level (Dirac fermions).
After Gutzwiller projection this wave function turn out to have a very low energy, even though it has no adjustable parameter.
Many correlation functions have an algebraic decay with distance.

The present DMRG results\cite{Yan2011,DepenbrockSchollwock2012} strongly indicate a spin gap and therefore dismiss this
Dirac $U(1)$ scenario but the later may be  a good candidate to describe a critical point or even an extended gapless phase
nearby. A full study of this interesting phase can be found in Refs.~\cite{ran2007,hermele2005,hermele2008,Kolee2010,Kolee2011,Tay_kagome,Iqbalb}

From a completely different starting point (easy plane model), Ryu \textit{et al.}\cite{ryu2007} have  developed a theory of the singlet sector of a gapless algebraic {\it vortex liquid} theory in a
XY model on the kagome lattice.

\subsubsection{Spontaneously breaking the time-reversal symmetry, ``chiral'' spin liquids}
The simplest $\mathbb{Z}_2$ topological SL does not break any lattice symmetries nor time reversal. A chiral spin liquid does not break $SU(2)$ symmetry nor translations,
but it breaks time reversal symmetry and usually  some discrete lattice  point-group symmetry. The idea  that the  kagome lattice can harbour a chiral spin liquid goes back to the end of the eighties.\cite{kl89,ywg93} This idea has revived recently,
Wen \textit{et al.}\cite{wen2010} have studied  chiral spin liquids  as instability of the Hubbard model and
Chua  \textit{et al.}\cite{Chu2011} have exhibited  an exact gapless spin liquid with stable spin Fermi surface in  a Kitaev model on the kagome lattice.

This year Messio \textit{et al.}\cite{Messiocuboc12012} have suggested that  a chiral $\mathbb{Z}_2$ spin liquid could be stabilized  in the pure
$J_1$ Heisenberg model and in the $J_1$-$J_3$ model on the kagome lattice. In the classical phase diagram of the $J_1$-$J_3$ model an infinitesimal $J_3$,
3rd neighbor coupling across the hexagons, lifts the degeneracy of the pure $J_1$ model in favor of a twelve sublattice LRO named \textit{cuboc1},
where the order parameter has the symmetry of a cuboctahedron and the neighboring spins are at $120^\circ$.
 This magnetic structure and its time-reversal counter part (spin inversion) cannot be transformed into each other by a global spin rotation in $SO(3)$.
In that sense it is a {\it chiral} magnetic order.

Three spins on a nearest neighbor triangle form a plane, but the spins of an
hexagon are not planar: the triple product of second neighbor spins around an hexagon (scalar chirality) is non zero ($ \langle\vec{S}_i\cdot(\vec{S}_j\times\vec{S}_k)\rangle = 0.0148 $).
Messio's  SL state,  the spin liquid descendant of this cuboc1  classical structure, is --
at the Schwinger boson mean-field level -- slightly more stable than the plain $\mathbb{Z}_2$ topological SL. It does not break any translation symmetry, but,
since its ordered parent is chiral,
it breaks time reversal (there is a non trivial flux on the hexagons -- {\i.e.} not $0$ nor $\pi$) and one reflection symmetry. The above-mentioned chirality has been computed  in ED in the ground state of
samples with an odd number of spins  (N=21 and 27) and found to be sizeable (0.817 of the classical value).\cite{web98}
The 48-site spectrum obtained by La\"uchli {\it et al.}\cite{LauchliKITP2012} may be compatible with the 8-fold quasi-degeneracy expected for such a chiral $\mathbb{Z}_2$
liquid.  DMRG or larger size ED calculations
of the associated  6-point correlation functions would be very valuable.
It has been conjectured that, due to gauge fluctuations (singlet bond fluctuations), the expectation value of the flux along large loops
would obey a perimeter law in a chiral spin liquid,\cite{wwz89} but this has not been tested numerically.\footnote{Such an hypothesis on the correlations
could perhaps explain the  puzzling  structure of the exact spectra of small samples as well as the presence of a non zero Chern
number for spin-1/2 excitations.~\cite{web98}}
Concerning the topological entanglement entropy, the values
measured by Jiang \textit{et al.}\cite{Jiangwangbalents2012}  and Depenbrock \textit{et al.}\cite{DepenbrockSchollwock2012}
are compatible with a (non-chiral) RVB $\mathbb{Z}_2$ SL but the values to be expected in the chiral SL scenario remains to
be worked out precisely.
If a chiral SL phase does exist at $T=0$, it should undergo a finite temperature a phase transition above which the time-reversal symmetry is  restored.
The classical problem  has been studied, showing a rich interplay of $\mathbb{Z}_2$ vortices and chirality domain walls,\cite{domenge2008,messio2008}
but the quantum case has not been studied.

\subsection{Experiments in compounds with kagome-like lattices}
\label{subsec:experiments}
The first experiments on compounds with kagome-like lattices   were done on compounds with spin 3/2, 5/2 or larger.
We have no place to recall all these results but it may be remembered that ``non-classical'' consequences of frustration were very early  observed in these compounds.
SrCr$_{9p}$Ga$_{12-9p}O_{19}$, with the spin-$\frac{3}{2}$ Cr ions on a kagome lattice (or  a bilayer of pyrochlore) has been one of the compounds which have been
studied through a large range of techniques. The magnetic excitations of this compound
as seen by muons spectroscopy were  described as  itinerant spins $\frac{1}{2}$  in
a ``sea of singlets''.\cite{ukkll94} The non-linear spin susceptibility of
SrCr$_9$Ga$_{12}$0$_{19}$ exhibits a very large increase at about 5 K,
reminiscent of  spin glasses, but  neutrons and muons show
that  a very significant fraction  of the spins  are  not frozen below
this   temperature     and       still exhibit    very       rapid
fluctuations.\cite{lbar96} The same  phenomena  have been observed  in
two jarosites that are equally  good models of kagome antiferromagnets
with half-odd-integer spin per unit cell.\cite{kklllwutdg96,whmmt98}
The low temperature specific heat of these spin systems  is  unusually large, with  a double
peak    structure.  In SrCr$_{9p}$Ga$_{12-9p}O_{19}$, it was shown that the low temperature peak  was insensitive to
large magnetic fields and apparently dominated by singlet states.\cite{rhw00}
This might have some relationship with the results of  numerical calculations performed on the  spin-$\frac{1}{2}$ model.\cite{smlbpwe00}

The theoretical interest in spin-1/2 compounds on the kagome lattice has lead the experimentalists to a world-wide effort to synthesize  new compounds.
Volborthite synthesized at the  ISSP by Hiroi's group has been the first success. Unhappily, it is not a ``perfect'' representative of the \textit{spatially isotropic} Hamiltonian and its modelization
is still disputed. This compound has attracted a lot of attention. For a partial  bibliography see Refs.~\cite{hh01,magnetization_plateaux_2011,Nilsenvolbortite2011,yavors'kii,wang07}
and references therein.

Herbersmithite, with the chemical formula ZnCu$_3$(OH)$_6$Cl$_2$, a rare mineral identified in 2004 and named after G.~F.~Herbertsmith (1872-1957) was first synthesized  in 2005 in MIT.\cite{Shores2005}
The spin-1/2  Cu$^{2+}$ ions form 2D kagome lattices. Inelastic neutron scattering,
NMR and $\mu$SR consistently show that, in zero or very low magnetic  field, Herbersmithite remains fluctuating down to 20 mK.\cite{mendels2007,helton07,olariu2008,imai2008,devries2009,jeong2011}
This compound is  undoubtedly in a SL  state but not a perfect realization of the nearest neighbor Heisenberg model as its gap, if it exists, is smaller than $J/8000$
(where $J$, the nearest neighbor coupling constant, is about 170K). In fact the lack of inversion center on the magnetic bounds allows for Dzyaloshinskii-Moriya interaction
of spin orbit origin.\cite{rigol2007,rigol2007b} These couplings have been estimated of the order of $0.08J$,\cite{Zorko2008,shawish2010}
not far from the theoretically expected quantum critical point.\cite{cepas2008}
The role of impurities in such a material has also been  theoretically discussed.\cite{Misguich2007,rousochatzakis09,devries2012}
Single crystals of Herbersmithite are now available and we can expect that future inelastic  neutron scattering experiments will uncover
new specificities of this spin liquid, and will check more deeply the different theoretical scenarios.\cite{ran2007,hermele2008,Tchernyshyov2010}

Many spin-1/2 compounds with the kagome geometry turn out to be ferromagnets.~\cite{kageyama2002,narumi2004,karaki2010} Domenge \textit{et al.} suggested in 2005 that competing
interactions (ferromagnetic between nearest neighbors and antiferromagnetic between second neighbors and farther) could lead classically to  a non-magnetic phase with a  non-planar twelve sublattice
magnetic unit cell (called cuboc2) as well as chiral properties.\cite{Domenge2005}
At $T=0$ this classical phase breaks  time reversal and spin inversion symmetry and a finite temperature is
needed to restore these discrete symmetries. Due to the presence of $\mathbb{Z}_2$ vortices, the phase transition is not Ising-like but weakly first order.\cite{domenge2008}
Messio has recently shown  that this phase is the ground state of $J_1$-$J_2$-$J_3$ model for a large domain of parameters.\cite{Messioregular2011}
The nature of the spin-1/2 phase of this model is for the moment unknown.

In 2008 a new compound called kapellasite, polymorphous to herbertsmithite, was synthesized in London.\cite{kapella1,kapella2}
In this compound the kagome planes are
well separated from each other by non magnetic Zn$^{2+}$ and Cl$^{-}$ ions. The absence of freezing down to very low temperature as well as NMR and inelastic neutron scattering spectra lead experimentalists to
the conclusion that it is a spin liquid.\cite{fakkapellasite2012}
\textit{Ab initio} calculations suggested that the $J_1$-$J_2$-$J_3$ parameters describing this compound were such that in the classical limit the ground state
was a chiral and non-planar magnetic structure with the symmetry of a cubocatedron  (in this \textit{cuboc1} state the spins are  at 120 degrees from each other
and define a plane on each small triangle, spins around an hexagon are however not coplanar (chirality)).\cite{Janson_2008}
The  inelastic neutron spectra
dismissed this hypothesis and suggested
that the parent classical phase is instead the \textit{cuboc2} phase (this state also has a three-dimensional order parameter with the symmetry of a cubocatedron but neighboring
spins are at 60 degrees from each other, there is chirality on the small triangles and the spins around distinct hexagons define distinct planes).\cite{Messioregular2011} A high temperature series
analysis of the spin susceptibility and specific heat confirms this hypothesis.~\cite{bernu2012} The model extracted from these  data allows a precise description of the $\mu$SR relaxation rate.
In spite of a strong disorder there is no  signature of ``defects''. A  theoretical explanation for the continuum of excitations seen in this compound is however still lacking.

Two other compounds have been synthesized recently:  vesignieite\cite{Okamoto2009}  and haydeeite.~\cite{kapella2} As herbertsmithite -- and in spite of its slight distorsion --
vesignieite seems a rather good model of the AF Heisenberg model but it is found to partially order at  $T_c \sim J/6$.\cite{Colman2011,Quilliam2011} This  might be mainly due to the presence of
DM interactions  relatively larger than in herbersmithite. It has been suggested that these two compounds are respectively below (herbersmithite) and above (vesignieite) the quantum critical point
separating the spin liquid phase from the $q=0$ Néel ordered phase.\cite{cepas2008}  Haydeeite, $ \alpha-Cu_3Mg(OD)_6Cl_2$,  is isostructural to kapellasite $ \alpha-Cu_3Zn(OD)_6Cl_2$, but the relative
weight of the first neighbor ferromagnetic coupling compared to second and third neighbor antiferromagnetic coupling is apparently stronger than in kapellasite and the system
is a ferromagnet.~\cite{bernu2012}

\section{Conclusions }
\setcounter{footnote}{0}

\begin{sidewaystable}[htb]
\tbl{Different phases encountered in $SU(2)$-symmetric frustrated
models in 2D \label{tab:conclusion} }{
\begin{tabular}{|c|c|c|c|c|c|c|c|}
\hline 
        &       &           &           &       &       &           &           \\
Phase       & $2S$/cell & Order         & Degeneracy        & Broken sym.   & Excitations   & Thermo.       & Examples      \\
        &       &           &           &       &       &           &           \\
\hline 
N\'eel AF   &       &           &           & $SU(2)$   &Gapless magnons& $C_v\sim T^2$ & Spin-$\frac{1}{2}$ triangular \\
p-sublattice    & any       & spin-spin LRO     &$\mathcal{O}(N^p)$ & Translations  &(spin waves)   & $\chi\sim {\rm cst}$  & Heisenberg AF     \\
        &       &           &           & Point group   &       &           &           \\
\hline 
VBC
(\S\ref{sec:VBC})& odd      & singlet-singlet  LRO  & $>1$          & Translations  &Gapped magnons & $C_v$ and $\chi$  & Honeycomb $J_1$--$J_2$\\
(spontaneous)   &           & {\tiny colli.
                    spin-spin SRO}  &           &  Point group  &               & activated     & Checkerboard      \\
        &       &           &           &       &       &           & Square $J_1$--$J_2$ ? \\
\hline 
VBC (\S\ref{sec:VBC})
        & even      &   (1)        & 1         & None      &Gapped magnons & $C_v$ and $\chi$  & SrCu$_2$(BO$_3$)$_2$  \\
(explicit)  &       &           &               &           &       & activated         & CaV$_4$O$_9$          \\
\hline VBS (\S \ref{ssec:VBS}) & even      & ``String'' LRO    & 1             & None      &Gapped magnons & $C_v$ and $\chi$  & AKLT  Hamiltonians    \\
        &           &           &               &           &Edge excitations& activated        & $S=1$ kagome AF ? \\
\hline 
        &           & Topological           & 4             &       &Gapped spinons & $C_v$ and $\chi$  & MSE (\S\ref{sec:MSE})\\
RVB SL      & odd       & {\tiny non-colli. SRO}& (torus)       & None      &Gapped visons  & activated         & Kagome AF \\
(\S\ref{ssec:QDMTri} \S\ref{ssec:QDMKag}, \S\ref{ssec:RVBMSE})        &       &           &           &       &       &           &   \\
\hline 
\end{tabular}
}
\begin{tabnote}
This  table   summarizes  the  properties  of   some  important
phases encountered in 2D frustrated magnets.  $S$ is the value of
the spin on each site.  ``Order''    refers  to the   nature of
the   long-range correlations (if any).  The ground state
degeneracy in the limit of an infinite system  (with periodic
boundary   conditions) is indicated in the  fourth column,  except
for   RVB \index{RVB}  SL   it  is  related  to   the
spontaneously broken symmetries  mentioned    in the  next
column. Elementary excitations  and  the  low-temperature
behavior of  the specific heat ($C_v$) and uniform susceptibility
($\chi$) are given in column six.  The  last column gives  some
examples of theoretical or experimental realizations of these
phases. The five families of systems presented here of course do
not exhaust all possibilities.   Some
authors classify all the systems  with gapped excitations  in a
loose category    of ``quantum  disordered systems'',  alluding to
the  absence of  N\'eel long-range order.  It is a rather
unhappy appellation for VBC (which obviously have some order) and
in fact for most of the quantum systems with   a   gap.  In
classical   statistical  physics ``disorder'' is associated to
some extensive entropy, which is not  the case in the quantum systems at $T=0$.
(1) The presence of order in an explicit VBC wave function is somewhat a question of definition:
while the singlet-singlet correlation $\langle (\vec S_1 \cdot \vec S_2) (\vec S_{1'} \cdot \vec S_{2'})\rangle$
should show some long-range modulations when the bond $(12)$ and $(1'2')$ are far apart, its {\it connected} counterpart
$\langle (\vec S_1 \cdot \vec S_2) (\vec S_{1'} \cdot \vec S_{2'})\rangle$-$\langle \vec S_1 \cdot \vec S_2\rangle\langle\vec S_{1'} \cdot \vec S_{2'}\rangle$
remains short-range.
\end{tabnote}
\end{sidewaystable}

\newpage

We  conclude  by summarizing  some
properties  --and  related  open questions-- of the different
phases  discussed in  this review.

The      properties  of       these   phases  are     summarized in Tab.~\ref{tab:conclusion}.  Semi-classical  phases with N\'eel
long-range order, magnons as gapless  excitations, do exist in spin-$\frac{1}{2}$ 2D systems  with  moderate frustration:  the
Heisenberg   model on  the triangular lattice is  the most explicit  example, with a  sublattice magnetization about one
half of the classical value.\cite{bllp94} On a large but finite system the ground state manifold is a tower of states including
a number of eigenstates which is a power of the number of sites.

An increased  frustration,  lower coordination number  or smaller spin lead  to  quantum phases, with a  ground state  of higher
symmetry, no long-range  order  in  spin-spin correlations,   a spin gap  and the restored $SU(2)$ symmetry. Two  main
alternatives are then opened: the VBC or VBS  phases on one hand, the RVB SL  on the other.  These states
first  require     the formation  of  short-range singlets.  When a particular local   resonance  pattern dominates  the dynamics of
the Hamiltonian the system will try to maximize the number of occurrences of this pattern.  This is usually  achieved by a
regular arrangement, that is a VBC. When  no such pattern dominates the system may form   a  translation invariant RVB   SL.
In  the  first case the ground state  can  be     qualitatively described   by   one   ordered configuration of singlets dressed
by small fluctuations. In the RVB SL the  amplitudes   of   the wave function   are  distributed  over  an exponentially large
number of configurations. These ground states lead to  very different excitations:  $\Delta  S=1$ gapped  magnons in the first
case (and  $\Delta S=0$ domain-wall excitations), gapped $\Delta S=0$ visons  and  gapped $\Delta S=\frac{1}{2}$  unconfined
spinons in the   second case.

The bosonic large-$N$ results (\S\ref{sec:largeN})  indicate that VBC are expected  in situations
where the short range magnetic order is collinear whereas SL appear when these correlation are non-collinear.
Although this is verified in many cases (kagome in particular), some exceptions seem to exist also. For instance,
the recent DMRG simulations  on the $J_1-J_2$ model on the square lattice\cite{Jiang2012} indicate a $\mathbb{Z}_2$ SL ground state in the vicinity of $J_2/J_1\sim 0.5$.

These states obey the 2D extension  of LSM theorem:\cite{o00,Hastings04,ns07}  if $2S$ (per unit cell) is odd and if excitations are  gapped there must be
ground state degeneracy in the thermodynamic limit (with periodic boundary conditions). However  the origin of the degeneracy
differs  in the two types   of  states.    In a   VBC   the degeneracy is associated to spontaneously broken
translation symmetry whereas in the RVB SL the   degeneracy  has a topological origin.\footnote{On the kagome lattice (three spins in the unit cell) a precursor
of this topological degeneracy might have been seen  in the $N=48$-sites spectrum.\cite{LauchliKITP2012}}
In  the VBS  (or explicit VBC) the
ground state is unique but  $2S$ is even (in one unit cell).
The situation of the hexagonal lattice with respect to this theorem and possible topological degeneracies is interesting:
since the lattice has 2 sites per unit cell, the LSM theorem does dot apply. Still, on this lattice, there is no obvious way to construct a gapped and $SU(2)$-symmetric spin-$\frac{1}{2}$
wave function without any conventional nor topological order  (see however Ref.~\cite{kptv12}).
Although not covered in this chapter, we mention that some recent quantum Monte Carlo simulations indicating an insulating
SL in the (unfrustrated!) Hubbard model at half-filling on the hexagonal lattice\cite{mlwam10} (see also Ref.~\cite{soy12}) have triggered a large number of studies concerning frustrated spin models on this lattice.\cite{wang10,ashcml11,clr11,yacls12}

These paradigms    are relatively well   understood,  at least  on the qualitative level.  They also appear naturally  in the broader
context of the   classification  of  Mott insulators.\cite{sachdev03} However several kinds of  2D
frustrated  magnets do  not fall in  these simple classes and many open  questions remain.

This review was restricted to $SU(2)$ invariant Hamiltonians. Whereas the Ising limit has been much studied, the differences
between quantum XY and Heisenberg models have received much less attention.

Chiral  SL   have  not been  thoroughly  discussed  in   this review.    They are characterized by a broken time-reversal symmetry. This
possibility has been    studied   intensely since the 80's.\cite{wwz89,kl87-89,ywg93,fradkinbook}
There might be a revival of this line of research in the future.\cite{Messiocuboc12012,fakkapellasite2012}

The issue of  quantum phase transitions in  frustrated antiferromagnet is also an active topic  that is not presented  in
this review.   Many properties of these critical points are  still unknown, not to mention the fascinating problems associated with
(quenched) disorder.

Limited by place (and competence) we have not discussed in details the works done   on spatially anisotropic  models. This  field
which is in between 1D (review   by P.~Lecheminant  in  this book) and 2D    is extremely flourishing tackled by bosonization
and large-$N$ methods.

Ten years ago we wrote in the first edition : ``To conclude  we  would like  to emphasize  that  new analytical and/or numerical methods  are highly desirable to  proceed in the
analysis of the two emblematic  problems by which  we have opened and closed this review:  the $J_1$--$J_2$    model  on the  square
lattice   and the spin-$\frac{1}{2}$  Heisenberg  model on   the kagome (and pyrochlore) lattices.   In  both  of these  problems a
consensus  remains  to be obtained.''

During the last few years two-dimensional DMRG studies have allowed a breakthrough which has lead to a complete revisiting of these two
problems.\cite{Yan2011}\cite{jiangsheng2008}\cite{DepenbrockSchollwock2012}\cite{Jiangwangbalents2012}\cite{Jiang2012}
As explained in this rewritten section  the full understanding and consensus is not yet
reached but a very important  step forward  has been achieved.



\begin{thebibliography}{100}

\bibitem{manousakis91}
E. Manousakis,
Rev. Mod. Phys. {\bf 63}, 1 (1991).

\bibitem{bllp94}
B. Bernu, P. Lecheminant, C. Lhuillier and L. Pierre,
Phys. Rev. B {\bf 50},  10048  (1994).

\bibitem{HFM2000}
Proceedings  of  the {\em Highly Frustrated Magnetism  2000} conference,
published in J. Can. Phys {\bf 79}, (2001).


\bibitem{fradkinbook}
E.   Fradkin, {\em   Field  Theories  of  Condensed Matter   Systems},
Addison-Wesley (1998).

\bibitem{auerbachbook}
A. Auerbach, {\em Interacting electrons and Quantum Magnetism},
Springer-Verlag, Berlin Heidelberg New York, 1994.

\bibitem{tsvelikbook}
A.~M.~Tsvelik, {\em Quantum Field Theory in Condensed Matter Physics}
Cambridge University Press  (1996).

\bibitem{sachdevbook}
S.~Sachdev, {\em Quantum Phase  Transitions},    Cambridge  U.  Press, New
York (1999).

\bibitem{lt47}
J. M. Luttinger and L.~Tisza, Phys. Rev. {\bf 70}, 954 (1946).

\bibitem{cd88}
P. Chandra and B. Dou{\c c}ot, Phys. Rev. B {\bf 38}, 9335 (1988).


\bibitem{vbcc80}
J. Villain, R. Bidaux, J. P. Carton and R. Conte,
J. Phys. (Paris) {\bf 41}, 1263 (1980).


\bibitem{shender82}
E.~Shender, Sov. Phys. JETP {\bf 56}, 178 (1982).

\bibitem{h89}
C.~L. Henley, Phys. Rev. Lett. {\bf 62}, 2056 (1989).

\bibitem{mdjr90}
A. Moreo, E.  Dagotto, T. Jolic{\oe}ur and  J. Riera,
Phys. Rev.  B {\bf 42}, 6283 (1990).

\bibitem{ccl90a}
P. Chandra, P. Coleman, and  A. Larkin,
Phys. Rev.  Lett. {\bf 64}, 88 (1990).

\bibitem{szosh03}
R.~R.~P.~Singh, W.~Zheng, J.~Oitmaa, O.~P.~Sushkov, C.~J.~Hamer,
Phys. Rev. Lett. {\bf 91}, 017201 (2003).

\bibitem{mbp03}
G.~Misguich,  B.~Bernu   and  L.~Pierre,
Phys. Rev. B {\bf 68}, 113409 (2003).

\bibitem{wm03}
C. Weber, L.~Capriotti, G.~Misguich, F.~Becca, M.~Elhajal and F.~Mila,
Phys. Rev. Lett. {\bf 91}, 177202 (2003).

\bibitem{jdgb90}
T. Jolic{\oe}ur,  E. Dagotto, E.   Gagliano and S.   Bacci,
Phys.  Rev. B {\bf 42}, 4800 (1990).

\bibitem{cj92}
A. Chubukov and T. Jolic{\oe}ur, Phys. Rev. B {\bf 46}, 11137 (1992).

\bibitem{k93}
S.~E.~Korshunov, Phys. Rev. B {\bf 47}, 6165 (1993).

\bibitem{lblp95}
P.~Lecheminant, B.~Bernu, C.~Lhuillier  and L.~Pierre,
Phys.  Rev. B {\bf 52}, 6647 (1995).



\bibitem{melzi00}
R.~Melzi {\it et   al.}, Phys.  Rev.  Lett.  {\bf 85}, 1318 (2000).


\bibitem{melzi01}
R. Melzi  {\it et   al.}, Phys. Rev.  B   {\bf 64}, 024409 (2001).


\bibitem{rosner02}
H.   Rosner    {\it et   al.},
Phys.  Rev.  Lett.  {\bf 88}, 186405 (2002).


\bibitem{rosner03}
H. Rosner   {\it et   al.},   Phys.   Rev. B    {\bf 67},   014416    (2003).


\bibitem{klein82}
D. J. Klein,
J. Phys. A: Math. Gen. {\bf 15}, 661 (1982).


\bibitem{mg69}
C.~K.~Majumdar and D.~K.~Ghosh, J. Math. Phys. {\bf 10}, 1399 (1969).


\bibitem{sgh88}
R.~R.~P.~Singh, M.~P.~Gelfand and D.~A.~Huse,
Phys. Rev. Lett. {\bf 61}, 2484 (1988).


\bibitem{gsh90}
M.~P.~Gelfand, R.~R.~P.~Singh and D.~A.~Huse, J. Stat. Phys. {\bf 59},
1093 (1990).


\bibitem{woh91}
Zheng Weihong, J. Oitmaa, and C. J. Hamer,
Phys. Rev. B {\bf 43}, 8321 (1991).


\bibitem{ow96}
J. Oitmaa and Zheng  Weihong, Phys. Rev. B {\bf 54}, 3022 (1996).


\bibitem{gsh89}
M. P. Gelfand, R. R. P.  Singh and D. A.  Huse,
Phys. Rev. B {\bf 40}, 10801 (1989).


\bibitem{gelfand90}
M. P. Gelfand.  Phys. Rev. B {\bf 42}, 8206 (1990).


\bibitem{swho99}
R.~R.~P.~Singh, Zheng Weihong, C.~J.~Hamer, and  J.~Oitmaa,
Phys. Rev.  B {\bf 60}, 7278 (1999).


\bibitem{kosw00}
V.~N.~Kotov, J.~Oitmaa, O.~Sushkov  and  Zheng Weihong,
Phil. Mag. B  {\bf 80}, 1483 (2000).


\bibitem{zu96}
M. E.   Zhitomirsky  and  K.  Ueda,
Phys.  Rev. B {\bf 54}, 9007 (1996).


\bibitem{cbps01}
L.  Capriotti,      F.     Becca,  A. Parola,    and      S.  Sorella,
Phys.  Rev. Lett.  {\bf 87},  097201  (2001).


\bibitem{sow02}
O.~P.~Sushkov, J.~Oitmaa, and   Zheng Weihong,
Phys.  Rev.  B {\bf 66}, 054401 (2002).


\bibitem{sow01}
O. P. Sushkov, J. Oitmaa, and Zheng Weihong,
Phys. Rev. B {\bf 63}, 104420 (2001).


\bibitem{cls00}
M.~S.~L. du~Croo~de  Jongh,  J.~M.~J.~Van   Leeuwen,  and  W.~Van  Saarloos,
Phys. Rev. B {\bf 62}, 14844 (2000).

\bibitem{dm89}
E.   Dagotto and  A.  Moreo,  Phys.   Rev.  B {\bf   39}, 4744 (1989),
Phys. Rev. Lett. {\bf 63}, 2148 (1989).

\bibitem{fkksrr90}
F.~Figueirido {\it et al.}, Phys. Rev. B {\bf 41}, 4619 (1990).


\bibitem{pgbd91}
D.~Poilblanc, E.~Gagliano, S.~Bacci and E.~Dagotto,
Phys. Rev. B {\bf 43}, 10970 (1991).

\bibitem{schulz}
H.~J.~Schulz, T.~A.~L.~Ziman, Europhys. Lett. {\bf 18}, 355 (1992).
T.~Einarsson and H.~J.~Schulz, Phys. Rev. B {\bf 51}, 6151 (1995).


H.~J.~Schulz, T.~A.~L.~Ziman, D.~Poilblanc, J. Physique I {\bf 6}, 675
(1996).



\bibitem{nlsm}
S. Chakravarty, B.~I.~Halperin and   D.~R.~Nelson,
Phys. Rev.  B {\bf 39}, 2344 (1989).
H.~Neuberger and T.~Ziman, Phys. Rev. B {\bf 39}, 2608 (1989).
D.~Fisher, Phys. Rev. B {\bf 39}, 11783 (1989).
T.~Einarsson and H.~Johannesson, Phys. Rev. B {\bf 43}, 5867 (1991).
P.  Hasenfratz and F.~Niedermayer,
Z.  Phys. B.  Condens. Matter {\bf 92}, 91 (1993).

\bibitem{s97}
A.~W. Sandvik, Phys. Rev. B {\bf 56},  11678  (1997).

\bibitem{s98}
S. Sorella, Phys. Rev. Lett. {\bf 80}, 4558 (1998).

\bibitem{s01}
S. Sorella, Phys. Rev. B {\bf 64},  024512  (2001).

\bibitem{w92}
S. White, Phys. Rev. Lett. {\bf 69}, 2863 (1992).

\bibitem{w93}
S. White, Phys. Rev. B {\bf 48}, 10345 (1993).

\bibitem{gl91}
T. Giamarchi and C. Lhuillier, Phys. Rev. B {\bf 43}, 12943 (1991).


\bibitem{ss81}
B. Shastry and B. Sutherland, Phys. Rev. Lett. {\bf 47},  964  (1981).

\bibitem{h82}
F. Haldane, Phys. Rev. B {\bf 25},  4925  (1982).

\bibitem{a89}
I. Affleck, J. Phys. Cond. Matt. {\bf 1},  3047  (1989).

\bibitem{ys97}
H. Yokoyama and Y. Saiga, J. Phys. Soc. Jpn. {\bf 66},  3617  (1997).

\bibitem{nt97a}
T. Nakamura and S. Takada, Phys. Rev. B {\bf 55},  14413  (1997).

\bibitem{asrp99}
D. Augier, E. Sorensen, J. Riera, and D. Poilblanc,
Phys. Rev. B {\bf 60}, 1075 (1999).

\bibitem{dr96}
E. Dagotto and T.~M. Rice, Science {\bf 271},  618  (1996).



\bibitem{nt97}
A.~A.~Nersesyan and A.~M.~Tsvelik,
Phys. Rev. Lett. {\bf 78}, 3939 (1997).

\bibitem{km98} A.~K.~Kolezhuk and H.-J.  Mikeska,
Int. J. Mod. Phys. B {\bf 12}, 2325 (1998).
Phys. Rev. Lett. {\bf 80}, 2709  (1998).
Phys. Rev.  B {\bf  56}, 11380  (1997).


\bibitem{aklt87}
I.~Affleck, T.~Kennedy, E.~Lieb, and H.~Tasaki,
Phys. Rev. Lett. {\bf 59}, 799 (1987).

\bibitem{aklt88}
I. Affleck, T. Kennedy, E.~H. Lieb, and H. Tasaki,
Commun. Math. Phys. {\bf  115},  477  (1988).

\bibitem{nr89}
M. den Nijs and K. Rommelse, Phys. Rev. B {\bf 40},  4709  (1989).

\bibitem{kt92}
T. Kennedy and H. Tasaki, Phys. Rev. B {\bf 45},  304  (1992).

\bibitem{j02}
T. Jolic{\oe}ur, private communication.

\bibitem{h00}
K. Hida, J. Phys. Soc. Jpn. {\bf 69},  4003  (2000).

\bibitem{yk00}
Y.~Yamashita and K.~Ueda,
Phys. Rev. Lett. {\bf 85},  4960  (2000).

\bibitem{CaV4O9}
N.~Katoh and M.~Imada, J. Phys. Soc. Jpn. {\bf 63},  4529  (1994).
M.~Troyer, H.~Tsunetsugu, and D.~Wuertz, Phys. Rev. B {\bf 50},  13515  (1994).
S.~Taniguchi {\it et~al.}, J. Phys. Soc. Jpn. {\bf 64},  2758  (1995).
K.~Kodama {\it et~al.}, J. Phys. Soc. Jpn. {\bf 65},  1941  (1996).
Y.~Fukumoto and A.~Oguchi, J. Phys. Soc. Jpn. {\bf 65},  1440  (1996).
K.~Kodama {\it et~al.}, J. Phys. Soc. Jpn. {\bf 66},  793  (1997).
T.~Miyasaki and D.~Yoshioka, J. Phys. Soc. Jpn. {\bf 65},  2370  (1996).
T.~Ohama, H.~Yasuoka, M.~Isobe, and Y.~Ueda,
J. Phys. Soc. Jpn. {\bf 66}, 23 (1997).

\bibitem{uksl96}
K. Ueda, H. Kontani, M. Sigrist, and P.~A. Lee, Phys. Rev. Lett. {\bf 76}, 1932  (1996).

\bibitem{am96b}
M.~Albrecht and F. Mila, Phys. Rev. B {\bf 53},  2945  (1996).

\bibitem{tku96}
M. Troyer, H. Kontani, and K. Ueda, Phys. Rev. Lett. {\bf 76},  3822  (1996).



\bibitem{sr96}
S. Sachdev and N. Read, Phys. Rev. Lett. {\bf 77},  4800  (1996).

\bibitem{szksu96}
O.~A. Starykh {\it et~al.}, Phys. Rev. Lett. {\bf 77},  2558  (1996).



\bibitem{wgsoh97}
Zheng Weihong {\it et~al.}, Phys. Rev. B {\bf 55},  11377  (1997).

\bibitem{k99}
H. Kageyama {\it et~al.}, Phys. Rev. Lett. {\bf 82},  3168  (1999).

\bibitem{nkoum99}
H. Nojiri {\it et~al.}, J. Phys. Soc. Jpn. {\bf 68},  2906  (1999).

\bibitem{mu99}
S. Miyahara and K. Ueda, Phys. Rev. Lett. {\bf 82},  3701  (1999).

\bibitem{k00}
H. Kageyama {\it et~al.}, Phys. Rev. Lett. {\bf 84},  5876  (2000).

\bibitem{kk00}
A. Koga and N. Kawakami, Phys. Rev. Lett. {\bf 84},  4461  (2000).

\bibitem{msku00}
E. M\"uller-Hartmann, R.~R.~P. Singh, C. Knetter, and G.~S. Uhrig,
Phys. Rev.  Lett. {\bf 84}, 1808 (2000).

\bibitem{tmu01}
K. Totsuka, S. Miyahara, and K. Ueda, Phys. Rev. Lett. {\bf 86},  520  (2001).

\bibitem{lws02}
A. L\"auchli, S. Wessel, and M. Sigrist, Phys. Rev. B {\bf 66},  014401  (2002).

\bibitem{kthb02}
K. Kodama {\it et al.}, Science {\bf 298},395 (2002).

\bibitem{mu03}
S.~Miyahara and K.~Ueda,
J. Phys.: Condens. Matter {\bf 15}, R327-R366 (2003).

\bibitem{ss81a}
B. Shastry and B. Sutherland, Physica B (Amsterdam) {\bf 108},  1069  (1981).

\bibitem{cz02} O.~Cépas, T.~Ziman, Proceedings of  the conference in
Fukuoka,   Nov.  2001. Fukuoka  University   Press
[cond-mat/0207191].     O.~Cepas,    T.~Sakai,     T.~Ziman,
Progr. Theor. Phys. Suppl. {\bf 145}, 43 (2002).

\bibitem{zoh02}
W. Zheng, J. Oitmaa, and C. J. Hamer
Phys. Rev. B {\bf 65}, 014408 (2002).

\bibitem{am96a}
M.~ Albrecht and F. Mila, Europhys. Lett. {\bf 34},  145  (1996).

\bibitem{cms01}
C.~H.~Chung, J.~B.~Marston, and S.~Sachdev,
Phys.  Rev.  B {\bf 64}, 134407 (2001).



\bibitem{fo99}
Y. Fukumoto and A. Oguchi, J. Phys. Soc. Jpn. {\bf 68},  3655  (1999).

\bibitem{mt00}
T. Momoi and K. Totsuka, Phys. Rev. B {\bf 61},  3231  (2000).

\bibitem{mt00b}
T. Momoi and K. Totsuka, Phys. Rev. B {\bf 62},  15067  (2000).

\bibitem{mu00}
S. Miyahara and K. Ueda, Phys. Rev. B {\bf 61},  3417  (2000).

\bibitem{mjg01}
G. Misguich, T. Jolic{\oe}ur, and S.~M.~Girvin,
Phys. Rev. Lett. {\bf 87}, 097203 (2001).

\bibitem{lm02}
C. Lhuillier and G. Misguich, in {\em High Magnetic Fields}, edited by
C.   Berthier,  L. Levy, and  G.  Martinez  (Springer,  Berlin, 2002),
pp.~161--190, [cond-mat/0109146].


\bibitem{ll96}
P.~W. Leung and N. Lam, Phys. Rev. B {\bf 53},  2213  (1996).

\bibitem{fsl01}  J.-B.~Fouet,  P.~Sindzingre,   and  C.~Lhuillier,
Eur. Phys. J. B {\bf 20}, 241 (2001).


\bibitem{canals02}
B.~Canals, Phys. Rev. B {\bf 65}, 184408 (2002).

\bibitem{fsl03}
J.-B.~Fouet,     M.~Mambrini,    P.~Sindzingre,  and     C.~Lhuillier,
Phys. Rev. B {\bf 67}, 054411 (2003).

\bibitem{sfl02}
P. Sindzingre, J.~B. Fouet, and C. Lhuillier,
Phys. Rev. B {\bf 66}, 174424 (2002).

\bibitem{bh02}
W. Brenig  and  A. Honecker,
Phys. Rev.   B {\bf 65},  140407R (2002).

\bibitem{baa03}
E.~Berg, E.~Altman, and A.~Auerbach,
Phys. Rev. Lett. {\bf 90}, 147204 (2003).

\bibitem{tsma03}
O.~Tchernyshyov, O.~A.~Starykh, R.~Moessner and A.~G.~Abanov,
Phys. Rev. B {\bf 68}, 144422 (2003).

\bibitem{mts01}
R. Moessner, Oleg Tchernyshyov and S.~L.~Sondhi,
cond-mat/0106286.

\bibitem{pc02}
S. Palmer and J.~.T.~Chalker, Phys. Rev. B {\bf 64},  094412  (2002).

\bibitem{mc98}
R. Moessner and J.~T. Chalker, Phys. Rev. B {\bf 58},  12049  (1998).

\bibitem{mc98a}
R. Moessner and J.~T. Chalker, Phys. Rev. Lett. {\bf 80},  2929  (1998).


\bibitem{pp01}
O.~A.~Petrenko and D.~McK.~Paul,
Phys. Rev. B {\bf 63}, 024409 (2001).


\bibitem{ka02}
H. Kawamura and T. Arimori, Phys. Rev. Lett. {\bf 88}, 077202 (2002)

\bibitem{henley2000}
C.~L.~Henley, Can. J. Phys. (Canada) {\bf 79}, 1307 (2001).
[cond-mat/0009130]

\bibitem{f03}
J.-B.~Fouet, Ph.D. thesis, Universit\'e Cergy Pontoise, 2003.

\bibitem{gwsoh96}
M.~P.~Gelfand {\it et~al.}, Phys. Rev. Lett. {\bf 77},  2794  (1996).


\bibitem{bm02}
F. Becca and F. Mila, Phys. Rev. Lett. {\bf 89}, 037204 (2002).

\bibitem{mlbw99}
G. Misguich, C. Lhuillier, B. Bernu and C. Waldtmann,
Phys. Rev. B {\bf 60}, 1064 (1999).


\bibitem{a85}
I. Affleck, Phys. Rev. Lett. {\bf 54},  966  (1985).

\bibitem{am88}
I. Affleck and J. Marston, Phys. Rev. B {\bf 37},  3774  (1988).


\bibitem{aa88}
D. Arovas and A.~Auerbach, Phys. Rev. B {\bf 38},  316  (1988).

\bibitem{rs89}
N. Read and S. Sachdev, Phys. Rev. Lett. {\bf 62},  1694  (1989).

\bibitem{rs90}
N.~Read and S.~Sachdev, Phys. Rev. B {\bf 42},  4568  (1990).

\bibitem{rs91}
N. Read and S. Sachdev,
Phys. Rev. Lett. {\bf 66}, 1773 (1991).

\bibitem{sr91}
S. Sachdev and N. Read, Int. J. Mod. Phys. {\bf 5}, 219 (1991).

\bibitem{s93}  S.  Sachdev  in  {\it  Low  Dimensional  Quantum  Field
Theories   for  Condensed   Matter  Physicists}   edited  by   Y.  Lu,
S. Lundqvist, and G. Morandi, World Scientific, Singapore, 1995.
cond-mat/9303014.


\bibitem{rs89b}
N. Read and S. Sachdev, Nucl. Phys. B {\bf 316}, 609 (1989).

\bibitem{rokhsar90}
D. Rokhsar, Phys. Rev. B {\bf 42}, 2526 (1991).

\bibitem{wilson-74}
K. G. Wilson,
Phys. Rev. D {\bf 10}, 2445 (1974).


\bibitem{bdi74}

R. Balian, J. M. Drouffe, and C. Itzykson,
Phys. Rev. D {\bf 10}, 3376 (1974),
Phys.  Rev. D {\bf 11},  2098 (1975),
Phys.  Rev. D {\bf 11}, 2098 (1975).


\bibitem{schwingerboson}
D.~P.~Arovas and A.~Auerbach, Phys.  Rev.  B {\bf 38}, 316 (1988).
A.~Auerbach and D.~P.~Arovas, Phys. Rev.  Lett. {\bf 61}, 617 (1988).
J.~E.~Hirsch and S.~Tang, Phys. Rev. B {\bf 39}, 2850 (1989).

\bibitem{ceccatto}
H.~A.~Ceccatto, C.~J.~Gazza, and A.~E.~Trumper,
Phys. Rev. B {\bf 47}, 12329 (1993).
A.~E.~Trumper, L.~O.~Manuel, C.~J.~Gazza, and
H.~A.~Ceccatto Phys. Rev. Lett. {\bf 78}, 2216 (1997).

\bibitem{ot03}
We thank O.~Tchernyshyov  for pointing us this physical interpretation
of the $U(1)$ flux.


\bibitem{sachdev92}
S. Sachdev, Phys. Rev. B {\bf 45}, 12377 (1992).

\bibitem{cmm01}
C.~H.~Chung and J.~B.~Marston and Ross H.~McKenzie,
J. Phys.: Condens. Matter {\bf 13}, 5159  (2001).

\bibitem{haldane83}
F.~D.~M.~Haldane, Phys. Lett. {\bf 93A}, 464 (1983);
%

Phys. Rev. Lett. {\bf 50}, 1153 (1983).
%

\bibitem{haldane88}
F.~D.~M.~Haldane, Phys. Rev. Lett. {\bf 61}, 1029 (1988).



\bibitem{noHopfTerm2D88}
X.~G.~Wen and A.~Zee, Phys. Rev. Lett. {\bf 61}, 1025 (1988).
E.~Fradkin and M.~Stone, Phys. Rev. B {\bf 38}, 7215 (1988).
T.~Dombre and N.~Read, Phys. Rev. B {\bf 38}, 7181 (1988).


\bibitem{hkt03}
K.~Harada, N.~Kawashima, and M.~Troyer,
Phys. Rev. Lett. {\bf 90}, 117203 (2003).



\bibitem{fs79}
E. Fradkin and S. H. Shenker, Phys. Rev. D {\bf 19}, 3682 (1979).

\bibitem{k61}
P. W. Kasteleyn, Physica {\bf 27}, 1209 (1961).

\bibitem{f61}
M. E. Fisher, Phys. Rev. {\bf 124}, 1664 (1961).


\bibitem{k63}
P.~W.~Kasteleyn,
J. of Math. Phys. {\bf 4}, 287 (1963).

\bibitem{msf02}
R. Moessner, S.~L. Sondhi, E.~Fradkin,
Phys. Rev. B {\bf 65}, 024504 (2002).

\bibitem{msp02}
G. Misguich, D.~Serban, V.~Pasquier,
Phys. Rev. Lett. {\bf 89}, 137202 (2002).

\bibitem{ms02}
R. Moessner, S. L. Sondhi, Phys. Rev. B {\bf 68}, 054405 (2003).

\bibitem{rk88}
D. S. Rokhsar and S. A. Kivelson,
Phys. Rev. Lett. {\bf 61},  2376  (1988).


\bibitem{ms01}
R. Moessner and S.~L. Sondhi, Phys. Rev. Lett. {\bf 86},  1881  (2001).


\bibitem{msc00}
R. Moessner, S.~L. Sondhi, and P. Chandra,
Phys. Rev. Lett. {\bf 84}, 4457 (2000).

\bibitem{s88}
B. Sutherland, Phys. Rev. B {\bf 37}, 3786 (1988).

\bibitem{mz91} J. B. Marston, C. Zeng,
J. Appl. Phys. {\bf 69}, 5962 (1991).

\bibitem{lcr96}
P. W. Leung, K. C. Chiu and K. J. Runge,
Phys. Rev. B {\bf 54}, 12938 (1996).



\bibitem{fs63}
M. E. Fisher and J. Stephenson Phys. Rev. {\bf 132}, 1411 (1963).

\bibitem{ms03}
R. Moessner and  S.~L. Sondhi,
Phys. Rev. B {\bf 68}, 184512 (2003).
See  Appendix B concerning the square-lattice QDM.


\bibitem{msc01}
R.~Moessner, S.~L.~Sondhi  and P.~Chandra,
Phys.  Rev. B {\bf 64}, 144416 (2001).


\bibitem{hkms03}
D.~A.~Huse,  W.~Krauth, R.~Moessner and S.~L.~Sondhi,
Phys. Rev. Lett. 91, 167004 (2003)
and references therein.

\bibitem{height_representation}

L.~S.~Levitov, Phys. Rev. Lett. {\bf 64}, 92 (1990),

C.~L.~Henley, J. Stat. Phys. {\bf 89}, 483 (1997).


\bibitem{fms02}
P. Fendley, R.~Moessner and S.~L.~Sondhi,
Phys. Rev. B {\bf 66}, 214513 (2002).


\bibitem{iif02}
A.  Ioselevich, D.~A.~Ivanov  and M.~V.~Feigelman,
Phys. Rev.  B {\bf 66}, 174405 (2002).

\bibitem{ifiitb02}
L.~B.~Ioffe, M.~V.~Feigel'man, A.~Ioselevich, D.~Ivanov,  M.~Troyer,
G.~Blatter, Nature {\bf 415}, 503 (2002).

\bibitem{mlms02}
G.~Misguich,   C.~Lhuillier,   M.~Mambrini,   P.~Sindzingre,
Eur. Phys. J. B {\bf 26}, 167 (2002).



\bibitem{wen91}
X. G. Wen, Phys. Rev. B {\bf 44}, 2664 (1991).


\bibitem{k89}
S. Kivelson, Phys. Rev. B {\bf 39}, 259 (1989).

\bibitem{rc89}
N. Read and B. Chakraborty, Phys. Rev. B {\bf 40}, 7133 (1989).

\bibitem{sf00}
T. Senthil and M.~P.~A.~Fisher,
Phys. Rev. B {\bf 62}, 7850 (2000).

\bibitem{sf01}
T. Senthil and M. P. A. Fisher,
Phys. Rev. Lett. {\bf 86}, 292 (2001);
Phys. Rev. B {\bf 63}, 134521 (2001).

\bibitem{feynman52-53}
R.~P.~Feynman, Phys. Rev. {\bf 90}, 1116 (1952),
Phys. Rev. {\bf 91}, 1291 (1953).

\bibitem{ez93}
V. Elser and C. Zeng.
Phys. Rev. B {\bf 48}, 13647 (1993).

\bibitem{hw88}
A.~J.~Phares and   F.~J.~Wunderlich,
Nuovo Cimento   B {\bf  101}, 653 (1988).
\bibitem{msp03b}
See {\S}V.E.6 of Ref.\cite{msp03}

\bibitem{msp03c}  This  follows from  the  independence  of the  arrow
variables, see {\S}V.B of Ref.\cite{msp03}


\bibitem{polyakov87}
A.~M.~Polyakov, {\it Gauge Fields and Strings}, (Harwood Academic, New York, 1987).



\bibitem{kogut79}
J. B. Kogut, Rev. Mod. Phys. {\bf 51}, 659 (1979).

\bibitem{msp03} G.~Misguich,  D.~Serban and V.~Pasquier,
Phys. Rev. B {\bf 67}, 214413 (2003).

\bibitem{ze95}
C.~Zeng    and   V.~Elser.
Phys.  Rev. B {\bf 51}, 8318 (1995).

\bibitem{thouless65}
D. J. Thouless, Proc. Phys. Soc. London {\bf 86}, 893 (1965).

\bibitem{rhd83}  M. Roger,  J.  H. Hetherington,  and  J. M.  Delrieu,
Rev. Mod. Phys. {\bf 55}, 1 (1983).

\bibitem{cf85} M. C. Cross and D.  S. Fisher,
Rev. Mod. Phys. {\bf 57}, 881 (1985).
\bibitem{greywall}   D.~S.~Greywall   and   P.~A.~Busch,
Phys.  Rev.  Lett.  {\bf 62}, 1868 (1989).
D.~S.~Greywall,  Phys. Rev. B  {\bf 41}, 1842  (1990).
\bibitem{gr95}
H. Godfrin and R.~E.~Rapp, Adv. Phys. {\bf 44}, 113 (1995).

\bibitem{roger84}
M.~Roger, Phys. Rev. B {\bf 30}, 6432 (1984).

\bibitem{rbbcg98} M.  Roger, C. B\"auerle,  Yu.~M.~Bunkov, A.-S.~Chen, and
H.~Godfrin, Phys. Rev.  Lett. {\bf 80}, 1308 (1998).



\bibitem{cj87}
D.~M.~Ceperley and G.  Jacucci,    Phys. Rev.  Lett. {\bf  58},   1648 (1987).


\bibitem{bcl92}  B.  Bernu,  D.  Ceperley,  and  C.  Lhuillier,  J.  Low
Temp. Phys. {\bf 89}, 589 (1992).


\bibitem{ceperley95}
D.~M.~Ceperley, Rev. Mod. Phys. {\bf 67}, 279 (1995).

\bibitem{bc99} B.  Bernu and D.  Ceperley, in {\it Quantum Monte Carlo
Methods in  Physics and Chemistry},  edited by M. P.   Nightingale and
C. J. Umrigar (Kluwer, Dordrecht, The Netherlands, 1999).

\bibitem{ah00}
H.~Ashizawa and D.~S.~Hirashima, Phys. Rev. B {\bf 62}, 9413 (2000).

\bibitem{drh80}  J.~M.~Delrieu, M.~Roger,  J.~H.~Hetherington,  J. Low
Temp.  Phys.  {\bf 40}, 71 (1980).


\bibitem{frg86}   H.~Franco,   R.~E.~Rapp   and   H.~Godfrin,
Phys.  Rev.  Lett.  {\bf 57}, 1161 (1986).


\bibitem{mblw98}   G.~Misguich,   B.~Bernu,  C.~Lhuillier,   and
C.~Waldtmann, Phys. Rev. Lett. {\bf 81}, 1098 (1998).

\bibitem{lmsl00}
W.~LiMing,  G.~Misguich, P.~Sindzingre, and  C.~Lhuillier,
Phys. Rev. B {\bf 62}, 6372 (2000).



\bibitem{kmyf97}  K.~Ishida,  M.~Morishita, K.~Yawata,  and  H.~Fukuyama,
Phys.  Rev.  Lett. {\bf 79}, 3451 (1997).


\bibitem{cthrbbg01}
E.~Collin {\it et al.}, Phys. Rev. Lett. {\bf 86}, 2447 (2001).


\bibitem{kh00}
M. Katano and D.~S.~Hirashima, Phys. Rev. B {\bf 62}, 2573 (2000).

\bibitem{hk01}          D.~S.~Hirashima          and          K.~Kubo,
Phys. Rev. B {\bf 63}, 125340 (2001).

\bibitem{bcc01}   B.   Bernu,  L.   Candido,   and   D.  M.   Ceperley
Phys.  Rev. Lett.  {\bf 86}, 870  (2001).


\bibitem{ok98} T. Okamoto  and S. Kawaji,
Phys. Rev.  B {\bf 57}, 9097 (1998).


\bibitem{rd89} M. Roger and J. M. Delrieu,
Phys. Rev. B {\bf 39}, 2299 (1989).


\bibitem{sugai90}
S.~Sugai {\it et al.},
Phys.  Rev. B {\bf 42},  1045 (1990).


\bibitem{coldea01}
R.~Coldea  {\it  et   al.},
Phys.  Rev.  Lett.  {\bf 86}, 5377 (2001).


\bibitem{mr02}
E.~M\"uller-Hartmann and A.~Reischl,
Eur.  Phys.  J. B {\bf 28}, 173 (2002).


\bibitem{kk03} A.~A.~Katanin, and A.~P.~Kampf,
Phys. Rev. B {\bf 67}, 100404R (2003).


\bibitem{mkebm00}

M.   Matsuda {\it et al.},
Phys. Rev. B {\bf 62}, 8903 (2000).

\bibitem{bmmnu99}
S.~Brehmer {\it et al.},
Phys. Rev. B {\bf 60}, 329 (1999).



\bibitem{sku01}  K.   P.  Schmidt,  C.   Knetter  and  G.   S.  Uhrig,
Europhys. Lett. {\bf 56}, 877 (2001).


\bibitem{gkt03}
A.~G\"oßling {\it et al.},
Phys.  Rev. B {\bf 67}, 052403 (2003).

\bibitem{mvm02}
M. M\"uller, T. Vekua, and H.-J. Mikeska,
Phys. Rev. B {\bf 66}, 134423 (2002).

\bibitem{hmh03}
T. Hikihara, T. Momoi, and X. Hu,
Phys. Rev. Lett. {\bf 90}, 087204 (2003).


\bibitem{lst03} A. L\"auchli,  G. Schmid, and M. Troyer,
Phys. Rev. B {\bf 67}, 100409R (2003).

\bibitem{honda01}
Y.  Honda and T.  Horiguchi, cond-mat/0106426.


\bibitem{hn02}
K. Hijii and K.  Nomura Phys.  Rev. B {\bf 65}, 104413 (2002).


\bibitem{mhnh03}  T. Momoi,  T.~Hikihara,  M.~Nakamura,
Xiao      Hu,   Phys. Rev. B {\bf 67}, 174410 (2003).


\bibitem{lauchli03}
A.  L\"auchli,   talk given at  the   {\em Highly Frustrated Magnetism
2003} conference, Grenoble, France (August 2003).


\bibitem{km97}  K. Kubo  and T.  Momoi, Z.  Phys. B  {\bf 103},  485 (1997).

\bibitem{mkn97}
T.  Momoi, K.   Kubo, and K.  Niki,
Phys.  Rev.  Lett. {\bf 79}, 2081 (1997).


\bibitem{mbl98}   G.~Misguich,   B.~Bernu,   and   C.~Lhuillier,
J.   Low  Temp. Phys.  {\bf 110},  327 (1998).


\bibitem{ksmn98}  K.  Kubo,  H.  Sakamoto,  T.~Momoi,  and  K.~Niki,
J.                 Low Temp. Phys.  {\bf 111}, 583 (1998).



\bibitem{StagVBC}
A 16- and  a 32-site triangular  lattices which  do not frustrate  the
staggered VBC were    investigated.    In both  cases   some  of   the
irreducible representations of the  space group required to break  the
appropriate     lattice    symmetries    are   very   high    in   the
spectrum. C. Lhuillier and G. Misguich (unpublished).

\bibitem{lsm61}
E.~H.~Lieb, T.~D.~Schultz,  D.~C.~Mattis., Ann. Phys. (N.Y)  {\bf 16},
407 (1961).

\bibitem{al86}
I. Affleck and E. Lieb, Lett. Math. Phys. {\bf 12},  57  (1986).

\bibitem{o00}
M. Oshikawa, Phys. Rev. Lett. {\bf 84},  1535  (2000).


\bibitem{o03}
M.~Oshikawa,
Phys. Rev. Lett. {\bf 90}, 236401 (2003).

\bibitem{nge98}
A.~A. Nersesyan, A.~O. Gogolin, and F.~H.~L. Essler,
Phys. Rev. Lett. {\bf 81}, 910  (1998).

\bibitem{ahllt98}
P. Azaria {\it et ~al.}, Phys. Rev. Lett. {\bf 81},  1694  (1998).

\bibitem{efkl00}
V.~J.~Emery, E. Fradkin, S.~A.~Kivelson and T.~C.~Lubensky,
Phys. Rev. Lett. {\bf 85}, 2160 (2000).

\bibitem{be01}
M.  Bocquet, F. Essler, A.~M.~Tsvelik and A.~O.~Gogolin,
Phys. Rev. B {\bf 64},094425 (2001),
M.~Bocquet, Phys. Rev. B  {\bf 65}, 1884415 (2001).


\bibitem{vc01}
A. Vishwanath and D. Carpentier, Phys. Rev. Lett. {\bf 86},  676 (2001).

\bibitem{sp02}
S. Sachdev and K. Park, Annals of Physics (N.Y.), {\bf 58}, 298 (2002).

\bibitem{ssl02}
O.~A. Starykh and R.~R.~P. Singh and  G.~C. Levine,
Phys. Rev. Lett.  {\bf 88}, 167203  (2002).

\bibitem{cttt01}
R. Coldea, D.~A. Tennant, A.~M. Tsvelik, and Z. Tylczynski,
Phys. Rev. Lett.  {\bf 86}, 1335 (2001).

\bibitem{kitaev97}
A.  Kitaev, Annals Phys. {\bf 303}, 2 (2003).
[quant-ph/9707021]

\bibitem{ns01}  C. Nayak  and  K.  Shtengel,
Phys.  Rev.  B {\bf 64}, 064422 (2001).


\bibitem{bfg02}
L.  Balents, M.  P.  A. Fisher, and S.  M.  Girvin,
Phys.   Rev. B  {\bf   65},  224412  (2002).


\bibitem{pbf02}  A.~Paramekanti, L.~Balents, and  M.~P.~A.~Fisher
Phys.  Rev.  B  {\bf 66}, 054526  (2002).


\bibitem{sdss02}  A.~W.~Sandvik,  S.~Daul,  R.~R.~P.~Singh,  and
 D.~J.~Scalapino, Phys. Rev. Lett. {\bf 89}, 247201 (2002).


\bibitem{sm02} T.  Senthil and O.  Motrunich,
Phys. Rev. B  {\bf 66}, 205104 (2002).

\bibitem{p38}
L. Pauling,  in {\em The nature of the chemical bond} (Cornell University
  Press, Ithaca, 1938).

\bibitem{kn53}
K. Kano and S. Naya, Prog. in Theor. Phys. {\bf 10},  158  (1953).

\bibitem{hr92}
D. Huse and A. Rutenberg, Phys. Rev. B {\bf 45},  7536  (1992).



\bibitem{ms01b}
R. Moessner and S.~L. Sondhi, Phys. Rev. B {\bf 63},  224401  (2001).

\bibitem{chs92}
J. Chalker, P.~C.~W.~Holdsworth, and E.~F.~Shender,
Phys. Rev. Lett. {\bf 68}, 855 (1992).

\bibitem{baxter70}
R. J. Baxter, J. Math. Phys. {\bf 11}, 784 (1970).

\bibitem{rcc93}
I. Richtey, P.~Chandra, and P.~Coleman, Phys. Rev. B {\bf 47},  15342  (1993).

\bibitem{rb93}
J. Reimers and A.~Berlinsky, Phys. Rev. B {\bf 48},  9539  (1993).

\bibitem{e02}
M.  Elhajal,           Ph.~D.     thesis,     Universit\'e      Joseph
Fourier. Grenoble. France, 2002.

\bibitem{ecl02}
M.~Elhajal, B.~Canals, and C.~Lacroix,
Phys. Rev. B {\bf 66}, 014422 (2002).

\bibitem{k94}
A. Keren, Phys. Rev. Lett. {\bf 72},  3254  (1994).

\bibitem{e89}
V. Elser, Phys. Rev. Lett. {\bf 62},  2405  (1989).

\bibitem{ce92}
J. Chalker and J. Eastmond, Phys. Rev. B {\bf 46},  14201  (1992).

\bibitem{s92}
S. Sachdev, Phys. Rev. B {\bf 45},  12377  (1992).

\bibitem{le93}
P. Leung and V. Elser, Phys. Rev. B {\bf 47},  5459  (1993).


\bibitem{lblps97}
P. Lecheminant {\it et~al.}, Phys. Rev. B {\bf 56},  2521  (1997).

\bibitem{web98}
C. Waldtmann {\it et~al.}, Eur. Phys. J. B {\bf 2},  501  (1998).

\bibitem{m98}
F. Mila, Phys. Rev. Lett. {\bf 81},  2356  (1998).

\bibitem{smlbpwe00}
P. Sindzingre {\it et~al.}, Phys. Rev. Lett. {\bf 84},  2953  (2000).

\bibitem{LauchliKITP2012}
A. M.  L\"auchli, oral communication at the
KITP Program: ``Frustrated Magnetism and Quantum Spin Liquids:
From Theory and Models to Experiments'' (Aug 13 - Nov 9, 2012).
Available online at:
\href{http://online.kitp.ucsb.edu/online/fragnets12/laeuchli}{\tt http://online.kitp.ucsb.edu/online/fragnets12/laeuchli}. Unpublished.

\bibitem{mm01}
M. Mambrini and F. Mila, Eur. Phys. J. B {\bf 17},  651  (2001).

\bibitem{dmnm03}
S.~Dommange, M.~Mambrini, B.~Normand and F.~Mila,
Phys. Rev. B {\bf 68}, 224416 (2003).

\bibitem{ze90}
C. Zeng and V. Elser, Phys. Rev. B {\bf 42},  8436  (1990).

\bibitem{sh92}
R. Singh and D. Huse, Phys. Rev. Lett. {\bf 68},  1766  (1992).

\bibitem{h01}
K. Hida, J. Phys. Soc. Jpn. {\bf 70},  3673  (2001).

\bibitem{cghp02}
D.~C.~Cabra, M.~D.~Grynberg, P.~C.~W.~Holdsworth,  P.~Pujol,
Phys. Rev. B {\bf 65}, 094418 (2002).

\bibitem{ey94}
N. Eltsner and A.~P. Young, Phys. Rev. B {\bf 50},  6871  (1994).

\bibitem{nm95}
T. Nakamura and S. Miyashita, Phys. Rev. B {\bf 52},  9174  (1995).

\bibitem{tr96}
P.~Tomczak and J.~Richter, Phys. Rev. B {\bf 54}, 9004 (1996).

\bibitem{rhw00}
A.~P. Ramirez, B. Hessen, and M.~Winkelmann, Phys. Rev. Lett. {\bf 84},  2957 (2000).

\bibitem{mklmch00}
P. Mendels {\it et~al.}, Phys. Rev. Lett. {\bf 85},  3496  (2000).

\bibitem{sma02}
A.~V.~Syromyatnikov and S.~V.~Maleyev,
Phys. Rev. B {\bf 66}, 132408 (2002).

\bibitem{ns03}
P. Nikolic and T. Senthil,
Phys. Rev. B {\bf 68}, 214415 (2003).



\bibitem{sh2007}
R.~R.~P. Singh and D.~A. Huse, Phys. Rev. B {\bf 76}, 180407(R) (2007).


\bibitem{singh2008}
R.~R.~P. Singh and D.~A. Huse, Phys. Rev. B {\bf 77}, 144415 (2008).


\bibitem{mmpfa00}
C. Mondelli {\it et~al.}, Physica B {\bf 284},  1371  (2000).

\bibitem{gsf01}
A. Georges, R. Siddhartan and S. Florens, Phys. Rev. Lett. {\bf 87}, 277203 (2001).

\bibitem{ls02}
C. Lhuillier and P. Sindzingre,  in {\em Quantum properties of Low dimensional
  antiferromagnets}, edited by Y.~Ajiro and J.~P.~Boucher (Kyushu University
  Press, Fukuoka, Japan, 2002), p.\ 111.

\bibitem{sv00}
S.~Sachdev and M.~Vojta,
Proceedings of the XIII International Congress on Mathematical Physics, July 2000, London.
A.~Fokas, A.~Grigoryan, T.~Kibble, and B.~Zegarlinski eds, International Press, Boston (2001)
[cond-mat/0009202].

\bibitem{ukkll94}
Y. Uemura {\it et~al.}, Phys. Rev. Lett. {\bf 73},  3306  (1994).



\bibitem{lbar96}
S.-H. Lee {\it et~al.}, Europhys. Lett. {\bf 35},  127  (1996).

\bibitem{kklllwutdg96}
A. Keren {\it et~al.}, Phys. Rev. B {\bf 53},  6451  (1996).

\bibitem{whmmt98}
A.~S. Wills {\it et~al.}, Europhys. Lett. {\bf 42},  325  (1998).

\bibitem{aoyhimw94}
K. Awaga {\it et~al.}, Phys. Rev. B {\bf 49},  3975  (1994).

\bibitem{wkyoya97}
N. Wada {\it et~al.}, J. Phys. Soc. Jpn. {\bf 66},  961  (1997).

\bibitem{wwyoaonn98}
I. Watanabe {\it et~al.}, Phys. Rev. B {\bf 58},  2438  (1998).

\bibitem{cl98}
B. Canals and C. Lacroix, Phys. Rev. Lett. {\bf 80},  2933  (1998).

\bibitem{hbb91}
A.~B.~Harris, A.~J.~Berlinsky and C.~Bruder,
J. Appl. Phys. {\bf 69}, 5200 (1991).

\bibitem{HarrisKallinBerlinsky92}
A.B. Harris, C. Kallin and A.J. Berlinsky, Phys. Rev. B {\bf 45},  2899  (1992).

\bibitem{t01}
H. Tsunetsugu, J. Phys. Soc. Jpn. {\bf 70},  640  (2001).

\bibitem{t02}
H. Tsunetsugu, Phys. Rev. B {\bf 65},  024415  (2002).

\bibitem{vdh92-93}
J.~V. Delft and C.~L. Henley, Phys. Rev. Lett. {\bf 69},  3236  (1992),
Phys. Rev. B {\bf 48},  965  (1993).

\bibitem{sachdev03}
S.~Sachdev, Annals Phys. {\bf 303}, 226 (2003) [cond-mat/0211027]
and  Rev. Mod. Phys. {\bf 75}, 913 (2003).

\bibitem{wwz89}
X. Wen, F. Wilczek, and A. Zee, Phys. Rev. B {\bf 39},  11413  (1989).

\bibitem{kl87-89}
V. Kalmeyer and R. Laughlin, Phys. Rev. Lett. {\bf 59},  2095  (1987),
Phys. Rev. B {\bf 39},  11879  (1989).

\bibitem{ywg93}
K. Yang, L. Warman, and S.~M.~Girvin, Phys. Rev. Lett. {\bf 70},  2641  (1993).

\bibitem{Janson_2008}
O.~Janson, J.~Richter, and   H.~Rosner, Phys. Rev. Lett. {\bf101},  106403 (2008).

\bibitem{Messioregular2011}
L.~Messio,   C.~Lhuillier,   and G.~Misguich, Phys. Rev. B {\bf 83}, 184401 (2011).

\bibitem{zhitomirsky2008}
M.~E. Zhitomirsky, Phys. Rev. B {\bf 78}, 094423 (2008).

\bibitem{Henley2009}
  C. L. Henley, Phys. Rev. B {\bf 80}, 180401 (2009).


\bibitem{Sindzingre2009}
P.~Sindzingre and   C.~Lhuillier,  EPL {\bf 88} 27009   (2009).



\bibitem{laeuchli2011}
A.~M. L\"auchli,  J.~Sudan, and  E.~S. S\o{}rensen,  Phys. Rev. B {\bf 83},  212401 (2011).


\bibitem{jiangsheng2008}
H.~C. Jiang,  Z.~Y. Weng, and  D.~N. Sheng,  Phys. Rev. Lett. {\bf 101},   117203  (2008).


\bibitem{Yan2011}
S.~Yan,   D.~A. Huse, and   S.~R. White,  Science Magazine {\bf 332},  1173 (2011).

\bibitem{Jiangwangbalents2012}
H.-C. Jiang,  Z.~Wang, and  L.~Balents,
Nature Physics {\bf 8}, 902 (2012).

\bibitem{DepenbrockSchollwock2012}
S.~Depenbrock,  I.~P. McCulloch,  and  U.~Schollwoeck, Phys. Rev. Lett. 109, 067201 (2012).
\bibitem{kl89}
V. Kalmeyer and R. B. Laughlin, Phys. Rev. B {\bf 39} 11879 (1989).


\bibitem{wen2010}
 J.  Wen,  A. R\"uegg,  C.-C.  Wang, G. A Fiete, Phys. Rev. B {\bf 82}, 075125,
  (2010).

\bibitem{Chu2011}
V. Chua, Victor H.  Yao and G.A. Fiete, Phys. Rev. B {\bf 83}, 180412 (2011).

\bibitem{Messiocuboc12012}
L.~Messio,   B.~Bernu, and   C.~Lhuillier,  Phys. Rev. Lett. {\bf 108},  207204 (2012).


\bibitem{Poilblanc2010}
D.~Poilblanc,  M.~Mambrini, and  D.~Schwandt,
  Phys. Rev. B {\bf 81},  180402 (2010).

\bibitem{Evenbly2010}
G.~Evenbly and   G.~Vidal,  Phys. Rev. Lett. {\bf 104},  187203 (2010).

\bibitem{Schwandt2010}
D.~Schwandt,   M.~Mambrini, and   D.~Poilblanc,  Phys. Rev. B {\bf 81},  214413 (2010).

\bibitem{Iqbal2011}
Y.~Iqbal,  F.~Becca, and  D.~Poilblanc,  Phys. Rev. B {\bf 83},  100404 (2011).

\bibitem{PoilblancMisguich2011}
D.~Poilblanc and  G.~Misguich,  Phys. Rev. B {\bf 84},  214401 (2011).

\bibitem{ran2007}
Y. Ran,  M. Hermele,   P.~A.Lee and   X.-G. Wen, Phys. Rev. Lett. {\bf 98}, 117205  (2007).

\bibitem{hermele2005}
M. Hermele,   T. Senthil and  M.~P.~A. Fisher, Phys. Rev. B {\bf 72},  104404 (2005).

\bibitem{hermele2008}
M.~Hermele,  Y.~Ran,  P.~A.~Lee and   X.-G.~Wen, Phys. Rev. B {\bf 77}, 224413  (2008).



\bibitem{wangVishwanath2006}
F. Wang and A. Vishwanath, Phys. Rev. B {\bf 74}, 174423  (2006).

\bibitem{SFMFT_kagome_PSG}
Y.-M. Lu,  Y.Ran and  P.~A.Lee, Phys. Rev. B {\bf 83}, 224413  (2011).

\bibitem{ms07}
G. Misguich and P. Sindzingre,
 J. Phys. Cond. Matt., {\bf 19}, 145202   (2007) and Errata: condmat/0607764v3.

\bibitem{Wen_PSG}
 X.-G. Wen, Phys. Rev. B {\bf 65}, 165113  (2002).

\bibitem{Kolee2010}
 W.-H. Ko,  Z.-X. Liu,   T-K Ng  and  P. A.Lee , Phys. Rev. B {\bf 81},  024414  (2010).

\bibitem{Kolee2011} W.-H. Ko  and  P. A.Lee , Phys. Rev. B {\bf 84},  125102  (2011).



 \bibitem{messioDM2010}
  L. Messio, O. C\'epas  and Lhuillier, C., Phys. Rev. B {\bf 81}, 064428, (2010).

\bibitem{luranSLhoneycomb2011}
Y.-M. Lu and Y. Ran, Phys. Rev. B {\bf 84},  024420  (2011).

\bibitem{hh01}  Z. Hiroi, M. Hanawa, N. Kobayashi, M. Nohara, H. Takagi, Y. Kato and M.~Takigawa,
  J. Phys. Soc. Japan, {\bf 70},  3377  (2001).

\bibitem{magnetization_plateaux_2011}
 Y. Okamoto,  M. Tokunaga, H. Yoshida, A. Matsuo, K. Kindo and Z. Hiroi,
 Phys. Rev. B {\bf 83}, 180407    (2011).


\bibitem{Nilsenvolbortite2011}
  G. J. Nilsen,  F. C.  Coomer,  M. A. de Vries, J. R. Stewart, P. P. Deen,  A. Harrison and H. M. R\o{}nnow,  Phys. Rev. B  {\bf 84},   172401, (2011).

  \bibitem{yavors'kii}
T. Yavors'kii, W. Apel and H.-U. Everts, Phys. Rev. B  {\bf 76}, 064430 (2007).

\bibitem{wang07}
Fa Wang,  A. Vishwanath and Y.B. Kim, Phys. Rev. B  {\bf 76}, 094421 (2007).

\bibitem{Huhvisonskagome2011}
Y. Huh M.  Punk,  and S. Sachdev, Phys. Rev. B {\bf 84}, 094419 (2011).

\bibitem{Tay_kagome}
 T. Tay and  O.-I. Motrunich,Phys. Rev. B {\bf 84}, 020404 (2011).

\bibitem{Iqbalb}
 Y. Iqbal, F.  Becca, D. Poilblanc,   Phys. Rev. B {\bf 84}, 020407 (2011).


\bibitem{ryu2007}
S. Ryu, O. I. Motrunich, J. Alicea and M. P. A. Fisher,
Phys. Rev. B {\bf 75}, 184406 (2007).



\bibitem{HanLee_synthese_2011}
T. H. Han, J. S. Helton,   S. Chu, A. Prodi,  D. K. Singh, C. Mazzoli, P.  M\"uller, D. G. Nocera and Y. S. Lee, Phys. Rev. B {\bf 83}, 100402 (2011).


\bibitem{Helton2010}
J.S. Helton, K. Matan, M. P. Shores, E.A. Nytko, B.M. Bartlett,Y. Qiu, D. G. Nocera Y.S. Lee, Phys. Rev. Lett. {\bf 104}, 147201 (2010).


\bibitem{Gotzekagome2011}
O. G\"otze, D. J. J. Farnell, R.F.  Bishop, P.H.  Li and  J. Richter, Phys. Rev. B {\bf 84}, 224428 (2011).


\bibitem{cepas2011}
O. C\'epas and A.  Ralko, Phys. Rev. B {\bf 84}, 020413 (2011).

 \bibitem{Misguich2005a}
G. Misguich and B. Bernu, Phys. Rev. B {\bf 71}, 014417 (2005).
	
\bibitem{Misguich2007}
G. Misguich and P. Sindzingre, Eur. Phys. J. B {\bf 59}, (2007).


\bibitem{rigol2007}
M. Rigol and R. R. P. Singh, Phys. Rev. Lett. {\bf 98}, 207204 (2007).


\bibitem{rigol2007b}
M. Rigol and R. R. P. Singh, Phys. Rev. B {\bf 76}, 184403 (2007).

\bibitem{chitra08}
R. Chitra and M. J. Rozenberg, Phys. Rev.B {\bf 77}, 052407 (2008).


\bibitem{rousochatzakis09}
I. Rousochatzakis, S. Manmana, A.M. L\"{a}uchli, B. Normand and F. Mila,
Phys. Rev. B {\bf 79} 214415 (2009).

\bibitem{laeuchli2009}
A. Laeuchli  and C. Lhuillier, arXiv:09011065v1.


\bibitem{Tchernyshyov2010}
Z. Hao and O. Tchernyshyov, Phys. Rev. B {\bf 81}, 214445 (2010).

\bibitem{bert2007}
F. Bert, S. Nakamae, F. Ladieu,D. L'H\^{o}te, P. Bonville, F. Duc, J.-C. Trombe and P. Mendels,
Phys. Rev. B {\bf 76}, 132411 (2007).



\bibitem{Kermarrec2011}
E. Kermarrec, P. Mendels, F. Bert R.H. Colman, A.S.  Wills, P. Strobel, P.  Bonville, A. Hillier and  A. Amato, Phys. Rev. B {\bf 84}, 100401 (2011)


\bibitem{jeong2011}
M. Jeong, F.  Bert, P.  Mendels, F. Duc, J. C. Trombe,  M. A. de Vries,  and A. Harrison,
Phys. Rev. Lett. {\bf 107}, 237201 (2011).


\bibitem{narumi2004}
 Y. Narumi, K. Katsumata,Z. Honda,J.-C. Domenge, P.	Sindzingre, C. Lhuillier, Y. Shimaoka, T. C. Kobayashi and 	K. Kindo,
Europhys. Lett. , {\bf 65} 705 (2004).

\bibitem{domenge2008}
J.-C. Domenge, C. Lhuillier, L. Messio, L. Pierre and P. Viot,
Phys. Rev. B {\bf 77}, 172413 (2008).

\bibitem{Shores2005}
M. P. Shores,  E.A. Nytko, B.M.  Bartlett and  D. G. Nocera, Journal of the American Chemical Society \textbf{127}, 13462 (2005).

\bibitem{mendels2007}
P. Mendels,F. Bert, M. A. de Vries, A. Olariu, A. Harrison, F. Duc, J. C. Trombe, J. S. Lord,  A. Amato and C. Baines,
Phys. Rev. Lett. \textbf{98}, 077204 (2007).

\bibitem{helton07}
J. S. Helton, K. Matan, M. P. Shores, E. A. Nytko,
B. M. Bartlett, Y. Yoshida, Y. Takano, A. Suslov, Y. Qiu, J.-H. Chung, D. G. Nocera and Y. S. Lee,
Phys. Rev. Lett. \textbf{98} 107204 (2007).

\bibitem{olariu2008}
A. Olariu, P. Mendels, F. Bert, F. Duc,J. C. Trombe, M. A. de Vries and A. Harrison,
Phys. Rev. Lett. \textbf{100} 087202 (2008)


\bibitem{imai2008}
T. Imai, E. A. Nytko, B. M. Bartlett, M. P. Shores and D. G. Nocera,
Phys. Rev. Lett. \textbf{100} 077203 (2008).

\bibitem{jeongkagome2011}
M. Jeong, F. Bert, P. Mendels, F. Duc, J. C.  Trombe, M. A. de Vries
 and Harrison, A.
  Phys.Rev. Lett. \textbf{ 107}, 237201 (2011).



\bibitem{Zorko2008}
A. Zorko, S. Nellutla, J. van Tol, L.C. Brunel, F. Bert, F.  Duc, J. C. Trombe, M.A.  de Vries, A. Harrison and  P. Mendels,
  Phys. Rev. Lett.\textbf{ 101}, 026405 (2008).

\bibitem{cepas2008}
O. Cépas and C. M. Fong and P. W. Leung and C. Lhuillier,
Phys. Rev. B.\textbf{ 78}, 140405 (2008).


\bibitem{devries2012}
M. A. de Vries,  D. Wulferding, P. Lemmens, J.S. Lord, A.Harrison, P. Bonville, F. Bert,  and  P. Mendels,
  Phys. Rev. B. \textbf{85}, 014422 (2012).

\bibitem{devries2009}
M.A. de Vries, J. R. Stewart,  P. P. Deen,  J. O. Piatek, G. J. Nilsen,H. M. R\o{}nnow,  A. Harrison,
  Phys. Rev. Lett. \textbf{103}, 237201 (2009).

\bibitem{shawish2010}
S. El Shawish, S. O. C\'epas, O. and S.  Miyashita,
Phys. Rev. B. \textbf{81}, 224421 (2010).

\bibitem{kageyama2002}
H.~Kageyama, T. Nakajima, M. Ichihara, F. Sakai and Y. Ueda, in {\em Spin frustration in two-dimensional compounds} edited by Y.~Ajiro and J.~P.~Boucher (Kyushu University
  Press, Fukuoka, Japan, 2002), p.\ 135.


\bibitem{karaki2010}
Y. Karaki, Y. M. Kou, A. Yamaguchi, M. Kubota, H. Ishimoto, Z. Honda,  and  K. Yamada,
J.  of Low Temp.  Phys. , \textbf{158}, 653 (2010).


\bibitem{Domenge2005}
J.-C. Domenge, P. Sindzingre, C. Lhuillier and L. Pierre,
Phys. Rev. B \textbf{72}, 024433 (2005).

\bibitem{kapella1}
R. H. Colman, C. Ritter, and A. S. Wills, Chem. Mater. \textbf{20}, 6897 (2008).
\bibitem{kapella2} R. H. Colman, A. Sinclair, and A. S.Wills, Chem. Mater.
{\bf 22}, 5774 (2010).

\bibitem{fakkapellasite2012}
B. Fak, E. Kermarrec, L. Messio,B. Bernu, B. C. Lhuillier, F.  Bert, P.  Mendels, B. Koteswararao, F. Bouquet, J.  Ollivier, A. D.  Hillier, A. Amato, R.H.  Colman, A. S. Wills,
Phys. Rev. Lett. \textbf{109}, 037208 (2012).



\bibitem{Hastings04}
M. B Hastings,
Phys. Rev. B \textbf{69}, 104431 (2004).

\bibitem{messio2008}
L. Messio, J.-C. Domenge, C. Lhuillier, L. Pierre, P. Viot and G. Misguich,
Phys. Rev. B \textbf{78}, 054435 (2008).

\bibitem{hiz05}
A. Hamma, R. Ionicioiu, and P. Zanardi, Phys. Lett. A {\bf 337}, 22 (2005).

\bibitem{kp06}
A. Kitaev and J. Preskill, Phys. Rev. Lett. {\bf 96}, 110404 (2006).
\bibitem{lw06}
M. Levin M and X.-G. Wen, Phys. Rev. Lett. {\bf 96}, 110405 (2006).

\bibitem{fm07}
S. Furukawa and G. Misguich,
Phys. Rev. B {\bf 75}, 214407 (2007).

\bibitem{ihm11}
S. V. Isakov, M. B. Hastings, R. G. Melko, Nature Physics {\bf 7}, 772 (2011).

\bibitem{zgv11}
Yi Zhang, T. Grover, and A. Vishwanath
Phys. Rev. B {\bf 84}, 075128 (2011).

\bibitem{lbh10}
A. M. Läuchli, E. J. Bergholtz, M. Haque,
New Journal of Physics {\bf 12}, 075004 (2010).


\bibitem{smp12}
J.-M. Stéphan, G. Misguich, V. Pasquier, J. Stat. Mech. P02003 (2012).


\bibitem{jyb12}
H.C. Jiang, H. Yao, and L. Balents,
Phys. Rev. B {\bf 86}, 024424 (2012).

\bibitem{zgtov12}
Yi Zhang, T. Grover, A. Turner, M. Oshikawa, A. Vishwanath,
Phys. Rev. B {\bf 85}, 235151 (2012).

\bibitem{Cepas2011}
O. Cépas and A. Ralko,
Phys. Rev. B, \textbf{84}, 020413 (2011).

\bibitem{Chern2012}
G.-W. Chern and R. Moessner,
arXiv:1207.4752.

\bibitem{messioPSG2012}
L. Messio, C. Lhuillier and G. Misguich, in preparation.
%
\bibitem{Colman2011}
R. H. Colman, F. Bert, D. Boldrin, A. D. Hillier,   P.	Manuel, P. Mendels and Wills, A. S.,
Phys. Rev. B \textbf{83}, 180416 (2011).

 \bibitem{Quilliam2011}
J. A. Quilliam, F. Bert, F. , R. H.  Colman, D. Boldrin,  A. S.	Wills and  P. Mendels,
Phys. Rev. B \textbf{84}, 180401 (2011).

 \bibitem{Okamoto2009}
 Y. Okamoto, H. Yoshida, and Z. Hiroi, J. Phys. Soc. Jpn. 78, 033701 (2009).


\bibitem{bernu2012}
B. Bernu, E. Kermarrec, L. Messio, C. Lhuillier, F. Bert, and P. Mendels,  arXiv:1210.2549.

\bibitem{Coldea2001a}
R. Coldea, D. A. Tennant, A. M.  Tsvelik and  Z. Tylczynski,
Phys. Rev. Lett.  \textbf{86}, 1335 (2001).
\bibitem{Jiang2012}
 Hong-Chen Jiang, Hong Yao  and Leon Balents,
Phys. Rev. B \textbf{86}, 024424 (2012).
\bibitem{Haldane1988b}
Haldane, F. D. M.
Phys. Rev. Lett. \textbf{61}, 2015 (1988).
\bibitem{Kane2005a}
C. L. Kane, C. L. and E.J. Mele, E. J.,
Phys. Rev. Lett. \textbf{95},  146802 (2005).

\bibitem{Hatsugai1993}
Y. Hatsugai,
Phys. Rev. Lett. \textbf{71}, 3697 (1993).

\bibitem{Qi2006}
 X.-L. Qi, Y.-S. Wu  and S.-C. Zhang,
Phys. Rev. B \textbf{74}, 085308 (2006).

\bibitem{Motomechiral2010}
M. Udagawa and  Y. Motome,
Phys. Rev. Lett. \textbf{104}, 106409 (2010).

\bibitem{ns07}
B. Nachtergaele, R. Sims, Commun. Math. Phys. {\bf 276}, 437 (2007).

\bibitem{kptv12}
I. Kimchi, S. A. Parameswaran, A. M. Turner, A. Vishwanath,  arXiv:1207.0498.


\bibitem{mlwam10}
Z. Y. Meng, T. C. Lang, S. Wessel, F. Assaad, A. Muramatsu,
Nature {\bf 464}, 847 (2010).


\bibitem{wang10}
F. Wang, Phys. Rev. B {\bf 82}, 024419 (2010).

\bibitem{ashcml11}
A. F. Albuquerque, D. Schwandt, B. Hetényi, S. Capponi, M. Mambrini and A. M. Läuchli, Phys. Rev. B {\bf 84} 024406 (2011).

\bibitem{clr11}
D. C. Cabra, C. A. Lamas, H. D. Rosales, Phys. Rev. B {\bf 83} 094506 (2011).

\bibitem{clark11}

Phys. Rev. Lett. {\bf 107}, 087204 (2011).


\bibitem{yacls12}
H-Y. Yang, A. F. Albuquerque, S. Capponi, A. M. Läuchli, K. P. Schmidt,
New J. Phys. {\bf 14}, 115027 (2012).

\bibitem{soy12}
S. Sorella, Y. Otsuka and S. Yunoki, arXiv:1207.1783.

\bibitem{Lauchli2005}
A. Laüchli, J. C. Domenge, C. Lhuillier, P. Sindzingre and	M. Troyer,
Phys. Rev. Lett. \textbf{95}, 137206 (2005).

 \bibitem{Momoi2006}
T. Momoi, P. Sindzingre and N. Shannon,
Phys. Rev. Lett., \textbf{97}, 257204 (2006)
  


\bibitem{FukuyamaKITP2012}
H. Fukuyama, 
oral communication at the
KITP  Conference: ``Exotic Phases of Frustrated Magnets'' (Oct 8-12, 2012).
Available online at:
\href{http://online.kitp.ucsb.edu/online/fragnets-c12/fukuyama/}{http://online.kitp.ucsb.edu/online/fragnets-c12/fukuyama/}. Unpublished.


\end{thebibliography}

\end{document}